\documentclass[a4paper,11pt]{article}
\pdfoutput=1
\usepackage{amsmath,amssymb,amsfonts,amsthm,latexsym}
\usepackage{jheppub}
\usepackage{bm}
\usepackage{bbold}
\usepackage{xcolor}
\usepackage[normalem]{ulem}
\usepackage[T1]{fontenc}

\allowdisplaybreaks

\title{Decay and electromagnetic production of strongly coupled quarkonia in pNRQCD}
\preprint{TUM-EFT 126/19}

\author[a,b]{Nora~Brambilla,}
\author[a,c]{Hee~Sok~Chung,}
\author[a]{Daniel~M\"uller}
\author[a]{and Antonio~Vairo}

\affiliation[a]{Physik-Department, Technische Universit\"at M\"unchen, James-Franck-Str. 1, 85748 Garching, Germany}
\affiliation[b]{Institute for Advanced Study, Technische Universit\"at M\"unchen, Lichtenbergstrasse 2~a, 85748 Garching, Germany}
\affiliation[c]{Excellence Cluster ORIGINS,
Boltzmannstrasse 2, D-85748 Garching, Germany}

\emailAdd{nora.brambilla@tum.de}
\emailAdd{heesok.chung@tum.de}
\emailAdd{dan.mueller@tum.de}
\emailAdd{antonio.vairo@tum.de}

\abstract{We improve the pNRQCD approach to annihilation processes of heavy
quarkonium and make first pNRQCD predictions for exclusive electromagnetic
production of heavy quarkonium.
  We consider strongly coupled quarkonia, i.e., quarkonia that are not Coulombic bound states.
  Possible strongly coupled quarkonia include excited charmonium and bottomonium states.
  For these, pNRQCD provides expressions for the decay and exclusive electromagnetic production NRQCD matrix elements that depend on the wavefunctions at the origin and few universal gluon field correlators.
  We compute electromagnetic decay widths and exclusive production cross sections,
  and inclusive decay widths into light hadrons for $P$-wave quarkonia at relative order $v^2$ and leading order, respectively.
  We also compute the decay widths of $2S$ and $3S$ bottomonium states into lepton pairs and their ratios with the inclusive widths into light hadrons at relative order $v^2$.}   

\begin{document} 
\maketitle
\flushbottom

\section{Introduction}
\label{sec:intro}
Exclusive electromagnetic processes involving heavy quarkonia are good probes of quarkonium production and decay mechanisms: 
the clean final states enable accurate measurements~\cite{Brambilla:2004wf,Brambilla:2010cs} 
and nonrelativistic effective field theories~\cite{Bodwin:1994jh,Brambilla:2004jw}
allow to express physical observables through systematic expansions and factorization formulas.
Systematic expansions guarantee a control over the accuracy of the theoretical expressions,
and factorization casts non perturbative contributions into few {\em long-distance matrix elements} (LDMEs).
LDMEs are eventually determined from data.
Heavy quarkonium annihilations into light hadrons are similarly well under control from the theoretical side,
but more difficult to determine experimentally as contributions from decay channels into leptons, photons or
heavy quarks have to be subtracted from the total width.
Heavy quarkonium production in hadron collisions is challenging both theoretically and experimentally.

Heavy quarkonia, like charmonia and bottomonia, are nonrelativistic bound states of a heavy quark and a heavy antiquark.
Nonrelativistic effective field theories exploit the typical hierarchy of energy scales characterizing such systems.
The energy scales are the heavy quark mass, $m$, the typical momentum and momentum transfer, $mv$, and
the typical kinetic energy, $mv^2$, where $v$ is the relative velocity of the heavy quark and antiquark.
Because $v \ll 1$, the above energy scales are hierarchically ordered.
Reference values for $v$ are $v^2\approx 0.3$ for charmonia and $v^2\approx 0.1$ for bottomonia. 

{\em Nonrelativistic QCD} (NRQCD) is the effective field theory, suited to describe states made of a heavy quark and a heavy antiquark, 
that follows from QCD by integrating out modes of energy and momentum of order $m$~\cite{Bodwin:1994jh}.
In NRQCD, heavy quarkonium annihilation rates into photons, leptons or inclusive annihilation rates into light hadrons, $\Gamma$,
and exclusive electromagnetic production cross sections, $\sigma$, are expressed by sums of products of
NRQCD LDMEs, $\langle {\cal O}_n \rangle$, with perturbative short distance coefficients, $c_n$:
\begin{equation}
\Gamma/\sigma = \sum_n \frac{c_n(\Gamma/\sigma)}{m^{d_n-4}} \, \langle {\cal O}_n \rangle,
\label{GammasigmaNRQCD}
\end{equation}
where $d_n$ is the mass dimension of the operator ${\cal O}_n$. 
The short distance coefficients are process dependent.
The LDMEs depend on the quarkonium state, but not on the process.
Whereas the short distance coefficients can be computed as a series expansion in the strong coupling constant $\alpha_{\rm s}$, this is guaranteed by $m \gg \Lambda_{\rm QCD}$,
the LDMEs are counted in powers of  $v$.
In practice, the factorization formula is truncated at a desired order in $\alpha_{\rm s}$ and $v$.
In the case of decay widths, the short distance coefficients $c_n$ are dimensionless, 
while in the case of exclusive electromagnetic production cross sections, they have mass dimension $-3$ and depend on $m$ and on the center of mass energy of the collision, $\sqrt{s}$.
The center of mass energy is, besides the heavy quark mass, the other large scale in production processes.

The NRQCD factorization formulas for quarkonium inclusive annihilation widths into light hadrons, electromagnetic annihilation widths
and exclusive electromagnetic production cross sections have been proved to all orders in the expansion parameters.
Early determinations of several of the short distance coefficients can be found in~\cite{Bodwin:1994jh,Petrelli:1997ge,MaltoniPhD:1999} and in the review~\cite{Vairo:2003gh}.
These have been constantly improved over the last years (see appendix~\ref{app:NRQCDfact} and references therein).

The NRQCD LDMEs entering quarkonium annihilation and exclusive electromagnetic production are expectation values of four-fermion operators on the quarkonium state. 
A list of four-fermion operators relevant for the present work is in appendix~\ref{app4fermion}.
One important feature of NRQCD is that the quarkonium state can contain contributions not only
from the leading color-singlet heavy quark-antiquark Fock state, but also from the subleading Fock states that include effects of dynamical gluons. 
Because gluons carry color, the heavy quark-antiquark pair in the subleading Fock states can be in a color octet state. 
Hence, four-fermion operators projecting on color octet states contribute to the observables.
Determinations of these contributions provide important verifications of the NRQCD factorization formalism. 

In order to make quantitative statements based on the NRQCD factorization formulas, it is important to be able to determine the LDMEs. 
This is a difficult task in the standard NRQCD approach, especially for the LDMEs of higher orders in $v$, which, as a result, are poorly known.
Also the power counting of the LDMEs is not unique as they depend on several, still dynamical, energy scales: $mv$, $mv^2$,  the typical hadronic scale $\Lambda_{\rm QCD}$, ...~.
Disentangling these energy scales in a suitable nonrelativistic effective field theory of lower energy than NRQCD 
provides a way to simplify and in some cases compute the NRQCD LDMEs.

\begin{figure}[ht]
\begin{center}
\includegraphics[width=.5\textwidth]{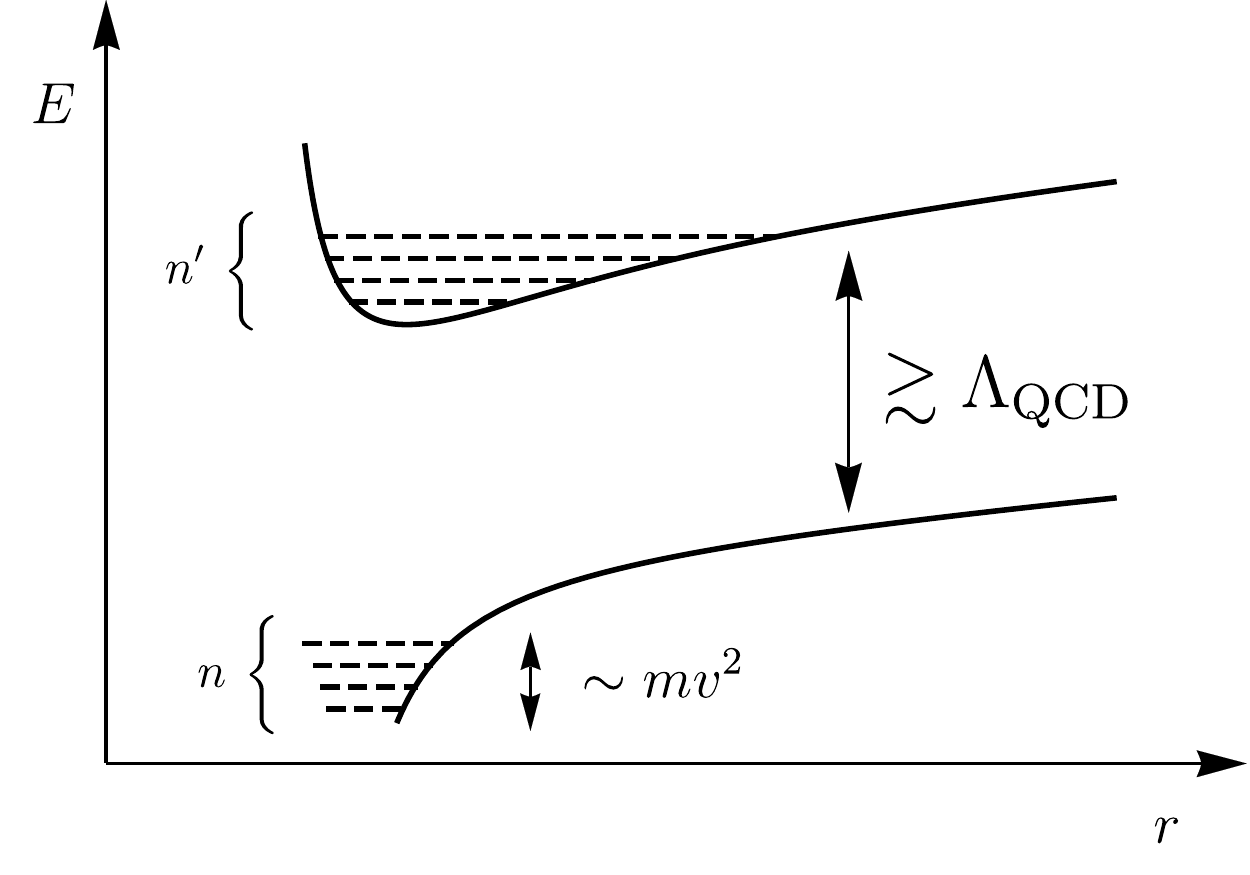}
\caption{Schematic distribution of energy levels associated with a heavy quark-antiquark pair and its first gluonic excitation with respect to the corresponding static potentials.
}
\label{fig:gap}
\end{center}
\end{figure}

{\em Potential NRQCD} (pNRQCD) follows from NRQCD by integrating out modes associated with energy scales larger than $mv^2$,
regardless of these energy scales being perturbative or non perturbative~\cite{Brambilla:1999xf}.
If all relevant energy scales are perturbative, the LDMEs can be expressed in pNRQCD as series in $\alpha_{\rm s}$~\cite{GarciaiTormo:2004jw}.
If $m v^2 \ll \Lambda_{\rm QCD}$, then the LDMEs are non perturbative.
In the non perturbative, confining, regime a heavy quark-antiquark pair may bind into
a quarkonium, i.e., a quark-antiquark pair in a color singlet configuration,
a hybrid, i.e., a quark-antiquark pair in a color octet configuration bound to gluons,
a quarkonium in the presence of glueballs,
a tetraquark, i.e., a heavy quark-antiquark pair bound in different combinations with a light quark-antiquark pair, and so on.
We will consider quarkonia that are well below the open flavor threshold, i.e., separated from it by an energy gap of order $\Lambda_{\rm QCD}$ or larger.
Moreover, lattice computations suggest that the quarkonium spectrum may be separated by an energy gap of order $\Lambda_{\rm QCD}$ or larger
from the spectrum of hybrids and quarkonia plus glueballs~\cite{Bali:2000vr,Juge:2002br,Capitani:2018rox}.
The distribution of energy levels would then be schematically the one shown in figure~\ref{fig:gap}.
If the kinematical condition $mv^2 \ll \Lambda_{\rm QCD}$ is realized, 
the higher gluonic excitations can be integrated out, and this leaves the
quark-antiquark pair in a color singlet configuration as the only dynamical 
degree of freedom. 
In this situation, the LDMEs can be factorized in a wavefunction contribution, which encodes information from the quarkonium state,
and some universal correlators of gluon fields, which encode contributions coming from higher excitations of the heavy quark-antiquark pair, 
those induced by gluons or light quarks and separated by an energy gap of order $\Lambda_{\rm QCD}$ from the quarkonium spectrum~\cite{Brambilla:2001xy,Brambilla:2002nu}.
The quarkonium wavefunction is obtained by solving the Schr\"odinger equation that is the equation of motion of pNRQCD.
The quarkonium potential may be expressed in terms of Wilson loops and gluon field insertions on it~\cite{Brambilla:2000gk,Pineda:2000sz}. 
It includes contributions coming from quarkonium modes of order $mv$ and from higher excitations of the heavy quark-antiquark pair
induced by gluons or light quarks and separated by an energy gap of order $\Lambda_{\rm QCD}$ from the quarkonium spectrum.
Under the kinematical condition $m v^2 \ll \Lambda_{\rm QCD}$, the potential is sensitive to distances of order $1/\Lambda_{\rm QCD}$, where it is non perturbative.
Hence, quarkonium satisfying the condition $m v^2 \ll \Lambda_{\rm QCD}$  is not a Coulombic bound state. 
To distinguish it from a Coulombic bound state, which is weakly coupled, a non Coulombic quarkonium state is referred to as {\em strongly coupled}.
Moreover, pNRQCD in the regime $m v^2 \ll \Lambda_{\rm QCD}$, is called {\em strongly coupled pNRQCD}.

In this work, we will focus on decay and exclusive electromagnetic production of quarkonium states for which the condition $m v^2 \ll \Lambda_{\rm QCD}$ is fulfilled. 
Under this condition, we will use pNRQCD to express the LDMEs in terms of quarkonium wavefunctions, binding energies and gluonic correlators. 
Ideally the quarkonium wavefunctions and binding energies should be determined from the solution of the pNRQCD Schr\"odinger equation, 
which requires knowing the quarkonium potential from lattice QCD.
Also the gluonic correlators should be computed on the lattice. 
In practice, however, due to the incomplete knowledge of these quantities, 
wavefunctions and binding energies are determined using potential models and the gluonic correlators from the data.
Since the gluonic correlators are non perturbative but universal quantities that do not depend on the heavy quark flavor, 
the non perturbative parameters needed in pNRQCD are in general fewer than the LDMEs needed in NRQCD.
When applicable, the pNRQCD factorization formulas have, therefore, more predictive power than the NRQCD ones.

The condition $m v^2 \ll \Lambda_{\rm QCD}$ is fulfilled by non Coulombic, strongly coupled, quarkonia.\footnote{
The quark-antiquark potential is Coulombic under the condition $mv^2 \gtrsim \Lambda_{\rm QCD}$, in which case, $v\sim \alpha_{\rm s}$.}
Charmonium and bottomonium states that are possibly non Coulombic bound states are higher excited states whose principal quantum number is greater than one. 
These states will be the subject of our phenomenological investigations in section~\ref{sec:results}.

The paper is organized in the following way. 
We start in section~\ref{sec:formalism} by briefly reviewing strongly coupled pNRQCD. 
In the following section~\ref{sec:LDME}, we compute in strongly coupled pNRQCD the relevant LDMEs and give their explicit expressions
in terms of quarkonium wavefunctions, binding energies and gluonic correlators.
Four-fermion operators are listed in appendix~\ref{app4fermion} and NRQCD factorization formulas in appendix~\ref{app:NRQCDfact}.
Details of the computation are in appendix~\ref{sec:LDMEapp}.
Some of the results presented in section~\ref{sec:LDME} correct previous findings of ref.~\cite{Brambilla:2002nu}.
Using these results and after having determined the wavefunctions and binding energies in several potential models,
we fit the unknown gluonic correlators and compute decay widths and exclusive electromagnetic production cross sections for charmonium $1P$
states and bottomonium $2S$, $3S$, $1P$, $2P$ and $3P$ states in section~\ref{sec:results}. 
We conclude in section~\ref{sec:summary}.

\section{Effective field theories for strongly coupled quarkonia}
\label{sec:formalism}
We compute the LDMEs of NRQCD assuming that the quarkonium states that we consider are well below the open flavor threshold and satisfy the condition $m v^2 \ll \Lambda_{\rm QCD}$.
We further assume that higher gluonic excitations of the heavy quark-antiquark pair are separated by an energy gap of order $\Lambda_{\rm QCD}$ or larger from the ground state;
this assumption is supported by lattice calculations that show the excitation spectrum of the gluon field around a static quark-antiquark pair
separated by a large energy gap from the ground state~\cite{Bali:2000vr,Juge:2002br,Capitani:2018rox}.
It follows that we can picture the distribution of the energy levels as illustrated in figure~\ref{fig:gap}.
Such picture allows us to describe the quarkonium spectrum in an effective field theory where all modes associated to the excitations of the heavy quark-antiquark pair
induced by gluons or light quarks and separated by an energy gap of order $\Lambda_{\rm QCD}$ from the quarkonium spectrum have been integrated out.
This effective field theory is strongly coupled pNRQCD~\cite{Brambilla:2004jw}.

In strongly coupled pNRQCD, the quarkonium potential and the LDMEs are computed by quantum mechanical perturbation theory, order by order in $1/m$, around the NRQCD static solution.
Each power of $1/m$ is suppressed by $v$ or $\Lambda_{\rm QCD}/m$.
The computation of the quarkonium potential in strongly coupled pNRQCD has been first performed in ref.~\cite{Brambilla:2000gk,Pineda:2000sz},
and the NRQCD LDMEs have been computed in ref.~\cite{Brambilla:2001xy,Brambilla:2002nu,Brambilla:2003mu}.
In this section, we briefly review the formalism and compute the potentials relevant for the present work.
We compute the LDMEs in section~\ref{sec:LDME}.

\subsection{NRQCD}
\label{sec:stcoNRQCD}
The degrees of freedom of NRQCD are heavy quark and antiquark fields, $\psi$ and $\chi$, describing modes of energy and momentum smaller than $m$, gluons and light quarks.
The NRQCD Hamiltonian, $H_{\rm NRQCD}$, can be organized as an expansion in $1/m$, so that $H_{\rm NRQCD} = H^{(0)}_{\rm NRQCD} + H^{(1)}_{\rm NRQCD}/m + \dots$, where  
\begin{eqnarray}
\label{eq:hamiltonian0}
H^{(0)}_{\rm NRQCD} &=& \frac{1}{2}\int d^{3}x\,(\bm{E}^{a}\!\cdot\bm{E}^{a}+\bm{B}^{a}\!\cdot\bm{B}^{a}) - \sum_{k=1}^{n_f} \int d^3x \,\bar{q_k} i\bm{D}\cdot\bm{\gamma} q_k,
\\
\label{eq:hamiltonian1}
H^{(1)}_{\rm NRQCD} &=&  
- \frac{1}{2} \int d^3x \,\psi^{\dagger}\bm{D}^2\psi -\frac{c_F}{2}\, \int d^{3}x\,\psi^{\dagger}\bm{\sigma}\cdot g\bm{B}\psi
\nonumber \\ && \hspace{6ex}
+ \frac{1}{2}\int d^{3}x\,\chi^{\dagger}\bm{D}^2\chi + \frac{c_F}{2}\, \int d^{3}x\,\chi^{\dagger}\bm{\sigma}\cdot g\bm{B}\chi.
\end{eqnarray}
Boldfaced characters indicate three-dimensional vectors.
The fields $\psi$ and $\chi$ respectively annihilate a heavy quark and create a heavy antiquark,    
the fields $q_k$ are $n_f$ massless quark fields, $\bm{D} = \bm{\nabla} -ig\bm{A}$ is the gauge covariant derivative,
$E^{i\,a}T^a = E^i =  G^{i0}$ and $B^{i\,a}T^a = B^i = -\epsilon_{ijk}G^{jk}/2$ are the chromoelectric and chromomagnetic fields, respectively,
$G^{\mu\nu\,a}T^a = G^{\mu\nu}$ is the gluon field strength tensor, $\sigma^i$ are the Pauli matrices  
and $c_F$ is a short distance coefficient, which is known up to three loops~\cite{Grozin:2007fh}.
The fields $\psi$ and $\chi$ satisfy the canonical equal time anticommutation relations:
$\{\psi_\alpha(\bm{x}),\psi^\dagger_\beta(\bm{y})\} = \{\chi_\alpha(\bm{x}),\chi^\dagger_\beta(\bm{y})\} =  \delta_{\alpha\beta}\delta^{(3)}(\bm{x}-\bm{y})$,
$\{\psi_\alpha(\bm{x}),\psi_\beta(\bm{y})\} = \{\psi^\dagger_\alpha(\bm{x}),\psi^\dagger_\beta(\bm{y})\} = 
\{\chi_\alpha(\bm{x}),\chi_\beta(\bm{y})\} = \{\chi^\dagger_\alpha(\bm{x}),\chi^\dagger_\beta(\bm{y})\} = 0$.
The mass $m$ is the heavy quark pole mass.

We restrict ourselves to the one-quark--one-antiquark sector of the NRQCD Fock space, where quarkonium states live. 
In this sector, we denote an energy eigenstate of the NRQCD Hamiltonian by $| \underline{\rm n}; \bm{x}_1, \bm{x}_2 \rangle$,
where $n$ represents a generic set of conserved quantum numbers, and $\bm{x}_1$ and $\bm{x}_2$ are the positions of the quark and antiquark, respectively.
The heavy quark and antiquark positions are conserved quantum numbers in the static limit.
We normalize the states as
$\langle \underline{\rm n}; \bm{x}_1, \bm{x}_2 | \underline{\rm m}; \bm{x}_1',\bm{x}_2' \rangle = \delta_{nm} \delta^{(3)} (\bm{x}_1- \bm{x}_1') \delta^{(3)} (\bm{x}_2- \bm{x}_2')$. 
The eigenstates satisfy the Schr\"odinger equation:
\begin{equation}
H_{\rm NRQCD} | \underline{\rm n}; \bm{x}_1, \bm{x}_2 \rangle = 
\int d^3x_1' d^3x_2' \, | \underline{\rm n}; \bm{x}_1', \bm{x}_2' \rangle E_n (\bm{x}_1', \bm{x}_2';\bm{\nabla}_1', \bm{\nabla}_2') \delta^{(3)} (\bm{x}_1'-\bm{x}_1) \delta^{(3)} (\bm{x}_2'-\bm{x}_2), 
\end{equation}
where $\bm{\nabla}_1 \equiv \bm{\nabla}_{\bm{x}_1}$, $\bm{\nabla}_2 \equiv \bm{\nabla}_{\bm{x}_2}$, $\bm{\nabla}_1' \equiv \bm{\nabla}_{\bm{x}_1'}$ and $\bm{\nabla}_2' \equiv \bm{\nabla}_{\bm{x}_2'}$.
The ground state, $n=0$, is the Fock state made of a heavy quark-antiquark pair
without excitations induced by gluons or light quarks and separated by an energy gap of order $\Lambda_{\rm QCD}$ from the quarkonium spectrum;
other values of $n$ identify heavy quark-antiquark pairs in the presence of such excitations.
The functions $E_n$ are the corresponding energies.
In the static limit the above equation becomes
\begin{equation}
  H^{(0)}_{\rm NRQCD} | \underline{\rm n}; \bm{x}_1, \bm{x}_2 \rangle^{(0)} = 
\int d^3x_1' d^3x_2' \, | \underline{\rm n}; \bm{x}_1', \bm{x}_2' \rangle^{(0)} E^{(0)}_n (\bm{x}_1', \bm{x}_2') \delta^{(3)} (\bm{x}_1'-\bm{x}_1) \delta^{(3)} (\bm{x}_2'-\bm{x}_2). 
\end{equation}
The static energies $E^{(0)}_n$ depend only on the quark-antiquark distance, $r = |\bm{x}_1-\bm{x}_2|$;  
$E^{(0)}_0$ identifies the ground state static energy between a quark and antiquark in a color singlet configuration,
which is well approximated by a Cornell-like potential (see section~\ref{sec:potentials}),
$E^{(0)}_n$ for $n\neq 0$ are the static energies of a quark-antiquark pair in the presence of the excitations described above.
The ground state energy and the first gluonic excitation are schematically shown in figure~\ref{fig:gap}.
The eigenstates  $| \underline{\rm n}; \bm{x}_1, \bm{x}_2\rangle^{(0)}$ are normalized as 
${}^{(0)} \langle \underline{\rm n}; \bm{x}_1, \bm{x}_2 | \underline{\rm m}; \bm{x}_1', \bm{x}_2'\rangle^{(0)}$ $ = \delta_{nm} \delta^{(3)} (\bm{x}_1- \bm{x}_1') \delta^{(3)} (\bm{x}_2- \bm{x}_2')$. 

Having assumed that the gap between the lowest-lying energy, $E^{(0)}_0$, and the higher ones is of order $\Lambda_{\rm QCD} \gg mv^2$,
we can compute $ | \underline{\rm 0}; \bm{x}_1, \bm{x}_2 \rangle$ by expanding in $1/m$ around the static solution $| \underline{\rm 0}; \bm{x}_1, \bm{x}_2 \rangle^{(0)}$:
\begin{equation}
  | \underline{\rm 0}; \bm{x}_1, \bm{x}_2 \rangle = | \underline{\rm 0}; \bm{x}_1, \bm{x}_2 \rangle^{(0)} +  \frac{1}{m} | \underline{\rm 0}; \bm{x}_1, \bm{x}_2 \rangle^{(1)} + \dots\,. 
\end{equation}
Corrections are obtained from quantum mechanical perturbation theory applied to the NRQCD Hamiltonian. 
In particular, $| \underline{\rm 0}; \bm{x}_1, \bm{x}_2 \rangle^{(1)}$, which is the correction to the eigenstate at order $1/m$, reads 
\begin{equation}
| \underline{\rm 0}; \bm{x}_1, \bm{x}_2 \rangle^{(1)} = - \sum_{n \neq 0} \int d^3x_1' d^3x_2'\, | \underline{\rm n} ; \bm{x}_1', \bm{x}_2' \rangle^{(0)} 
\frac{{}^{(0)} \langle \underline{\rm n}; \bm{x}_1', \bm{x}_2' | H^{(1)}_{\rm NRQCD} | \underline{\rm 0}; \bm{x}_1, \bm{x}_2 \rangle^{(0)}}{E_n^{(0)} (\bm{x}_1', \bm{x}_2') - E_0^{(0)} (\bm{x}_1, \bm{x}_2)}, 
\end{equation}
where $H^{(1)}_{\rm NRQCD}$ is given in eq.~\eqref{eq:hamiltonian1}.

The explicit expression for the correction term $| \underline{\rm 0}; \bm{x}_1, \bm{x}_2 \rangle^{(1)}$ has been obtained in refs.~\cite{Brambilla:2000gk,Pineda:2000sz}.
Here, we briefly list the main ingredients to obtain it, as we will use them to compute the LDMEs.
In order to obtain  $| \underline{\rm 0}; \bm{x}_1, \bm{x}_2 \rangle^{(1)}$, we need to evaluate the matrix elements 
${}^{(0)} \langle \underline{\rm n}; \bm{x}_1', \bm{x}_2' | H^{(1)}_{\rm NRQCD}|\underline{\rm 0}; \bm{x}_1, \bm{x}_2 \rangle^{(0)}$. 
The first step is to make the quark content of the eigenstates explicit: 
\begin{equation}
\label{eq:gluonstate}
| \underline{\rm n}; \bm{x}_1, \bm{x}_2 \rangle^{(0)} =  \psi^\dag (\bm{x}_1) \chi(\bm{x}_2) |n; \bm{x}_1, \bm{x}_2 \rangle^{(0)}, 
\end{equation}
where $|n; \bm{x}_1, \bm{x} \rangle^{(0)}$ encodes the light degrees of freedom content of $|\underline{\rm n}; \bm{x}_1, \bm{x}_2 \rangle^{(0)}$.
The states $|n; \bm{x}_1, \bm{x}_2 \rangle^{(0)}$ also diagonalize $H^{(0)}$:
\begin{equation}
H^{(0)} | n; \bm{x}_1, \bm{x}_2 \rangle^{(0)} = | n; \bm{x}_1, \bm{x}_2 \rangle^{(0)} E_n^{(0)} (\bm{x}_1, \bm{x}_2)\,.
\end{equation}
The states $|n; \bm{x}_1, \bm{x}_2 \rangle^{(0)}$ do not contain heavy (anti)quarks, hence they are annihilated by $\psi$ and $\chi^\dag$. 
This implies the normalization ${}^{(0)} \langle n; \bm{x}_1, \bm{x}_2 | m; \bm{x}_1, \bm{x}_2 \rangle^{(0)} = \delta_{nm}$. 
Then, the matrix elements ${}^{(0)} \langle \underline{\rm n}; \bm{z}_1, \bm{z}_2 | H^{(1)}_{\rm NRQCD}| \underline{\rm 0}; \bm{x}_1, \bm{x}_2 \rangle^{(0)}$ 
can be computed by using Wick's theorem, which removes the quark and antiquark fields leaving delta functions that constrain $\bm{x}_1 = \bm{x}_1'$ and $\bm{x}_2 = \bm{x}_2'$. 
After the quark and antiquark fields have been removed in this way, we can use the following shorthands without ambiguity:
$|n\rangle^{(0)} \equiv | n; \bm{x}_1, \bm{x}_2\rangle^{(0)}$, $E_n^{(0)} \equiv E_n^{(0)} (\bm{x}_1, \bm{x}_2)$, 
$\bm{D}_1 \equiv \bm{D} (\bm{x}_1)$, $\bm{D}_{c2} \equiv \bm{D}_c (\bm{x}_2)$, $\bm{E}_1 \equiv \bm{E} (\bm{x}_1)$, and $\bm{E}_2 \equiv \bm{E} (\bm{x}_2)$,
where $\bm{D}_c = \bm{\nabla} + i g \bm{A}^T$ is the charge conjugate of $\bm{D}$.
Finally, the following identities can be used to simplify the matrix elements
\begin{eqnarray}
\label{eq:Didentity}
&&
{}^{(0)}\langle n | \bm{D}_1 | n \rangle^{(0)}    = \bm{\nabla}_1, 
\qquad\qquad\qquad\quad 
{}^{(0)}\langle n | \bm{D}_{c2 } | n \rangle^{(0)} = \bm{\nabla}_2, 
\\
\label{eq:Eidentity}
&& {}^{(0)}\langle n | g \bm{E}_1 | n\rangle^{(0)} = - \left( \bm{\nabla}_1 E_n^{(0)}  \right) , 
\qquad\quad 
{}^{(0)}\langle n | g \bm{E}_2^T | n \rangle^{(0)} = \left(\bm{\nabla}_2 E_n^{(0)} \right), 
\end{eqnarray}
and, for $n \neq k$,  
\begin{equation}
\label{eq:Didentities_neq}
{}^{(0)}\langle n | \bm{D}_1 | k \rangle^{(0)} =  \frac{ {}^{(0)}\langle n | g \bm{E}_1 | k \rangle^{(0)}}{E_n^{(0)}  - E_k^{(0)} },
\qquad\quad 
{}^{(0)}\langle n | \bm{D}_{c2} | k \rangle^{(0)} = - \frac{ {}^{(0)}\langle n | g \bm{E}^T_2 | k \rangle^{(0)}}{E_n^{(0)}  - E_k^{(0)} }.
\end{equation}
Equations~\eqref{eq:Didentity} follows from symmetry considerations, and 
eqs.~(\ref{eq:Eidentity}) and  (\ref{eq:Didentities_neq}) may be derived from 
canonical commutation relations~\cite{Brambilla:2000gk,Brambilla:2004jw}. 
The parentheses on the right-hand sides of eqs.~\eqref{eq:Eidentity} imply that the derivatives act only on $E_n^{(0)}$. 
These ingredients are sufficient to derive the explicit expression of $| \underline{\rm 0}; \bm{x}_1, \bm{x}_2 \rangle^{(1)}$. 

In this work, the state $| \underline{\rm 0}; \bm{x}_1, \bm{x}_2 \rangle^{(1)}$  will turn out to be relevant only for $P$-wave states.
For $P$-wave states only terms containing derivatives acting on the wavefunctions give nonvanishing contributions.
For our purposes it is sufficient, therefore, to isolate from  $| \underline{\rm 0}; \bm{x}_1, \bm{x}_2 \rangle^{(1)}$ only this part, 
which we denote with $|\underline{\rm 0}; \bm{x}_1, \bm{x}_2 \rangle^{(1)}_{P\text{-wave}}$.
Its explicit expression reads
\begin{equation}
\label{eq:state_pwave}
| \underline{\rm 0}; \bm{x}_1, \bm{x}_2 \rangle^{(1)}_{P\text{-wave}} = - \sum_{n \neq 0} |\underline{\rm n}; \bm{x}_1, \bm{x}_2\rangle^{(0)}
\left[\frac{{}^{(0)}\langle n| g\bm{E}_{1}|0\rangle^{(0)}}{(E^{(0)}_{0}-E^{(0)}_{n})^2}\cdot \overleftarrow{\bm{\nabla}}_{1}
- \frac{{}^{(0)}\langle n| g\bm{E}_{2}^T|0\rangle^{(0)}}{(E^{(0)}_{0}-E^{(0)}_{n})^2}\cdot \overleftarrow{\bm{\nabla}}_{2}\right].
\end{equation}

Also the energy eigenstate $E_0$ of the NRQCD Hamiltonian can be organized as an expansion in $1/m$:
\begin{equation}
  E_0(\bm{x}_1, \bm{x}_2; \bm{\nabla}_1,\bm{\nabla}_2) =  E_0^{(0)}(\bm{x}_1, \bm{x}_2) +  \frac{1}{m}  E_0^{(1)}(\bm{x}_1, \bm{x}_2; \bm{\nabla}_1,\bm{\nabla}_2)  + \dots\,. 
\end{equation}
Following~\cite{Brambilla:2000gk,Pineda:2000sz}, the static energy $E_0^{(0)}$ can be identified with the exponential fall off at large times of a rectangular Wilson loop:  
\begin{equation}
E_0^{(0)}(\bm{x}_1, \bm{x}_2) = \lim_{T\to\infty}\frac{i}{T} \ln \langle W_{r\times T} \rangle,
\label{E0}
\end{equation}
where $\bm{r} = \bm{x}_1 - \bm{x}_2$ is the space extension and $T$ is the time extension of the rectangular Wilson loop $W_{r\times T}$;
$\langle \cdots \rangle$ stands for the normalized QCD functional integral.
The first correction $E_0^{(1)}$ reads 
\begin{eqnarray}
  E_0^{(1)}(\bm{x}_1, \bm{x}_2; \bm{\nabla}_1,\bm{\nabla}_2) &=& - \frac{\bm{\nabla}_1^2}{2}- \frac{\bm{\nabla}_2^2}{2}
                                                                 \nonumber\\
                                                             && \hspace{-3cm}
+ \frac{1}{2} \sum_{n\neq 0} \left\vert  \frac{ {}^{\,\,(0)} \langle n | g \bm{E}_1 | 0 \rangle^{(0)}}{E_0^{(0)} - E_n^{(0)}} \right\vert^2
+ \frac{1}{2} \sum_{n\neq 0} \left\vert  \frac{ {}^{\,\,(0)} \langle n | g \bm{E}_2^T | 0 \rangle^{(0)}}{E_0^{(0)} - E_n^{(0)}} \right\vert^2.
\label{E1}
\end{eqnarray}
The second line of eq.~\eqref{E1} may be conveniently reexpressed in terms of functional integrals of Wilson loop operators as 
\begin{eqnarray}
  E_0^{(1)}(\bm{x}_1, \bm{x}_2; \bm{\nabla}_1,\bm{\nabla}_2) &=&
                                                                 - \frac{\bm{\nabla}_1^2}{2}- \frac{\bm{\nabla}_2^2}{2} \nonumber\\
&&\hspace{-3cm}
    - \int_{0}^{\infty} dt \, t \left[ \frac{\langle g\bm{E}_1(t) \cdot g\bm{E}_1(0) W_{r\times T}\rangle}{\langle W_{r\times T}\rangle}
    - \frac{\langle g\bm{E}_1(t) W_{r\times T}\rangle}{\langle W_{r\times T}\rangle} \cdot \frac{\langle g\bm{E}_1(0) W_{r\times T}\rangle}{\langle W_{r\times T}\rangle}
    \right],
\label{E1W}
\end{eqnarray}
where $\langle O_{1(2)}(t) W_{r\times T}\rangle$ means that the fields appearing on the left of $W_{r\times T}$ are inserted at a time $t$ on the quark (antiquark) line of the Wilson loop.
The Wilson loop and field insertions on it are traced in color space.

\subsection{Strongly coupled pNRQCD}
\label{sec:stcopNRQCD}
The degrees of freedom of strongly coupled pNRQCD, i.e., the degrees of freedom that are resolved at an energy scale of order $mv^2$ are only color singlet heavy quark-antiquark pairs, 
if we neglect the interaction with light hadrons of energy and momentum of order $mv^2$ or smaller (these are the Goldstone bosons of the chiral symmetry, for a discussion see ref.~\cite{Brambilla:2002nu}).
The reason is that, having assumed $mv^2 \ll \Lambda_{\rm QCD}$, the scale $mv^2$ is  below the confinement scale, $\Lambda_{\rm QCD}$. 
Hence, the pNRQCD Hamiltonian has the very simple form 
\begin{equation}
H_{\rm pNRQCD} = \int d^3x_1 d^3x_2 \, S^\dagger \, h(\bm{x}_1, \bm{x}_2; \bm{\nabla}_1,\bm{\nabla}_2) \,S\,,
\label{HpNRQCD}
\end{equation}
where $S$ annihilates a color singlet heavy quark-antiquark field. 
It satisfies the canonical equal time commutation relation: $[S(\bm{x}_1',\bm{x}_2'),S^\dagger(\bm{x}_1,\bm{x}_2)] = \delta(\bm{x}_1'-\bm{x}_1) \delta(\bm{x}_2'-\bm{x}_2)$.

Because of the assumed energy gap of order $\Lambda_{\rm QCD}$ between the ground state and the higher excitations of the heavy quark-antiquark pair, 
these have been integrated out when matching to strongly coupled pNRQCD, 
whereas the NRQCD ground state, $|\underline{\rm 0}; \bm{x}_1, \bm{x}_2 \rangle$, matches 
the pNRQCD state made of one color singlet heavy quark-antiquark pair, $S^\dagger(\bm{x}_1,$ $\bm{x}_2)|{\rm vac} \rangle$. 
If we neglect light hadrons of energy and momentum of order $mv^2$ or smaller, then $|{\rm vac} \rangle$ is the pNRQCD vacuum.

The function $h$ can be organized as an expansion in $1/m$: $h = h^{(0)} + h^{(1)}/m + \dots$~.
The terms $h^{(0)}$, $h^{(1)}$, ... can be determined by matching $h$ with the NRQCD ground state energy order by order in $1/m$:
\begin{eqnarray}
h^{(0)}(\bm{x}_1, \bm{x}_2) & \equiv& V^{(0)}(\bm{x}_1, \bm{x}_2) = E_0^{(0)}(\bm{x}_1, \bm{x}_2), \\
h^{(1)}(\bm{x}_1, \bm{x}_2; \bm{\nabla}_1,\bm{\nabla}_2) &\equiv&  - \frac{\bm{\nabla}_1^2}{2}- \frac{\bm{\nabla}_2^2}{2} + V^{(1)}(\bm{x}_1, \bm{x}_2) 
= E_0^{(1)}(\bm{x}_1, \bm{x}_2; \bm{\nabla}_1,\bm{\nabla}_2),
\end{eqnarray}  
and so on. Also in this case, like in the NRQCD case, the mass $m$ should be understood as the heavy quark pole mass.
The term $h^{(0)}$ is the static quark-antiquark potential $V^{(0)}$, which, according to \eqref{E0}, 
can be determined from the large time behavior of the static Wilson loop:
\begin{equation}
V^{(0)}(\bm{x}_1, \bm{x}_2) = \lim_{T\to\infty}\frac{i}{T} \ln \langle W_{r\times T} \rangle.
\label{V0}
\end{equation}
The term  $h^{(1)}$ contains the quark and antiquark kinetic energies, and the $1/m$ potential $V^{(1)}$, which,  
according to \eqref{E1W}, may be computed from a Wilson loop with two chromoelectric field insertions:
\begin{equation}
V^{(1)}(\bm{x}_1, \bm{x}_2) =
 - \int_{0}^{\infty} dt \, t \left[ \frac{\langle g\bm{E}_1(t) \cdot g\bm{E}_1(0) W_{r\times T}\rangle}{\langle W_{r\times T}\rangle}
    - \frac{\langle g\bm{E}_1(t) W_{r\times T}\rangle}{\langle W_{r\times T}\rangle} \cdot \frac{\langle g\bm{E}_1(0) W_{r\times T}\rangle}{\langle W_{r\times T}\rangle}
    \right]\,.
\label{V1}
\end{equation}

The eigenstates of $h$ are the solutions of the following Schr\"odinger equation in coordinate space
\begin{equation}
  h\, \langle \bm{R}|\bm{P} \rangle\, \langle \bm{r}|nJLS\rangle  = \varepsilon_{nJLS} \, \langle \bm{R}|\bm{P} \rangle\, \langle \bm{r}|nJLS\rangle \,,
\label{schroedinger}
\end{equation}
where $\bm{R} = (\bm{x}_1 + \bm{x}_2)/2$ is the center of mass coordinate, $\bm{r} = \bm{x}_1 - \bm{x}_2$ is the quark-antiquark distance, 
$\bm{P}$ is the center of mass momentum and $\varepsilon_{nJLS}$ is the binding energy. 
The states, $|\bm{P} \rangle\,|nJLS\rangle$, are classified  according to the center of mass momentum, the principal quantum number, $n$, the total angular momentum, $J$, 
the orbital angular momentum, $L$, and the spin, $S$.
At leading order, $h$ contains the kinetic energy, $ - \bm{\nabla}_1^2/(2m) - \bm{\nabla}_2^2/(2m)$, and the static potential, $V^{(0)}$, 
which both count like $mv^2$, as a consequence of the scale hierarchy in a nonrelativistic bound state and the virial theorem.\footnote{
The momenta $-i\bm{\nabla}_1$ and $-i\bm{\nabla}_2$ may be decomposed in a center of mass momentum and a relative momentum:
$-i\bm{\nabla}_1 = -i\bm{\nabla}_{\bm{r}} -i\bm{\nabla}_{\bm{R}}/2$ and $-i\bm{\nabla}_2 =  i\bm{\nabla}_{\bm{r}} - i\bm{\nabla}_{\bm{R}}/2 $.
The relative momentum, $-i\bm{\nabla}_{\bm{r}}$, scales like $mv$, while the center of mass momentum, $-i\bm{\nabla}_{\bm{R}}$, 
scales like the momentum of the dynamical low-energy degrees of freedom of the effective theory. 
Hence, the center of mass momentum scales at most like $mv^2$.
In the effective field theory of eq.~\eqref{HpNRQCD} that does not contain dynamical low-energy degrees of freedom besides 
the quark-antiquark color singlet, the reference frame may be always chosen so that the center of mass momentum is set to zero.
The kinetic energy entering the leading order pNRQCD Hamiltonian is, therefore, only the kinetic energy associated with the relative momentum: $-\bm{\nabla}_{\bm{r}}^2/m$.}
Also $\varepsilon_{nJLS}$ is of order $mv^2$.
For a strongly coupled bound state, the potential $V^{(1)}/m$ of eq.~\eqref{V1} can be of the same order as the static potential 
if $mv \sim \Lambda_{\rm QCD}$~\cite{Brambilla:2000gk,Pineda:2000sz}. Under this condition it should be included in the leading order~$h$.
Whatever the specific regime that we are describing is, the leading order potential is a central potential, i.e., it depends only on $r = |\bm{x}_1-\bm{x}_2|$. 
Hence, the leading order binding energy may be classified in terms of $n$ and $L$ only. 
The corresponding Schr\"odinger equation reads 
\begin{equation}
  h^{(0)} \, \langle \bm{R}|\bm{P} \rangle\, \langle \bm{r}|nJLS\rangle^{(0)}  = \varepsilon_{nL}^{(0)} \, \langle \bm{R}|\bm{P} \rangle\, \langle \bm{r}|nJLS\rangle^{(0)} \,,
\label{schroedingerLO}
\end{equation}
where $h^{(0)}$, $\varepsilon_{nL}^{(0)}$ and $|nJLS\rangle^{(0)}$ are respectively $h$, the binding energy and the eigenstate at leading order.
The center of mass wavefunction is a plane wave: $\langle \bm{R}|\bm{P} \rangle = \exp(-i\bm{R}\cdot\bm{P})$.

In summary, a generic strongly coupled quarkonium state $|H\rangle$ with quantum numbers $n$, $J$, $L$ and $S$ and center of mass momentum $\bm{P}$ is described in pNRQCD 
by a state 
\begin{equation} 
\frac{1}{\sqrt{\langle \bm{P} = \bm{0}| \bm{P} = \bm{0} \rangle}} 
\int d^3x_1 d^3x_2 \, \langle \bm{R}|\bm{P} \rangle\, \langle \bm{r}|nJLS\rangle \, S^\dagger(\bm{x}_1,\bm{x}_2)|{\rm vac}\rangle\,.
\label{pNRQCDHstate}
\end{equation}
The wavefunction $\langle \bm{R}|\bm{P} \rangle$ is equal to 1 in the center of mass frame, $\bm{P}=\bm{0}$.
The factor $1/\sqrt{\langle \bm{P} = \bm{0}| \bm{P} = \bm{0} \rangle}$ normalizes the state.
The wavefunction $\langle \bm{r}|nJLS\rangle$ is the solution of the Schr\"odinger equation \eqref{schroedinger}, 
whose static potential is given by \eqref{V0}, $1/m$ potential by \eqref{V1} and so on~\cite{Pineda:2000sz}. 
The field $S^\dagger(\bm{x}_1,\bm{x}_2)$ creates a heavy quark-antiquark pair in a color singlet configuration.

\section{LDMEs}
\label{sec:LDME}
Four-fermion operators show up in NRQCD at order $1/m^2$ or higher.
Some of them are listed in appendix~\ref{app4fermion}.
They match into contact terms of pNRQCD. 
The matching condition reads
\begin{eqnarray}
&&\langle\underline{\rm 0};\bm{x}_{1},\bm{x}_{2}|\int d^3x \, {\cal O}_n(\bm{x})|\underline{\rm 0};\bm{x}_1',\bm{x}_2'\rangle  
\nonumber\\
&&
= \langle {\rm vac}|S(\bm{x}_1,\bm{x}_2) \, \int d^3x d^3y \, S^\dagger(\bm{x},\bm{y})  [-V_{{\cal O}_n}^{(d_n-4)}(\bm{x}, \bm{y};  \bm{\nabla}_{\bm{x}}, \bm{\nabla}_{\bm{y}})]
S(\bm{x},\bm{y}) \, S^\dagger(\bm{x}_1',\bm{x}_2')|{\rm vac}\rangle
\nonumber\\
&&
= -V_{{\cal O}_n}^{(d_n-4)}(\bm{x}_1, \bm{x}_2;  \bm{\nabla}_1, \bm{\nabla}_2) \delta^{(3)} (\bm{x}_1 - \bm{x}_1') \delta^{(3)} (\bm{x}_2 - \bm{x}_2'),
\label{matchinglocal}
\end{eqnarray}
where ${\cal O}_n$ is a four-fermion operator in the NRQCD Lagrangian, $d_n$ is its dimension and $V_{{\cal O}_n}^{(d_n-4)}$ is a dimension $d_n-3$ contact term, i.e.,
a function of $\delta^3(\bm{r})$ or its derivatives.\footnote{
This is a generic feature that follows from the heavy quark and antiquark content of the state $|\underline{\rm 0};\bm{x}_1,\bm{x}_2\rangle$
and the structure of the four-fermion operator. For instance, for ${\cal O}_n(\bm{x}) = \psi^\dagger K \chi\,\chi^\dagger K \psi$ we have 
\begin{eqnarray}
&& ^{(0)}\langle\underline{\rm 0};\bm{x}_{1},\bm{x}_{2}|\int d^3x \, \psi^\dagger(\bm{x}) K(\bm{x},\bm{\nabla})\chi(\bm{x})\,\chi^\dagger(\bm{x}) K(\bm{x},\bm{\nabla}) \psi(\bm{x})
|\underline{\rm 0};\bm{x}_1',\bm{x}_2'\rangle^{(0)} 
\nonumber\\
&& \qquad \qquad \qquad \quad =
^{(0)}\langle0;\bm{x}_1,\bm{x}_2| K(\bm{x}_1,\bm{\nabla}_1)
\delta^{(3)} (\bm{x}_1 - \bm{x}_2) K(\bm{x}_1,\bm{\nabla}_1)
|0;\bm{x}_1,\bm{x}_2\rangle^{(0)}\delta^{(3)}(\bm{x}_1 - \bm{x}_1')
\delta^{(3)} (\bm{x}_2 - \bm{x}_2')\,.
\nonumber
\end{eqnarray}}

From eqs.~\eqref{pNRQCDHstate} and \eqref{matchinglocal} it follows that the LDME of a generic four-fermion operator of the type listed 
in appendix~\ref{app4fermion}, which includes decay and exclusive electromagnetic production LDMEs but excludes hadronic production LDMEs,  
for a strongly coupled quarkonium $H$ of quantum numbers $n$, $J$, $L$ and $S$, at rest ($\bm{P}=\bm{0}$), can be expressed in strongly coupled pNRQCD  by means 
of the master formula~\cite{Brambilla:2002nu}: 
\begin{eqnarray}
\langle H | {\cal O}_n | H \rangle 
&=& \frac{1}{\langle \bm{P} = \bm{0}| \bm{P} = \bm{0} \rangle} \int d^3x_1 d^3x_2 \, d^3x_1' d^3x_2'\,  \langle nJLS | \bm{r} \rangle  
\nonumber\\
&& \hspace{-1cm} 
\times \bigg[  -V_{{\cal O}_n}^{(d_n-4)}(\bm{x}_1, \bm{x}_2;  \bm{\nabla}_1, \bm{\nabla}_2) \delta^{(3)} (\bm{x}_1 - \bm{x}_1') \delta^{(3)} (\bm{x}_2 - \bm{x}_2')\bigg]
                \langle \bm{r}' | nJLS \rangle\,,
\label{eq:master}
\end{eqnarray}
where $\bm{r} = \bm{x}_1 - \bm{x}_2$ and $\bm{r}' = \bm{x}_1' - \bm{x}_2'$.

Because $V_{{\cal O}_n}^{(d_n-4)}$ is a contact term, the wavefunctions $\langle \bm{r} | nJLS \rangle$ and their derivatives 
contribute to eq.~\eqref{eq:master} only at the origin $\bm{r}= \bm{r}' = 0$.
In the particular case of $P$-wave states, since their wavefunctions vanish at the origin,
the contact term  $V_{{\cal O}_n}^{(d_n-4)}$ must contain a sufficient number of derivatives acting on the wavefunctions in order to give a nonvanishing contribution.
In appendix~\ref{sec:LDMEapp} we compute some relevant LDMEs in strongly coupled pNRQCD from the master formula \eqref{eq:master}.
The results are listed and discussed in the following section.

\subsection{LDMEs in strongly coupled pNRQCD}
\label{sec:LDMEpNRQCD}
The LDMEs appearing in the NRQCD factorization formulas for the quarkonium decay widths and electromagnetic production cross sections, see appendix~\ref{app:NRQCDfact},  
involve four-fermion operators of the type listed in appendix~\ref{app4fermion}.
These LDMEs can be evaluated from the master formula~\eqref{eq:master}, which holds when strongly coupled pNRQCD is valid, 
i.e., for quarkonium states satisfying the condition $mv^2 \ll \Lambda_{\rm QCD}$.

Equation~\eqref{eq:master} requires first the computation of the contact term $V_{{\cal O}_n}^{(d_n-4)}$ from the matching condition~\eqref{matchinglocal}.
This can be done straightforwardly using the same ingredients listed in section~\ref{sec:stcoNRQCD}.
Eventually one gets $V_{{\cal O}_n}^{(d_n-4)}$ as a function of $\delta^{(3)}(\bm{r})$ or derivatives of it 
multiplying matrix elements of gluonic fields computed in $\bm{r}=\bm{0}$ ($\bm{x}_1=\bm{x}_2$). 
These are, for instance in the case of $P$-wave quarkonia, of the type 
\begin{equation}
{\cal E}_n \frac{\delta_{ij}}{3} \equiv (-i)^{n+1} n! \sum_{k \neq 0} 
\left.\frac{^{(0)}\langle 0 | g E^i | k \rangle^{(0)} {}^{(0)} \langle k| g E^j | 0 \rangle^{(0)}}{(E_k^{(0)} - E_0^{(0)})^{n+1}}\right|_{\bm{x}_1=\bm{x}_2}. 
\label{corr1}
\end{equation}
The chromoelectric fields are evaluated at the same location $\bm{x}_1=\bm{x}_2$.
The quantity ${\cal E}_n$ is a gluonic matrix element that can be conveniently expressed in terms of a correlator of two chromoelectric fields~\cite{Brambilla:2002nu}:
\begin{equation}
{\cal E}_n = \frac{T_F}{N_c} \int_0^\infty dt \, t^n \langle {\rm vac} | g E^{i,a} (t,\bm{0}) \Phi_{ab}(t,0) g E^{i,b} (0,\bm{0}) | {\rm vac} \rangle, 
\label{corr2}
\end{equation}
where $\Phi_{ab}(t,0)$ is a straight Wilson line in the adjoint representation connecting the points $(t,\bm{0})$ and $(0,\bm{0})$ 
and $T_F=1/2$ is the normalization of the color matrices.
The Wilson line ensures the gauge invariance of ${\cal E}_n$.
We have used that $| 0\rangle^{(0)}|_{\bm{x}_1=\bm{x}_2} = \mathbb{1}_c | {\rm vac} \rangle/\sqrt{N_c}$, $\mathbb{1}_c$ being the SU$(N_c)$ identity matrix and 
$N_c=3$ the number of colors.
Note that the correlator $\langle {\rm vac} | g E^{i,a} (t,\bm{0}) \Phi_{ab}(t,0) g E^{i,b} (0,\bm{0}) | {\rm vac} \rangle$ 
may be understood as the $r\to0$ limit of a Wilson loop with two chromoelectric field insertions.
As we have seen in section~\ref{sec:stcopNRQCD}, Wilson loops with field insertions are related to the quarkonium potential. 
We will exploit this observation in the following.

The power counting of gluon field correlators and their time integrals is obvious, for they depend on only one scale: $\Lambda_{\rm QCD}$.
Hence they scale like $\Lambda_{\rm QCD}$ to their dimension. For instance, we have that ${\cal E}_n \sim \Lambda_{\rm QCD}^{3-n}$.

Once the relevant contact terms have been computed and expressed in terms of delta functions at $\bm{r}=\bm{0}$ and field strength correlators, 
eq.~\eqref{eq:master} allows to compute the LDMEs.
Because of the delta functions at $\bm{r}=\bm{0}$ the LDMEs will depend on the  wavefunction $\langle \bm{r} | nJLS \rangle$
or its derivatives computed at the origin.
If the four-fermion operator and/or corrections to the state (see eq.~\eqref{eq:state_pwave}) contain derivatives, 
one can generate Laplacian operators acting on the wavefunction at the origin.
Such terms can be rewritten in terms of the binding energy of the state by using the Schr\"odinger equation~\eqref{schroedinger}:
\begin{equation}
\label{eq:laplacian}
\left. \bm{\nabla}_{\bm{r}}^2 \langle \bm{r} | nJLS \rangle\right|_{\bm{r}=\bm{0}}
= \left. \left( m V^{(0)}(r) + V^{(1)}(r)  + ...  - m \varepsilon_{nJLS} \right) \langle \bm{r} | nJLS \rangle\right|_{\bm{r}=\bm{0}}, 
\end{equation}
where the dots stand for higher order terms in the $1/m$ expansion of $h$.
In dimensional regularization, $V^{(0)}(r=0)$ vanishes as the static potential is purely perturbative at short distances~\cite{Pineda:2003jv}
(for a more recent analysis with the same conclusion see~\cite{Bazavov:2014soa}) and, therefore, its Fourier transform is scaleless.
The situation is different for $V^{(1)}(r)$, which at $r=0$ reduces to  $V^{(1)}(r=0) = - {\cal E}_1$ 
(compare eq.~\eqref{V1} with eq.~\eqref{corr2} in the $r \to 0$ limit, taking into account that $\langle W_{r\times T}\rangle|_{r=0} = 1$).
Therefore we have in dimensional regularization 
\begin{equation}
\label{eq:laplacian2}
\left. \bm{\nabla}_{\bm{r}}^2 \langle \bm{r} | nJLS \rangle\right|_{\bm{r}=\bm{0}}
= \left(- {\cal E}_1  + ... - m \varepsilon_{nJLS} \right) \langle \bm{r}=\bm{0}| nJLS \rangle.
\end{equation}
The first neglected correction in the right-hand side is suppressed by a factor of order $\Lambda_{\rm QCD}/m$ with respect to ${\cal E}_1$, which is of order $\Lambda_{\rm QCD}^2$.
Equation~\eqref{eq:laplacian2} corrects an analogous expression obtained and used in ref.~\cite{Brambilla:2002nu}, 
where the contribution from $V^{(1)}(r=0)$ was set to zero. 
We will show in section~\ref{sec:swave} how this modifies some of the LDMEs obtained in~\cite{Brambilla:2002nu}.
Note that, since $m \varepsilon_{nJLS}$ is of order $(mv)^2$, ${\cal E}_1$ is smaller than $m \varepsilon_{nJLS}$ if $mv \gg \Lambda_{\rm QCD}$.
Therefore, one can neglect at leading order the term ${\cal E}_1$ in the right-hand side of~\eqref{eq:laplacian2}, 
if the examined quarkonium state fulfills the kinematical condition $mv \gg \Lambda_{\rm QCD} \gg mv^2$.

After matching the contact terms, evaluating the LDMEs with the master formula \eqref{eq:master} 
and rewriting the Laplacian of the wavefunction by means of \eqref{eq:laplacian2},
the LDMEs are expressed in terms of the quarkonium wavefunctions at the origin, correlators of field strength tensors and the quarkonium binding energies.
Because the angular dependence of the wavefunctions is know, we will use the radial parts of the wavefunctions, rather than the wavefunctions. 
We will denote them with $R_{nJLS}(r)$ ($R_{nL}^{(0)}(r)$ at leading order).
We note that the correlators of field strength tensors are universal non perturbative parameters, 
since they do not depend neither on the heavy quark nor on the quarkonium state.
Hence they may be fixed on some set of observables and used in some other one, even involving heavy quarks of different flavor.
As it has been noted in ref.~\cite{Brambilla:2002nu}, this leads eventually to a reduction in the number of 
non perturbative parameters needed to describe quarkonium decay widths and electromagnetic production cross sections in strongly coupled pNRQCD 
in comparison to the number of LDMEs required in NRQCD.

\subsubsection[$P$-wave LDMEs]{\boldmath $P$-wave LDMEs}
\label{sec:pwave}
Here we list some relevant $P$-wave LDMEs computed in strongly coupled pNRQCD. 
We consider a generic spin one quarkonium state that is a $P$-wave state with principal quantum number $n$ and total angular momentum $J$
made of a heavy quark-antiquark pair of flavor $Q$: $\chi_{QJ}(nP)$.\footnote{
  Following the Particle Data Group notation~\cite{Tanabashi:2018oca},
  in the paper we will write $\chi_{QJ}((n-1)P)$ instead of $\chi_{QJ}(nP)$ when identifying a specific state,
  so that a $1P$ state ($\chi_{cJ}(1P)$, $\chi_{bJ}(1P)$) is a state with principal quantum number 2,
  a $2P$ state ($\chi_{cJ}(2P)$, $\chi_{bJ}(2P)$) a state with principal quantum number 3 and so on.}  
Details can be found in appendix~\ref{sec:LDMEapp}.

The hadronic LDME $\langle \chi_{QJ}(nP) | {\cal O}_1({}^3P_J) | \chi_{QJ}(nP) \rangle$ reads in pNRQCD at leading order in the $v$ and $\Lambda_{\rm QCD}/m$ expansion: 
\begin{equation}
\label{eq:o1mathad}
\langle \chi_{QJ}(nP) | {\cal O}_1({}^3P_J) | \chi_{QJ}(nP) \rangle  = \frac{3 N_c}{2 \pi} |R'_{nJ11}(0)|^2 ,
\end{equation}
where the hadronic operators ${\cal O}_1({}^3P_J)$ have been defined in eqs.~\eqref{3P0}-\eqref{3P2} and $R'_{nJLS}$ stands for the derivative of $R_{nJLS}$.
We have computed the corresponding electromagnetic LDME $\langle \chi_{QJ}(nP) | {\cal O}_1^{\rm em} ({}^3P_J) | \chi_{QJ}(nP) \rangle$ at next-to-leading order: 
\begin{equation}
\label{eq:o1mat}
\langle \chi_{QJ}(nP) | {\cal O}_1^{\rm em} ({}^3P_J) | \chi_{QJ}(nP) \rangle  = \frac{3 N_c}{2 \pi} |R'_{nJ11}(0)|^2 
\left[ 1 + \frac{2}{3} \frac{i{\cal E}_2}{m} + O\left(v^2\right) \right],
\end{equation}
where the electromagnetic operators ${\cal O}_1^{\rm em} ({}^3P_J)$ have been defined in eqs.~\eqref{EM3P0}-\eqref{EM3P2}.
The expressions \eqref{eq:o1mat} and \eqref{eq:o1mathad} agree at leading order.
The leading order expressions are known since ref.~\cite{Bodwin:1994jh}.
Instead, the correction proportional to $i{\cal E}_2/{m}$ in eq.~\eqref{eq:o1mat} is new.
This is the dominant correction to the pure wavefunction contribution. It is of order $\Lambda_{\rm QCD}/m$.
As we detail in appendix~\ref{sec:LDMEapp} it originates from the $1/m$ correction to the quarkonium Fock state given in eq.~\eqref{eq:state_pwave}.

The electromagnetic LDME $\langle \chi_{QJ}(nP) | {\cal T}_8^{\rm em} ({}^3P_J) | \chi_{QJ}(nP) \rangle$ reads in pNRQCD
at leading order in the $v$ and $\Lambda_{\rm QCD}/m$ expansion: 
\begin{equation}
\label{eq:t8mat}
\langle \chi_{QJ}(nP) | {\cal T}_8^{\rm em} ({}^3P_J) | \chi_{QJ}(nP) \rangle =  \frac{3 N_c}{2 \pi} |R'_{nJ11}(0)|^2 \frac{4}{3} \frac{{\cal E}_1}{m}, 
\end{equation}
where the  electromagnetic operators ${\cal T}_8^{\rm em} ({}^3P_J)$ have been defined in eqs.~\eqref{T8em3P0}-\eqref{T8em3P2}.
The result \eqref{eq:t8mat} agrees with the result of ref.~\cite{Brambilla:2002nu} for the $J=0$ case.
Note that the operator ${\cal T}_8^{\rm em} ({}^3P_J)$ has no overlap with the color singlet component of the heavy quark-antiquark pair.
Hence, the expression of its matrix element in terms of the quarkonium wavefunction is a specific feature of strongly coupled pNRQCD: 
the above expression has no equivalent in~ref.~\cite{Bodwin:1994jh}.

The electromagnetic LDME $\langle \chi_{QJ} (nP) | {\cal P}_1^{\rm em} ({}^3P_J) | \chi_{QJ} (nP) \rangle$ reads in pNRQCD
in dimensional regularization: 
\begin{eqnarray}
\label{eq:p1mat}
  \langle \chi_{QJ} (nP) | {\cal P}_1^{\rm em} ({}^3P_J) | \chi_{QJ} (nP) \rangle = \frac{3 N_c}{2 \pi} | R'_{nJ11} (0) |^2 
\left[ m \varepsilon_{n1}^{(0)} - \frac{2}{3} {\cal E}_1 + O\left(v^3\right) \right], 
\end{eqnarray}
where the  electromagnetic operators ${\cal P}_1^{\rm em} ({}^3P_J)$ have been defined in eqs.~\eqref{EMP3P0}-\eqref{EMP3P2}.
The term $m \varepsilon_{n1}^{(0)}$ is of order $(mv)^2$, while $ {\cal E}_1$ is of order $\Lambda_{\rm QCD}^2$.
The expression follows from having used eq.~\eqref{eq:laplacian2}. The result is new. 

Since $|R'_{nJLS}(0)|^2 = |R_{nL}^{(0)\,\prime}(0)|^2(1 + O(v^2))$,
the right-hand sides of eqs.~\eqref{eq:o1mathad}-\eqref{eq:p1mat} are independent of $J$ up to the computed corrections.
This verifies the heavy-quark spin symmetry.
In particular, the symmetry is realized also for the octet matrix element \eqref{eq:t8mat}.

\subsubsection[$S$-wave LDMEs]{\boldmath $S$-wave LDMEs}
\label{sec:swave}
The inclusion of the term ${\cal E}_1$ in eq.~\eqref{eq:laplacian2} modifies the expression of some of the $S$-wave color singlet LDMEs computed in ref.~\cite{Brambilla:2002nu}. 
The modification is relevant at relative order $(\Lambda_{\rm QCD}/m)^2$. 
The corrected $S$-wave color singlet LDMEs read
\begin{eqnarray}
&& \langle V_Q(nS)|{\cal O}_1 ({}^3S_1)|V_Q(nS)\rangle
\nonumber \\ 
&& \hspace{1.5cm} 
= \frac{N_c}{2 \pi} |R_{n101}(0)|^2
\left[ 1 - \frac{\varepsilon_{n0}^{(0)}}{m} \frac{2 {\cal E}_3}{9} -\frac{2 {\cal E}_1 {\cal E}_3}{9 m^2} + \frac{2 {\cal E}_3^{(2,t)}}{3 m^2} + \frac{c_F^2 {\cal B}_1}{3 m^2}
+ O\left(v^3\right) \right],
\label{O3S1}\\
&& \langle P_Q(nS)|{\cal O}_1({}^1S_0)|P_Q(nS)\rangle
\nonumber \\ 
&& \hspace{1.5cm} 
= \frac{N_c}{2 \pi} |R_{n000}(0)|^2
\left[ 1 - \frac{\varepsilon_{n0}^{(0)}}{m} \frac{2 {\cal E}_3}{9} -\frac{2 {\cal E}_1 {\cal E}_3}{9 m^2} + \frac{2 {\cal E}_3^{(2,t)}}{3 m^2} + \frac{c_F^2 {\cal B}_1}{m^2} 
+ O\left(v^3\right) \right],
\label{O1S0}\\
&& \langle V_Q(nS)|{\cal O}_1^{\rm em} ({}^3S_1)|V_Q(nS)\rangle
\nonumber \\ 
&& \hspace{1.5cm} 
= \frac{N_c}{2\pi} |R_{n101}(0)|^2
\left[ 1 - \frac{\varepsilon_{n0}^{(0)}}{m} \frac{2 {\cal E}_3}{9} -\frac{2 {\cal E}_1 {\cal E}_3}{9 m^2} + \frac{2 {\cal E}_3^{(2,{\rm em})}}{3 m^2} + \frac{c_F^2 {\cal B}_1}{3 m^2}
+ O\left(v^3\right) \right],
\label{Oem3S1}\\
&& \langle P_Q(nS)|{\cal O}_1^{\rm em} ({}^1S_0)|P_Q(nS)\rangle
\nonumber \\ 
&& \hspace{1.5cm} 
= \frac{N_c}{2\pi} |R_{n000}(0)|^2
\left[ 1 - \frac{\varepsilon_{n0}^{(0)}}{m} \frac{2 {\cal E}_3}{9} -\frac{2 {\cal E}_1 {\cal E}_3}{9 m^2}  + \frac{2 {\cal E}_3^{(2,{\rm em})}}{3 m^2}+ \frac{c_F^2 {\cal B}_1}{m^2} 
+ O\left(v^3\right) \right],
\label{Oem1S0}\\
&& \langle V_Q(nS)|{\cal P}_1 ({}^3S_1)|V_Q(nS)\rangle = \langle P_Q(nS)|{\cal P}_1 ({}^1S_0)|P_Q(nS)\rangle =  \langle V_Q(nS)|{\cal P}_1^{\rm em} ({}^3S_1)|V_Q(nS)\rangle 
\nonumber\\
&& \hspace{1.5cm}
= \langle P_Q(nS)|{\cal P}_1^{\rm em} ({}^1S_0)|P_Q(nS)\rangle = \frac{N_c}{2\pi} |R_{n0}^{(0)}(0)|^2 \left[ m \varepsilon_{n0}^{(0)} + O\left(v^3\right) \right],
\label{P3S1}
\end{eqnarray}
where $V_Q(nS)$ ($P_Q(nS)$) is an $S$-wave vector (pseudoscalar) quarkonium state made of a heavy quark and antiquark of flavor $Q$.
The operators ${\cal O}_1 ({}^3S_1)$, ${\cal O}_1 ({}^1S_0)$, ${\cal O}_1^{\rm em} ({}^3S_1)$ and ${\cal O}_1^{\rm em} ({}^1S_0)$
can be found in eqs.~\eqref{3S1}, \eqref{1S0}, \eqref{EM3S1} and \eqref{EM1S0}, respectively.
The operators ${\cal P}_1({}^3S_1)$ and ${\cal P}_1({}^1S_0)$ are defined in eqs.~\eqref{PP3S1} and~\eqref{PP1S0},
and the operators ${\cal P}_1^{\rm em}({}^3S_1)$ and ${\cal P}_1^{\rm em}({}^1S_0)$ in eqs.~\eqref{EMP3S1} and~\eqref{EMP1S0}.
The correlator ${\cal B}_1$ is analogous to the correlator ${\cal E}_1$ but with the chromoelectric fields replaced by chromomagnetic ones.
The correlators ${\cal E}_3^{(2,{\rm em})}$ and ${\cal E}_3^{(2,t)}$ are four-chromoelectric field correlators of mass dimension two, 
whose definition can be found in ref.~\cite{Brambilla:2002nu} but is irrelevant for the present work.

The (leading order) binding energy $\varepsilon_{n0}^{(0)}$ scales like $mv^2$, the correlator ${\cal E}_3$ is a scaleless constant, whereas all other correlators 
in eqs.~\eqref{O3S1}-\eqref{P3S1} scale like $\Lambda_{\rm QCD}^2$. Hence the term proportional to the binding energy is the dominant 
correction in eqs.~\eqref{O3S1}-\eqref{Oem1S0} if the quarkonium state satisfies the condition $mv \gg \Lambda_{\rm QCD} \gg mv^2$.
All corrections are of the same order if $mv \sim \Lambda_{\rm QCD}$.

The inclusion of the term ${\cal E}_1$ in eq.~\eqref{eq:laplacian2} has modified eqs.~\eqref{O3S1}-\eqref{Oem1S0} with respect to  
ref.~\cite{Brambilla:2002nu} by adding the term proportional to  $- 2 {\cal E}_1 {\cal E}_3 /(9 m^2)$. This term is of order $(\Lambda_{\rm QCD}/m)^2$.
It has also modified eq.~\eqref{P3S1}. 
Differently from the version in ref.~\cite{Brambilla:2002nu}, eq.~\eqref{P3S1} does not contain the term $-N_c |R_{n0}^{(0)}(0)|^2/(2\pi)\, {\cal E}_1$.

We note that eqs.~\eqref{O3S1}-\eqref{Oem1S0} are accurate up to relative order $v^2$, hence also the wavefunctions at the origin include corrections of relative order $v^2$.
These corrections distinguish the vector from the pseudoscalar radial wavefunctions. 
On the contrary, eq.~\eqref{P3S1} is accurate only at leading order, hence the wavefunction appearing there needs not to be more accurate than that.
At leading order the radial parts of the vector and pseudoscalar wavefunctions are equal.

For completeness we reproduce here also the $S$-wave color octet LDMEs of the
operators ${\cal O}_8(^1S_0)$ and ${\cal O}_8(^3S_1)$, defined in eqs.~\eqref{Oct1S0} and~\eqref{Oct3S1} respectively,
and of the operators ${\cal O}_8(^3P_J)$ and ${\cal O}_8(^1P_1)$, defined in eqs.~\eqref{Oct3P0}-\eqref{Oct3P2} and~\eqref{Oct1P1} respectively, computed in ref.~\cite{Brambilla:2002nu}
\begin{eqnarray}
&& \langle V_Q(nS)|{\cal O}_8(^1S_0)|V_Q(nS)\rangle= \frac{\langle
P_Q(nS)|{\cal O}_8(^3S_1)|P_Q(nS)\rangle}{3}
\nonumber\\
&& \hspace{4.5cm}
= \frac{N_c}{2\pi} |R^{(0)}_{n0}({0})|^2 \left[-\frac{(N_c/2-C_F) c_F^2 {\cal B}_1}{3 m^2 }\right],
\label{O83S1}\\
&& \langle V_Q(nS)|{\cal O}_8(^3P_J)|V_Q(nS)\rangle= \frac{\langle
P_Q(nS)|{\cal O}_8(^1P_1)|P_Q(nS)\rangle}{3}
\nonumber\\
&& \hspace{4.5cm}
=(2J+1)\,  \frac{N_c}{2\pi} |R^{(0)}_{n0}({0})|^2 \left[-\frac{(N_c/2-C_F) {\cal E}_1}{9}\right],
\label{O83PJ}
\end{eqnarray}
where $C_F=(N_c^2-1)/(2N_c) =4/3$.
These relations are valid at leading order in the velocity and $\Lambda_{\rm QCD}/m$ expansion.

\subsection{Gremm--Kapustin relations}
\label{sec:GK}
The NRQCD equations of motion imply that some LDMEs are related at leading order in the velocity expansion.
These relations are often referred to as {\em Gremm--Kapustin relations}~\cite{Gremm:1997dq}.
Over the years, following the same method, more relations have been derived, see, for instance, refs.~\cite{Ma:2002eva,Bodwin:2002hg,Brambilla:2017kgw}.

The Gremm--Kapustin relations are automatically satisfied by the expressions of
the LDMEs derived in strongly coupled pNRQCD in sections~\ref{sec:pwave} and~\ref{sec:swave}.
The reason is that at the level of pNRQCD the information encoded in the equations of motion has been implemented through the Schr\"odinger equation, 
or, more specifically through eq.~\eqref{eq:laplacian2}.

In particular, for electromagnetic $P$-wave LDMEs, from eqs.~\eqref{eq:o1mat}, \eqref{eq:t8mat} and~\eqref{eq:p1mat} it follows that
\begin{eqnarray}
\label{eq:gremm-kapustin}
\langle\chi_{QJ}(nP)|{\cal P}_1^{\rm em} ({}^3P_J)|\chi_{QJ}(nP)\rangle &=&m\varepsilon^{(0)}_{n1} \langle\chi_{QJ}(nP)|{\cal O}_1^{\rm em} ({}^3P_J)|\chi_{QJ}(nP)\rangle
\nonumber\\
&& \hspace{1cm}
-\frac{m}{2}\langle\chi_{QJ}(nP)|{\cal T}_8^{\rm em} ({}^3P_J)|\chi_{QJ}(nP)\rangle.
\end{eqnarray}
This relation was first derived in ref.~\cite{Ma:2002eva}. 
According to the expressions of the LDMEs in strongly coupled pNRQCD, it holds at the orders $|R'_{nJ11}(0)|^2 \times (mv)^2$ and $|R'_{nJ11}(0)|^2 \times \Lambda_{\rm QCD}^2$.
This is the leading order if $mv\sim \Lambda_{\rm QCD}$, but goes beyond it if $mv \gg \Lambda_{\rm QCD} \gg mv^2$.

For $S$-wave LDMEs, from eqs.~\eqref{O3S1}-\eqref{P3S1} it follows that at leading order in both regimes, $mv\sim \Lambda_{\rm QCD}$ and $mv \gg \Lambda_{\rm QCD} \gg mv^2$, 
\begin{eqnarray}
&& \hspace{-0.5cm} 
\frac{\langle V_Q(nS)|{\cal P}_1 ({}^3S_1)|V_Q(nS)\rangle}{\langle V_Q(nS)|{\cal O}_1({}^3S_1)|V_Q(nS)\rangle} = 
\frac{\langle P_Q(nS)|{\cal P}_1 ({}^1S_0)|P_Q(nS)\rangle}{\langle P_Q(nS)|{\cal O}_1({}^1S_0)|P_Q(nS)\rangle} 
\nonumber\\
&& \hspace{0.5cm}
= \frac{\langle V_Q(nS)|{\cal P}_1^{\rm em} ({}^3S_1)|V_Q(nS)\rangle}{\langle V_Q(nS)|{\cal O}_1^{\rm em}({}^3S_1)|V_Q(nS)\rangle} = 
\frac{\langle P_Q(nS)|{\cal P}_1^{\rm em} ({}^1S_0)|P_Q(nS)\rangle}{\langle P_Q(nS)|{\cal O}_1^{\rm em}({}^1S_0)|P_Q(nS)\rangle} = m \varepsilon_{n0}^{(0)}.
\label{eq:gremm-kapustin2}
\end{eqnarray}
This relation was first derived in ref.~\cite{Gremm:1997dq}.
We note that ref.~\cite{Brambilla:2002nu} could reproduce this relation only in the regime $mv \gg \Lambda_{\rm QCD} \gg mv^2$.

\section{Fits, analyses and results}
\label{sec:results}
We apply now the pNRQCD factorization of the LDMEs computed in the previous section to the analysis of some quarkonium decay and production observables, in particular electromagnetic ones.
The strategy that we will pursue is the following: first we determine the quarkonium wavefunctions and binding energies by means of models, 
then we fit the relevant chromoelectric field correlators on charmonium data and finally we compute the observables.
Because the correlators are universal we are in the position to predict observables in the bottomonium sector. 

We focus on two sets of observables that depend on two distinct sets of correlators.
In the first part of the section, we compute quarkonium $P$-wave electromagnetic decay widths and production cross sections.
In the second part, we analyze quarkonium $P$-wave widths for inclusive decays into light hadrons, and bottomonium $S$-wave widths for decays into lepton pairs. 
We use the strongly coupled pNRQCD factorization formulas in their regime of validity, i.e., for non Coulombic quarkonium states.
For this reason we limit ourselves to states with principal quantum number greater than one.

The section is organized as follows. In section~\ref{sec:potentials} we establish some reasonable values for the 
quarkonium wavefunctions at the origin and the binding energies by comparing several potential models.
In section~\ref{sec:e1e2} we fit the correlators ${\cal E}_1$ and $i{\cal E}_2$ on the $P$-wave charmonium electromagnetic decay widths 
and the recently measured cross section $\sigma(e^+ e^- \rightarrow \chi_{c1}(1P) + \gamma)$. 
We also compute these quantities within our framework.
In section~\ref{sec:Pdecayproduction}, we compute the $P$-wave bottomonium electromagnetic decay widths and electromagnetic cross sections from the determined correlators.
In section~\ref{sec:e3}, we compute the $P$-wave charmonium widths for inclusive decays into light hadrons and fit the correlator ${\cal E}_3$.
We also compute with this information $P$-wave bottomonium widths for inclusive decays into light hadrons.
Finally, in section~\ref{sec:upsilon}, we use the determination of the correlator ${\cal E}_3$ to compute, under some assumptions, 
the leptonic decay widths of the bottomonium $S$-wave states $\Upsilon(2S)$ and $\Upsilon(3S)$.

\subsection{Potential models}
\label{sec:potentials}
The first ingredients entering the LDMEs are the quarkonium wavefunctions at the origin and the binding energies.
In particular the wavefunctions at the origin are very important as they affect all LDMEs in the pNRQCD formulation and contribute to widths and cross sections at leading order.
The uncertainty of the wavefunction at the origin is typically the major source of uncertainty for these observables.

Ideally, quarkonium wavefunctions and binding energies should follow from the solution of the Schr\"odinger equation~\eqref{schroedinger} with the potentials 
computed within lattice QCD from the corresponding Wilson loops, like the one in eq.~\eqref{V0} for the static potential
or the one in eq.~\eqref{V1} for the $1/m$ potential. The knowledge of the lattice potentials beyond the static one is, however, 
incomplete and sometimes poor~\cite{Koma:2006si,Koma:2006fw,Koma:2007jq}.
In practice, therefore, one uses potential models, with the idea that tuning the potential model parameters on some observables
may provide enough input to mimic the full real dynamics, the one that lattice computations are not yet in the position to provide.
Clearly, the use of potential models introduces possibly large and, to some extent, uncontrolled uncertainties. 
Nevertheless, it has also proved to be successful in many cases, besides being the only available solution at present.

We employ four different potential models to compute wavefunctions at the origin and binding energies for charmonium states.
We label these potential models A, B, C, and D.
A common feature of them is that they reduce to a liner rising potential at long distances. 

{\em Model A} is the Cornell potential model of refs.~\cite{Eichten:1978tg,Eichten:1995ch}, where $V^{(0)}(r) = -\kappa/r + \sigma r$, with $\kappa=0.52$ and $\sigma = 0.1826$~GeV${}^{2}$;
$\sigma$ may be identified with a string tension.
In this model, the charm and bottom quark mass parameters are taken to be $m_c = 1.84$~GeV and $m_b = 5.18$~GeV, respectively. 

{\em Model B} is the frozen $\alpha_{\rm s}$ model of refs.~\cite{Eichten:2019gig,Eichten:2019hbb}.
It is similar to model A, except that now $\kappa$ depends on $r$. 
The $r$-dependent values of $\kappa$ are tabulated in ref.~\cite{Eichten:2019gig}.

{\em Model C} is the Buchm\"uller--Tye potential model~\cite{Buchmuller:1980su}.
Here the charm and bottom quark mass parameters are taken to be $m_c = 1.48$~GeV and $m_b = 4.88$~GeV, respectively. 

{\em Model D} is the Cornell potential model in the version of ref.~\cite{Bodwin:2007fz}.
In this version, the parameters are set to be $\kappa=0.538$, $\sigma=0.1682$~GeV${}^{2}$ 
and $m_c = 1.44$~GeV in order to reproduce the mass difference of the $J/\psi$ and $\psi(2S)$, and the leptonic width of the $J/\psi$.
Similarly, the bottom quark mass parameter is taken to be $m_b = 3.98$~GeV in order to reproduce the mass difference of the $\Upsilon(1S)$ and $\Upsilon(2S)$, 
and the leptonic width of the $\Upsilon(3S)$~\cite{Chung:2010vz}. 

\begin{table}[ht]
\centering
\begin{tabular}{|c|c|c|c|c|}
\hline
Potential model & A & B & C & D \\
\hline
$|R_{21}^{(0)\,\prime}(0)|^2$~(GeV${}^5$) & $0.131$ & $0.1296$ & $0.075$ & $0.0682$ \\
$\varepsilon_{21}^{(0)}$~(GeV) &0.68591 &0.68764 & 0.55219 & 0.72544 \\
$\Delta_{LS} (1P)$ & 0.0164 & 0.0167 & 0.0190 & 0.0216 \\
\hline
\end{tabular}
\caption{\label{tab:chicpotential}
  For the four potential models described in the text, we list the squared derivative of the radial wavefunction at the origin, $|R_{21}^{(0)\,\prime}(0)|^2$, 
  the binding energy, $\varepsilon_{21}^{(0)}$, and the spin-dependent correction defined in eq.~\eqref{eq:lscorrection}, $\Delta_{LS}(1P)$, of the charmonium $1P$ state.}
\end{table}

We have determined from these potential models the binding energy, $\varepsilon_{21}^{(0)}$,
and squared derivative of the radial wavefunction at the origin, $|R_{21}^{(0)\,\prime}(0)|^2$, at leading order in $v$
for charmonium $1P$ states. They are listed in Table~\ref{tab:chicpotential}. 
The values of $|R_{21}^{(0)\,\prime}(0)|^2$ for the models A, B, and C are taken from refs.~\cite{Eichten:1995ch,Eichten:2019hbb}.

\begin{table}[ht]
\centering
\begin{tabular}{|c|c|c|c|c|c|}
\hline
Potential model & A & B & C & D & E \\
\hline
$|R_{21}^{(0)\,\prime} (0)|^2$~(GeV${}^5$) & $2.067$ & $1.6057$ & $1.417$ &
$0.932$ & $1.342$ \\
$\varepsilon_{21}^{(0)}$~(GeV)& 0.32792& 0.33804& 0.13173& 0.3763& 0.33133 \\
$\Delta_{LS}(1P)$ & 0.00379& 0.00390& 0.00347& 0.00522& 0.00417 \\
\hline 
$|R_{31}^{(0)\,\prime} (0)|^2$~(GeV${}^5$) & $2.440$ & $1.8240$ & $1.653$ &
$1.147$ & $1.624$ \\
$\varepsilon_{31}^{(0)}$~(GeV)& 0.68206& 0.68956& 0.49168& 0.7321& 0.68264 \\
$\Delta_{LS}(2P)$ & 0.00286& 0.00302& 0.00269& 0.00397& 0.00316 \\  
\hline
$|R_{41}^{(0)\,\prime} (0)|^2$~(GeV${}^5$) & $2.700$ & $1.9804$ & $1.794$ &
$1.296$ & $1.817$ \\
$\varepsilon_{41}^{(0)}$~(GeV)& 0.96466& 0.97115& 0.76921& 1.01933& 0.95681 \\
$\Delta_{LS}(3P)$ & 0.00233& 0.00250& 0.00222& 0.00325& 0.00258 \\
\hline
\end{tabular}
\caption{\label{tab:chibpotential} Similar to Table~\ref{tab:chicpotential} but for bottomonium $nP$ states.
In the bottomonium case, we also consider the model E described in the text.}
\end{table}

Furthermore, we have used models A, B, C, and D to compute the wavefunctions at the origin and binding energies of some $nP$ bottomonium states. 
The results are listed in Table~\ref{tab:chibpotential}. 
The values of the squared derivatives of the radial wavefunctions at the origin for the models A, B and C are taken from refs.~\cite{Eichten:1995ch,Eichten:2019hbb}.
For bottomonium we employ also a potential model~E.

{\em Model E} is similar to model D. The only difference is that we take
$\kappa = 0.508$ and $m_b = 4.68$~GeV so to reproduce the mass
difference of the $\Upsilon(2S)$ and $\Upsilon(3S)$, and the leptonic width of
the $\Upsilon(3S)$ in the calculation of section~\ref{sec:upsilon}.

In order to compute electromagnetic decay widths and production cross sections of $P$-wave quarkonia at relative order $v^2$,
we would need to include also order $v^2$ corrections to the wavefunctions at the origin.
Since the model dependence of the values of $|R_{n1}^{(0)\,\prime}(0)|^2$ that we employ exceeds $v^2$, 
their inclusion would not improve, however, the accuracy of the model determinations.
Hence, we account for the order $v^2$ corrections to the wavefunctions at the origin only in our final error budget.
The sole corrections that we add explicitly are those that depend on the total angular momentum, 
since they contribute to distinguish between decay and production of $\chi_{QJ}(nP)$ states with different $J$.
Nevertheless, also in this case the corrections are smaller than the systematic uncertainty due to the models.
The total angular momentum corrections originate from one single $1/m^2$ spin-orbit potential, $V_{LS}^{(2)}/m^2$.
In the short range, this potential generates corrections to the wavefunction at the origin of relative order $\alpha_{\rm s}^2$ that are divergent. 
The renormalization of these divergences requires introducing order $\alpha_{\rm s}^2$ short distance coefficients for $P$-wave quarkonium production and decay processes.
Since these are unknown at present, in this work we will not include corrections due to the short distance part of the spin-orbit potential.
In the long range, the behavior of the spin-orbit potential is entirely described by the string tension $\sigma$ and fixed by Lorentz symmetry~\cite{Gromes:1984ma,Brambilla:2003nt}.
This behavior is confirmed by lattice calculations~\cite{Koma:2006fw}.
The spin-orbit potential in the long range reads:
\begin{equation}
V_{LS}^{(2)} = - \frac{\sigma}{2 r} \bm{L}\cdot \bm{S} ,
\end{equation}
where $\bm{L}$ and $\bm{S}$ are the total orbital angular momentum and spin, respectively.
For $P$-wave spin-triplet states, the corrections to $|R_{n1}^{(0)\,\prime}(0)|^2$ from $V_{LS}^{(2)}/m^2$ have the following form 
\begin{equation}
\label{eq:lscorrection}
|R'_{nJ11} (0)|^2 = |R^{(0)}_{n1}{}' (0)|^2 \, \left[1+ \left(\frac{3}{2} J (J+1) - 6\right) \Delta_{LS}(nP) \right], 
\end{equation}
where $\Delta_{LS}(nP)$ depends on the radial excitation. 
We have listed the values of $\Delta_{LS}(nP)$ in the different models for the charmonium $1P$ state in Table~\ref{tab:chicpotential}, 
and for the bottomonium $1P$, $2P$, and $3P$ states in Table~\ref{tab:chibpotential}.
We see explicitly that the correction induced by the spin-orbit potential is smaller than the intrinsic potential model uncertainty,
which we infer from the spread of the different wavefunction determinations.
Finally, we observe that uncalculated corrections of relative order $v^2$ coming from the quantum-mechanical 
$1/m$ expansion of the quarkonium Fock state, in particular for the LDME of eq.~\eqref{eq:o1mat}, can be spin and angular momentum dependent as well.
These uncalculated corrections are included in the error budget of the LDME, although the tuning of the potential model parameters may effectively reduce their size.

\begin{table}[ht]
\centering
\begin{tabular}{|c|c|c|c|c|c|}
\hline
Potential model & A & B & C & D & E \\
\hline
$|R_{20}^{(0)}(0)|^2$~(GeV$^3$) & 5.668 & 2.8974 & 3.234 & 3.47 & 4.36\\
$\varepsilon_{20}^{(0)}$~(GeV)& 0.421& 0.463& 0.258& 0.478& 0.435 \\
\hline
$|R_{30}^{(0)}(0)|^2$~(GeV$^3$) & 4.271 & 2.2496 & 2.474 & 2.67 & 3.32 \\
$\varepsilon_{30}^{(0)}$~(GeV)& 0.767& 0.795& 0.597& 0.823& 0.767 \\
\hline
\end{tabular}
\caption{\label{tab:sbpotential}
For the five potential models described in the text, we list the squared radial wavefunctions at the origin, $|R_{n0}^{(0)}{}(0)|^2$, 
and the binding energies, $\varepsilon_{n0}^{(0)}$, of the bottomonium $2S$ and $3S$ states.
}
\end{table}

With the same five potential models described above we have also determined at leading order in $v$ the squared radial wavefunctions at the origin and the binding energies 
of the $2S$ and $3S$ bottomonium states. 
The results are listed in Table~\ref{tab:sbpotential}.
The values of the wavefunctions at the origin for the models A, B, and C are taken from refs.~\cite{Eichten:1995ch,Eichten:2019hbb}.

\subsection[$P$-wave charmonium electromagnetic decay and production]
{\boldmath $P$-wave charmonium electromagnetic decay and production}
\label{sec:e1e2}
In this section, we compute the charmonium decay widths $\Gamma(\chi_{cJ}(1P) \rightarrow \gamma \gamma)$ and the cross sections $\sigma(e^+ e^- \rightarrow \chi_{cJ}(1P) + \gamma)$ 
using the NRQCD factorization formulas \eqref{factNRQCDPem} and \eqref{factNRQCDPempro}, which are valid up to order $v^2$,
and rewriting the LDMEs according to the strongly coupled pNRQCD factorization formulas \eqref{eq:o1mat}-\eqref{eq:p1mat}.
We determine the gluonic correlators ${\cal E}_1$ and $i{\cal E}_2$ by fitting the available data.

The experimental inputs that we use are the $\chi_{c0}(1P)$ and $\chi_{c2}(1P)$ two photon decay widths and the cross section $\sigma(e^+ e^- \to \chi_{c1}(1P) + \gamma)$. 
The BESIII measurements for the former give~\cite{Ablikim:2012xi}
\begin{eqnarray}
\Gamma(\chi_{c0}(1P) \rightarrow \gamma \gamma)\big|_{\rm BESIII} &=& 2.33 \pm 0.20 \pm 0.22\; \textrm{keV}\,,
\label{Gammachic0exp}\\
\Gamma(\chi_{c2}(1P) \rightarrow \gamma \gamma)\big|_{\rm BESIII} &=& 0.63 \pm 0.04 \pm 0.06\; \textrm{keV}\,.
\label{Gammachic2exp}
\end{eqnarray}
For the latter, very recently Belle has observed the process $e^+ e^- \to \chi_{c1}(1P) + \gamma$ and measured at $\sqrt{s} = 10.6$~GeV~\cite{Jia:2018xsy}
\begin{equation}
\sigma(e^+ e^- \to \chi_{c1}(1P)+\gamma)\big|_{\rm Belle} = 17.3 {}^{+4.2}_{-3.9} \pm 1.7\; \textrm{fb}\,.
\label{sigmachic1exp}
\end{equation}

From the theoretical side, rather than using the NRQCD factorization formulas for electromagnetic processes in their original form (see appendix~\ref{app:NRQCDfact})
we prefer using NRQCD factorization formulas at the amplitude level. So that the matching, the velocity expansion and the power counting are done for the amplitudes
rather than for the decay widths or cross sections. In practice, one moves from the original factorization formulas to the ones at the amplitude level
through replacements of the type
\begin{eqnarray}
&&\langle H| \psi^\dagger K \chi | {\rm vac}\rangle \langle {\rm vac}|\chi^\dagger K\psi |H \rangle \left(1 + c^{\rm NLO}\frac{\alpha_{\rm s}}{\pi} + ...\right)
\nonumber\\
&& \hspace{5cm}
\to \left|\langle H| \psi^\dagger K \chi | {\rm vac}\rangle  \left(1 + \frac{c^{\rm NLO}}{2}\frac{\alpha_{\rm s}}{\pi} + ...\right)\right|^2.
\end{eqnarray}
The advantage is that in this way, without losing any systematicity, one is effectively including some potentially large contributions of order $v^4$, $\alpha_{\rm s}v^2$ and $\alpha_{\rm s}^2$
in the expressions of the observables.
Moreover, we replace the uncertain heavy quark pole mass with the spin average of the masses of the $P$ states.
This is defined in the case of charmonium $1P$ states as 
\begin{equation}
  M_{1P_c} = \frac{M_{h_c(1P)} + M_{\chi_{c0}(1P)} + 3 M_{\chi_{c1}(1P)} + 5 M_{\chi_{c2}(1P)}}{10}\,.
\label{Mspinaverage}
\end{equation}
We take the $1P$ charmonium masses from ref.~\cite{Tanabashi:2018oca}.
At our accuracy it is sufficient to use the following relation between $M_{1P_c}$ and the charm pole mass $m_c$:  
\begin{equation}
M_{1P_c} = 2m_c + \varepsilon^{(0)}_{21}\,,
\label{Mmrelation}
\end{equation}
which is valid up to order $v^2$. 
Afterwards we expand in powers of the binding energy up to relative order $v^2$ accuracy in the amplitude. 
Eventually, the theoretical expressions for the two photon decay widths that we use in the numerical analyses are  
\begin{eqnarray}
\Gamma(\chi_{c0}(1P) \rightarrow \gamma \gamma)
&=& 
    \frac{96 \pi e_c^4 \alpha^2}{M_{1P_c}^4} \, \frac{3 N_c}{2 \pi} |R'_{n011} (0)|^2 
    \left[ 1+ \frac{3 \pi^2-28}{24} C_F \frac{\alpha_{\rm s}}{\pi} \right.
\nonumber\\
  && \hspace{4.5cm}
     \left.     - \frac{ \varepsilon^{(0)}_{21}}{3 M_{1P_c}} + \frac{16 {\cal E}_1}{9 M_{1P_c}^2} + \frac{2}{3} \frac{i {\cal E}_2}{M_{1P_c}}  \right]^2\!\!,
\label{Gammachi0final}\\
\Gamma(\chi_{c2}(1P) \rightarrow \gamma \gamma)
&=& 
    \frac{128 \pi e_c^4 \alpha^2}{5 M_{1P_c}^4} \, \frac{3 N_c}{2 \pi} |R'_{n211} (0)|^2 
    \left[ 1 - 2 C_F \frac{\alpha_{\rm s}}{\pi} \right.
\nonumber\\
  && \hspace{4.5cm}
\left . + \frac{8 {\cal E}_1 }{3 M_{1P_c}^2}  + \frac{2}{3} \frac{i {\cal E}_2}{M_{1P_c}}  \right]^2, 
\label{Gammachi2final}
\end{eqnarray}
where $e_c=2/3$.
Note that $\Gamma(\chi_{c2}(1P) \rightarrow \gamma \gamma)$ does not depend explicitly on the binding energy $\varepsilon^{(0)}_{21}$.
Similarly, the expression for the cross sections $\sigma(e^+ e^- \to \chi_{cJ}(1P)+\gamma)$ at the center of mass energy $\sqrt{s}$ that we use is 
\begin{eqnarray}
&& \hspace{-8mm}
   \sigma(e^+ e^- \rightarrow \chi_{cJ}(1P) + \gamma) = \sigma_{cJ}^{(0)}\left(\frac{M_{1P_c}}{2},s,r\right) \, \frac{3 N_c}{2 \pi} |R'_{nJ11} (0)|^2 \left[ 1 + \frac{c_J^{(O_1){\rm NLO}} (r)}{2}\frac{\alpha_{\rm s}}{\pi}
   \right.
\nonumber\\
&&  \hspace{-4mm}
\left.\left.   
   + \frac{3 \varepsilon^{(0)}_{21}}{2 M_{1P_c}}  + \frac{2i {\cal E}_2}{3M_{1P_c}}  
+ \frac{c_J^{(T)}(r)}{2} \frac{16 {\cal E}_1}{3 M_{1P_c}^2}
   + \frac{c_J^{(P)}(r)}{2} \left( \frac{2 \varepsilon^{(0)}_{21}}{M_{1P_c}} - \frac{8 {\cal E}_1}{3 M_{1P_c}^2} \right) \right]^2\right|_{r= (M_{1P_c}-\varepsilon^{(0)}_{21})^2/s}
   \hspace{-10mm}.
\label{sigmachiJfinal}
\end{eqnarray}
The factor $\sigma_{cJ}^{(0)}$ has been defined in eqs.~\eqref{sigmaQ0}-\eqref{sigmaQ2}, and $c_J^{(O_1){\rm NLO}}$, $c_J^{(T)}$ and $c_J^{(P)}$ in eqs.~\eqref{c0O1}-\eqref{c2P}.

In eqs.~\eqref{Gammachi0final}-\eqref{sigmachiJfinal} the corrections proportional to $\alpha_{\rm s}$ come from the short distance coefficients.
Hence $\alpha_{\rm s}$ should be understood as evaluated at a high energy scale and counting parametrically like a $v^2$ correction in the velocity expansion.
For the decay widths, we use $\alpha_{\rm s} = 0.282$, which is evaluated at the scale $M_{1P_c}/2$,
and $\alpha = 1/137$ reflecting the fact that the photons in the final state are on-shell.
For the cross section at $\sqrt{s} = 10.6$~GeV, we evaluate $\alpha_{\rm s}$ at the scale $\sqrt{s}/2$ and take $\alpha_{\rm s} = 0.200$.
Of the fine structure constants appearing in eqs.~\eqref{sigmaQ0}-\eqref{sigmaQ2}, two originate from the virtual production mechanism, and are evaluated at virtuality $\sqrt{s}$,
and one originates from the real photon emission, and is evaluated at zero virtuality.
For $\alpha$ at the scale $\sqrt{s}$ we take $\alpha = 1/131$, while for $\alpha$ at zero virtuality we take $\alpha= 1/137$.
In the fit, we take the uncertainties in the decay rates and in the cross section to be $0.3$ times the central values, for the order $v^2$ corrections that we have not included, 
and $\alpha_{\rm s}^2$ times the central values, for the uncalculated corrections of higher orders in $\alpha_{\rm s}$.
We also add the experimental errors.

\begin{table}[ht]
\centering
\begin{tabular}{|c|c|c|c|c|}
\hline
Potential model& A & B & C & D \\
\hline
${\cal E}_1$~(GeV$^2$) & $-0.06 \pm 0.75$ & $-0.07 \pm 0.76$ & $-0.34 \pm 1.00$ & $-0.33 \pm 1.08$ \\
$i {\cal E}_2$~(GeV) &$-0.09{} \pm 0.72$ & $-0.06 \pm 0.72$ & $1.49 \pm 0.94$ & $1.75 \pm 1.01$ \\
\hline
\end{tabular}
\caption{\label{tab:e1e2} 
Results of the fit of the gluonic correlators ${\cal E}_1$ and $i{\cal E}_2$, when the wavefunctions and binding energies 
are computed within the potential models described in section~\ref{sec:potentials}.}
\end{table}

We determine ${\cal E}_1$ and $i{\cal E}_2$ by a least squares fit.
The results for ${\cal E}_1$ and $i{\cal E}_2$, when wavefunctions and binding energies 
are computed by means of the potential models described in section~\ref{sec:potentials}, are shown in Table~\ref{tab:e1e2}. 
The errors include the theoretical errors due to higher order corrections in $v^2$ and $\alpha_{\rm s}$ as described above, and the experimental errors in the data.
Taking the averages over the different models, we get 
\begin{eqnarray}
{\cal E}_1  &=& -0.20 {}^{+0.14}_{-0.14} \pm 0.90 \textrm{ GeV}^2\,, \label{E1final}\\
i{\cal E}_2 &=& 0.77 {}^{+0.98}_{-0.86} \pm 0.85 \textrm{ GeV} \,, \label{E2final}
\end{eqnarray}
where the first uncertainty comes from the potential model dependence 
and the second one is the average of the uncertainties in each potential model determination. 
The correlators ${\cal E}_1$ and $i{\cal E}_2$ have a size that is consistent, within uncertainties, with their naive scaling in powers of $\Lambda_{\rm QCD}$.
The uncertainties are, however, large, reflecting the large uncertainty carried by the potential models.
Vanishing small correlators are also consistent with our determinations.

One may wonder if it would not be possible to fit also the quarkonium wavefunctions eliminating in this way a major source of uncertainty. 
For the considered observables, see eqs.~\eqref{Gammachi0final}-\eqref{sigmachiJfinal}, it is not possible to disentangle 
the contribution of the wavefunction from the one of the correlator $i{\cal E}_2$. 
Hence, a fit would be able to determine a combination of the wavefunction and $i{\cal E}_2$, but not each of the two.
Since in section~\ref{sec:Pdecayproduction} we aim at making some predictions for $P$-wave bottomonium electromagnetic decay and production, 
we have chosen to add the information on the wavefunction coming from potential models and gain some insight in the universal correlator $i{\cal E}_2$.

The correlators undergo renormalization and therefore depend on a subtraction scheme and a renormalization scale.
The considered observables, at their present accuracy, are however insensitive to the renormalization of ${\cal E}_1$ and $i{\cal E}_2$, 
and, in particular, they are insensitive to the renormalization scale of the correlators. 
We may reasonably expect that the obtained values refer to a renormalization scale that is of the order of the typical hadronic scale, but at this point further specifications are not possible.
In section~\ref{sec:e3}, we will see instead a case where the observable is sensitive to the renormalization of the involved correlator, 
so that a proper renormalization scale can be fixed, at least at leading logarithmic accuracy.

\begin{table}[ht]
\centering
\begin{tabular}{|c|c|c|c|c|c|}
\hline
Potential model& A & B & C & D\\
\hline 
$\!\!\langle{\cal O}_1^{\rm em} ({}^3P_0)\rangle_{ \chi_{c0}}$~(GeV$^5$) \!\!\!& $\!\!0.163 \pm 0.046\!$ & $\!\!0.163 \pm 0.046\!$ & $\!\!0.149 \pm 0.034\!$ & $\!\!0.141 \pm 0.032\!$  \\
$\!\!\langle{\cal O}_1^{\rm em} ({}^3P_1)\rangle_{ \chi_{c1}}$~(GeV$^5$) \!\!\!& $\!\!0.172\pm 0.048\!$ & $\!\!0.172 \pm 0.048\!$ & $\!\!0.158 \pm 0.036\!$ & $\!\!0.152 \pm 0.035\!$ \\
$\!\!\langle{\cal O}_1^{\rm em} ({}^3P_2)\rangle_{ \chi_{c2}}$~(GeV$^5$) \!\!\!& $\!\!0.190 \pm 0.053\!$ & $\!\!0.190 \pm 0.053\!$ & $\!\!0.177 \pm 0.040\!$ & $\!\!0.173 \pm 0.040\!$ \\
$\!\!\langle{\cal T}_8^{\rm em} ({}^3P_J)\rangle_{ \chi_{cJ}}$~(GeV$^6$) \!\!\!& $\!\!-0.009 \pm 0.107\!$ & $\!\!-0.009 \pm 0.106\!$ & $\!\!-0.028 \pm 0.081\!$ & $\!\!-0.024 \pm 0.080\!$ \\
$\!\!\langle{\cal P}_1^{\rm em} ({}^3P_J) \rangle_{ \chi_{cJ}}$~(GeV$^7$) \!\!\!& $\!\!0.235 \pm 0.094\!$ & $\!\!0.233 \pm 0.094\!$ & $\!\!0.129 \pm 0.071\!$ & $\!\!0.146 \pm 0.070\!$ \\
\hline
\end{tabular}
\caption{\label{tab:charmmatrixelements} The LDMEs $\langle \chi_{cJ}(1P) | {\cal O}_1^{\rm em} ({}^3P_J) | \chi_{cJ}(1P) \rangle$
  ($\langle{\cal O}_1^{\rm em} ({}^3P_J)\rangle_{ \chi_{cJ}}$ for short), $\langle \chi_{cJ}(1P) |$ ${\cal T}_8^{\rm em} ({}^3P_J)$ $| \chi_{cJ}(1P) \rangle$
  ($\langle{\cal T}_8^{\rm em} ({}^3P_J)\rangle_{ \chi_{cJ}}$ for short) and $\langle \chi_{cJ}(1P) | {\cal P}_1^{\rm em} ({}^3P_J) | \chi_{cJ}(1P)$
  ($\langle{\cal P}_1^{\rm em} ({}^3P_J) \rangle_{ \chi_{cJ}}$ for short)
  obtained from our numerical analysis within the potential models of section~\ref{sec:potentials}.}
\end{table}

Combining the values of ${\cal E}_1$ and $i{\cal E}_2$ with the potential model calculations of the wavefunctions and binding energies in Table~\ref{tab:chicpotential},
we obtain the $\chi_{cJ}(1P)$ LDMEs listed in Table~\ref{tab:charmmatrixelements}, 
whose errors are due to the uncertainties in the correlators ${\cal E}_1$ and $i{\cal E}_2$.
Here and in the following we take into account that the uncertainties in  ${\cal E}_1$ and $i{\cal E}_2$ are correlated.
The averages read
\begin{eqnarray}
\langle \chi_{c0}(1P) | {\cal O}_1^{\rm em} ({}^3P_0) | \chi_{c0}(1P) \rangle &=& 0.154 {}^{+0.009}_{-0.013} \pm 0.039\textrm{~GeV$^5$,} \\
\langle \chi_{c1}(1P) | {\cal O}_1^{\rm em} ({}^3P_1) | \chi_{c1}(1P) \rangle &=& 0.164 {}^{+0.009}_{-0.012} \pm 0.042\textrm{~GeV$^5$,} \\
\langle \chi_{c2}(1P) | {\cal O}_1^{\rm em} ({}^3P_2) | \chi_{c2}(1P) \rangle &=& 0.183 {}^{+0.008}_{-0.010} \pm 0.047\textrm{~GeV$^5$,} \\
\langle \chi_{cJ}(1P) | {\cal T}_8^{\rm em} ({}^3P_J) | \chi_{cJ}(1P) \rangle &=& -0.017 {}^{+0.009}_{-0.010} \pm 0.094\textrm{~GeV$^6$,} \\
\langle \chi_{cJ}(1P) | {\cal P}_1^{\rm em} ({}^3P_J) | \chi_{cJ}(1P) \rangle &=& 0.186 {}^{+0.049}_{-0.057} \pm 0.082\textrm{~GeV$^7$,}
\end{eqnarray}
where the first uncertainties come from the potential model dependence and the second ones from the uncertainties in ${\cal E}_1$ and $i{\cal E}_2$.
In $\langle \chi_{cJ}(1P) |$ ${\cal T}_8^{\rm em} ({}^3P_J)$ $| \chi_{cJ}(1P) \rangle$ and $\langle \chi_{cJ}(1P)|$ ${\cal P}_1^{\rm em} ({}^3P_J)$  $|\chi_{cJ}(1P) \rangle$,
we have ignored the total angular momentum dependent corrections to the wavefunction at the origin, because they are of higher order in $v$. 
The obtained value for the octet LDME is compatible with zero.

The above values for $\langle \chi_{cJ}(1P) |$ ${\cal O}_1^{\rm em} ({}^3P_J)$ $| \chi_{cJ}(1P) \rangle$ are consistent with the determination in~\cite{Braaten:2002fi},
which uses a model very close to our model C, where the authors obtain 
$\langle \chi_{cJ}(1P) | {\cal O}_1^{\rm em} ({}^3P_J) | \chi_{cJ}(1P) \rangle = 0.107 \pm 0.032 \,\textrm{GeV}^5$.
The values for $\langle \chi_{cJ}(1P) |$ ${\cal T}_8^{\rm em} ({}^3P_J)$ $| \chi_{cJ}(1P) \rangle$ and $\langle \chi_{cJ}(1P) |$ ${\cal P}_1^{\rm em} ({}^3P_J)$ $| \chi_{cJ}(1P) \rangle$
are consistent, within errors, with the ones obtained in~\cite{Brambilla:2017kgw}:
$\langle \chi_{cJ}(1P) |$ ${\cal T}_8^{\rm em} ({}^3P_J)$ $| \chi_{cJ}(1P) \rangle$ $=$ $0.045 \pm 0.052 \pm 0.014 \pm 0.039 \, \textrm{GeV}^6$
and $\langle \chi_{cJ}(1P) |$ ${\cal P}_1^{\rm em} ({}^3P_J)$ $| \chi_{cJ}(1P) \rangle = 0.058 \pm 0.074 \pm 0.086 \pm 0.033 \, \textrm{GeV}^7$.

\begin{table}[ht]
\centering
\begin{tabular}{|c|c|c|c|c|}  
\hline
  Potential model & A & B & C & D\\
\hline
  $\Gamma_{\chi_{c0}(1P)}^{\gamma \gamma}$~(keV) & $2.92 \pm 0.54$ & $2.91 \pm
0.54$ & $2.76 \pm 0.52$ & $2.62 \pm 0.50$ \\
  $\Gamma_{\chi_{c2}(1P)}^{\gamma \gamma}$~(keV) & $0.58 \pm 0.16$ & $0.58 \pm 0.16$ & $0.58 \pm 0.16$ & $0.59 \pm 0.17$ \\
\hline
\end{tabular}
\caption{\label{tab:chicdecayrates}
  Results for the two photon decay widths of the states $\chi_{c0}(1P)$ and $\chi_{c2}(1P)$,
  indicated with  $\Gamma_{\chi_{c0}(1P)}^{\gamma \gamma}$ and
$\Gamma_{\chi_{c2}(1P)}^{\gamma \gamma}$ for short, for each of the potential
models of section~\ref{sec:potentials}.}
\end{table}

The results for the two photon decay widths of the charmonium $P$-wave states $\chi_{c0}(1P)$ and $\chi_{c2}(1P)$ for each potential model
determination of the wavefunction and binding energy are listed in Table~\ref{tab:chicdecayrates}. 
The errors are due to the uncertainties in the correlators  ${\cal E}_1$ and $i{\cal E}_2$.
The averages of these determinations read
\begin{eqnarray}
\Gamma(\chi_{c0}(1P) \rightarrow \gamma \gamma) &=& 2.80 {}^{+0.12}_{-0.19} \pm 0.52   \; \textrm{keV}\,,
\label{Gammachic0result}  \\
\Gamma(\chi_{c2}(1P) \rightarrow \gamma \gamma) &=& 0.58 {}^{+0.01}_{-0.00} \pm 0.16    \; \textrm{keV} \,,
\label{Gammachic2result}                                                  
\end{eqnarray}
where the first  uncertainty comes from the potential model dependence and the second one is the average of the uncertainties from each potential model. 

\begin{table}[ht]
\centering
\begin{tabular}{|c|c|c|c|c|}
\hline
Potential model & A & B & C & D \\
\hline
$\sigma(e^+ e^- \to \chi_{c0}(1P) + \gamma)$~(fb) & $2.10 \pm 0.80$ & $2.08 \pm
0.80$ & $1.58 \pm 0.71$ & $1.62 \pm 0.71$ \\
$\sigma(e^+ e^- \to \chi_{c1}(1P) + \gamma)$~(fb) & $16.2 \pm 6.3$ & $16.2 \pm 6.3$ & $16.4 \pm 6.4$ & $16.6 \pm 6.4$ \\
$\sigma(e^+ e^- \to \chi_{c2}(1P) + \gamma)$~(fb) & $3.19 \pm 1.97$ & $ 3.22 \pm 1.98$ & $4.18 \pm 2.29$ & $4.42 \pm 2.39$ \\
\hline
\end{tabular}
\caption{\label{tab:chicprod} 
  Results for the cross sections $\sigma(e^+ e^- \to \chi_{cJ}(1P) + \gamma)$
at $\sqrt{s} =10.6$~GeV for the potential models described in section~\ref{sec:potentials}.}
\end{table}

The determined values of ${\cal E}_1$ and $i{\cal E}_2$ allow us to make predictions for the cross sections $\sigma(e^+ e^- \to \chi_{cJ}(1P) + \gamma)$. 
In Table~\ref{tab:chicprod}, we list for each potential model the results at $\sqrt{s} =10.6$~GeV. 
The uncertainties in Table~\ref{tab:chicprod} are computed from the uncertainties of ${\cal E}_1$ and $i{\cal E}_2$, 
which already account for the uncertainties originating from the missing corrections of relative order $v^2$ and $\alpha_s^2$, and from adding in quadrature the uncertainty
that comes from varying $\alpha_{\rm s}$ between $\alpha_{\rm s} (\sqrt{s}) = 0.171$ and $\alpha_{\rm s} (\sqrt{s}/4) = 0.245$. 
From the averages of the results in Table~\ref{tab:chicprod} we obtain
\begin{eqnarray}
\sigma(e^+ e^- \to \chi_{c0}(1P) + \gamma) &=& 1.84 {}^{+0.25}_{-0.26} \pm 0.76 \textrm{~fb}\,,
\label{sigmachic0result}\\
\sigma(e^+ e^- \to \chi_{c1}(1P) + \gamma) &=& 16.4 {}^{+0.2}_{-0.2}  \pm 6.4 \textrm{~fb}\,, 
\label{sigmachic1result}\\
\sigma(e^+ e^- \to \chi_{c2}(1P) + \gamma) &=& 3.75 {}^{+0.67}_{-0.56} \pm 2.16 \textrm{~fb}\,,
\label{sigmachic2result}
\end{eqnarray}
where the first uncertainty is from the model dependence and the second one is the average of the uncertainties in Table~\ref{tab:chicprod}. 
The obtained cross sections are consistent, inside errors, with the results of ref.~\cite{Brambilla:2017kgw}.

It is worthwhile emphasizing that, although the measured two photon decay widths of the $\chi_{c0}(1P)$ and $\chi_{c2}(1P)$ states
and the cross section $\sigma(e^+ e^- \to \chi_{c1}(1P) + \gamma)$ have been used as an experimental input,  
the theoretical results for these quantities, eqs.~\eqref{Gammachic0result}, \eqref{Gammachic2result} and \eqref{sigmachic1result},
and their agreement with the data, eqs.~\eqref{Gammachic0exp}-\eqref{sigmachic1exp}, is nevertheless significant.
The reason is that the two correlators ${\cal E}_1$ and $i{\cal E}_2$ are the result of a least squares fit of three data 
and not of a fine tuning of some of them.

\subsection[$P$-wave bottomonium electromagnetic decay and production]
{\boldmath $P$-wave bottomonium electromagnetic decay and production}
\label{sec:Pdecayproduction}
With the values of ${\cal E}_1$ and $i{\cal E}_2$ determined in the previous section 
we can make predictions for $P$-wave electromagnetic decay widths and production cross sections of bottomonium states. 
Wavefunctions and binding energies are computed according to the potential model results listed in Table~\ref{tab:chibpotential}. 

\begin{table}[ht]
\centering
\begin{tabular}{|c|c|c|c|c|c|}
\hline
Potential model & A & B & C & D & E \\
\hline
$\Gamma_{\chi_{b0}(1P)}^{\gamma \gamma}$~(eV)& $58.3 \pm 7.8$ & $45.5 \pm 6.1$ & $49.1 \pm 7.0$ & $32.6 \pm 4.7$ & $47.3 \pm 6.9$ \\ 
$\Gamma_{\chi_{b2}(1P)}^{\gamma \gamma}$~(eV)& $11.2 \pm 1.5$ & $8.7 \pm 1.2$ & $9.6 \pm 1.4$ & $6.6 \pm 1.0$ & $9.5 \pm 1.4$ \\
\hline
$\Gamma_{\chi_{b0}(2P)}^{\gamma \gamma}$~(eV)& $58.6 \pm 7.8$ & $44.0 \pm 5.8$ & $48.6 \pm 6.8$ & $34.1 \pm 4.9$ & $48.7 \pm 7.0$ \\ 
$\Gamma_{\chi_{b2}(2P)}^{\gamma \gamma}$~(eV)& $11.4 \pm 1.5$ & $8.5 \pm 1.2$ & $9.6 \pm 1.4$ & $6.9 \pm 1.0$ & $9.8 \pm 1.4$ \\ 
\hline
$\Gamma_{\chi_{b0}(3P)}^{\gamma \gamma}$~(eV)& $57.9 \pm 7.6$ & $42.6 \pm 5.6$ & $47.0 \pm 6.6$ & $34.3 \pm 4.9$ & $48.5 \pm 7.0$ \\ 
$\Gamma_{\chi_{b2}(3P)}^{\gamma \gamma}$~(eV)& $11.4 \pm 1.5$ & $8.4 \pm 1.1$ & $9.4 \pm 1.3$ & $7.0 \pm 1.0$ & $9.9 \pm 1.4$ \\ 
\hline
\end{tabular}
\caption{\label{tab:chibdecay} Results for the two photon decay widths of the states $\chi_{b0}(nP)$ and $\chi_{b2}(nP)$,
  indicated with  $\Gamma_{\chi_{c0}(nP)}^{\gamma \gamma}$ and $\Gamma_{\chi_{c2}(nP)}^{\gamma \gamma}$ respectively, 
  for each of the potential models described in section~\ref{sec:potentials}.}
\end{table}

We consider, first, the two photon decay rates of the states $\chi_{b0}(nP)$ and  $\chi_{b2}(nP)$ with $n=1$, $2$ and $3$.
Following the same procedure discussed in section~\ref{sec:e1e2} for charmonium states, we replace in our theoretical expressions for the $\chi_{bJ}(nP)$ decay widths 
the bottom pole mass with the spin average of the $nP$ bottomonium masses.
We use the equivalent of eq.~\eqref{Mspinaverage} and eq.~\eqref{Mmrelation}.
The masses of the $1P$, $2P$ and $3P$ bottomonium states are taken from ref.~\cite{Tanabashi:2018oca}.\footnote{
  Since only the $3^3P_2$ and $3^3P_1$ states have been observed among the $n=4$, $L=1$ bottomonium states,
  we include only them when computing the spin average of the $3P$ bottomonium masses and normalize accordingly.}
Moreover, we take $\alpha = 1/137$ reflecting the fact that the photons in the final state are on shell,
and we take $\alpha_{\rm s} = 0.200$ at the scale of half the spin averaged masses.
The results for each choice of potential model used to compute wavefunctions and binding energies are shown in Table~\ref{tab:chibdecay}.
The uncertainties in Table~\ref{tab:chibdecay} come from the correlated uncertainties in ${\cal E}_1$ and $i{\cal E}_2$,
as well as from the uncertainties stemming from uncalculated corrections of order $v^2$ and $\alpha_{\rm s}^2$ in the bottomonium sector,
which we estimate to be $0.1$ and $\alpha_{\rm s}^2$ times the central values, respectively.
The uncertainties are added in quadrature.

After averaging over the determinations from the different potential models, we obtain the following predictions
\begin{eqnarray}
\Gamma(\chi_{b0}(1P) \rightarrow \gamma \gamma) &=& 46.6{}^{+11.7}_{-14.0} \pm 6.5 \textrm{~eV},
\label{Gammachib01Presult}\\
\Gamma(\chi_{b2}(1P) \rightarrow \gamma \gamma) &=& 9.1{}^{+2.1}_{-2.5} \pm 1.3 \textrm{~eV}, 
\label{Gammachib21Presult}\\
\nonumber\\  
\Gamma(\chi_{b0}(2P) \rightarrow \gamma \gamma) &=& 46.8{}^{+11.8}_{-12.7} \pm 6.5 \textrm{~eV}, 
\label{Gammachib02Presult}\\
\Gamma(\chi_{b2}(2P) \rightarrow \gamma \gamma) &=& 9.3{}^{+2.1}_{-2.3} \pm 1.3 \textrm{~eV}, 
\label{Gammachib22Presult}\\
\nonumber\\
\Gamma(\chi_{b0}(3P) \rightarrow \gamma \gamma) &=& 46.1{}^{+11.9}_{-11.8} \pm 6.3 \textrm{~eV},
\label{Gammachib03Presult}\\ 
\Gamma(\chi_{b2}(3P) \rightarrow \gamma \gamma) &=& 9.2{}^{+2.2}_{-2.2} \pm 1.3 \textrm{~eV},
\label{Gammachib23Presult}                                                    
\end{eqnarray}
where the first uncertainty comes from the potential model dependence, and the second one is the average of the uncertainties in Table~\ref{tab:chibdecay}.

\begin{figure}[ht]
  \centering
\includegraphics[width=.6\textwidth]{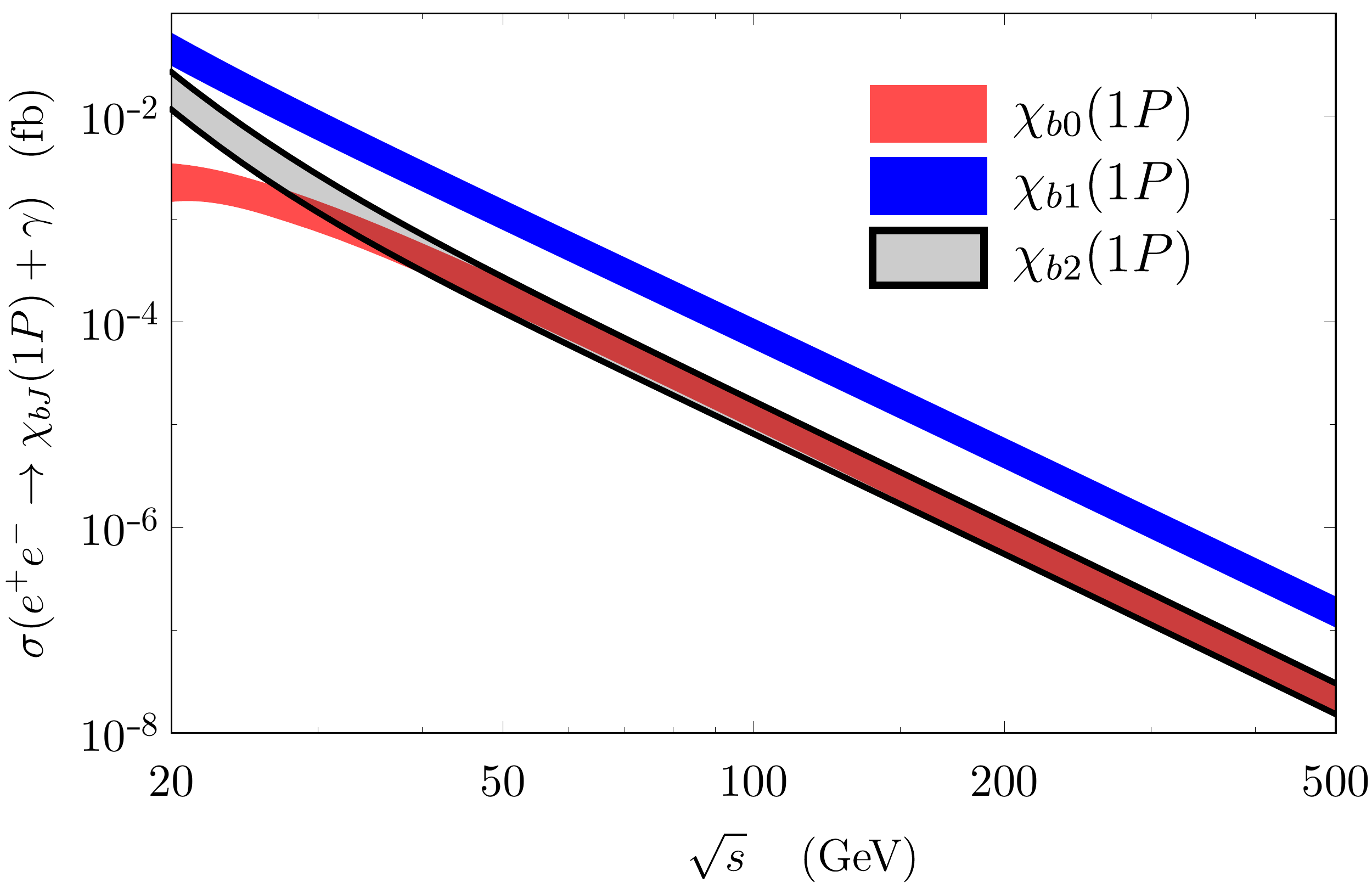}
\caption{\label{fig:chib1Pcrosssection} 
  Predicted cross sections $\sigma(e^+ e^- \to \chi_{bJ}(1P) + \gamma)$ for $J=0$ (red band), $J=1$ (blue band), and $J=2$ (grey band with black lines).}
\end{figure}

\begin{figure}[ht]
  \centering
\includegraphics[width=.6\textwidth]{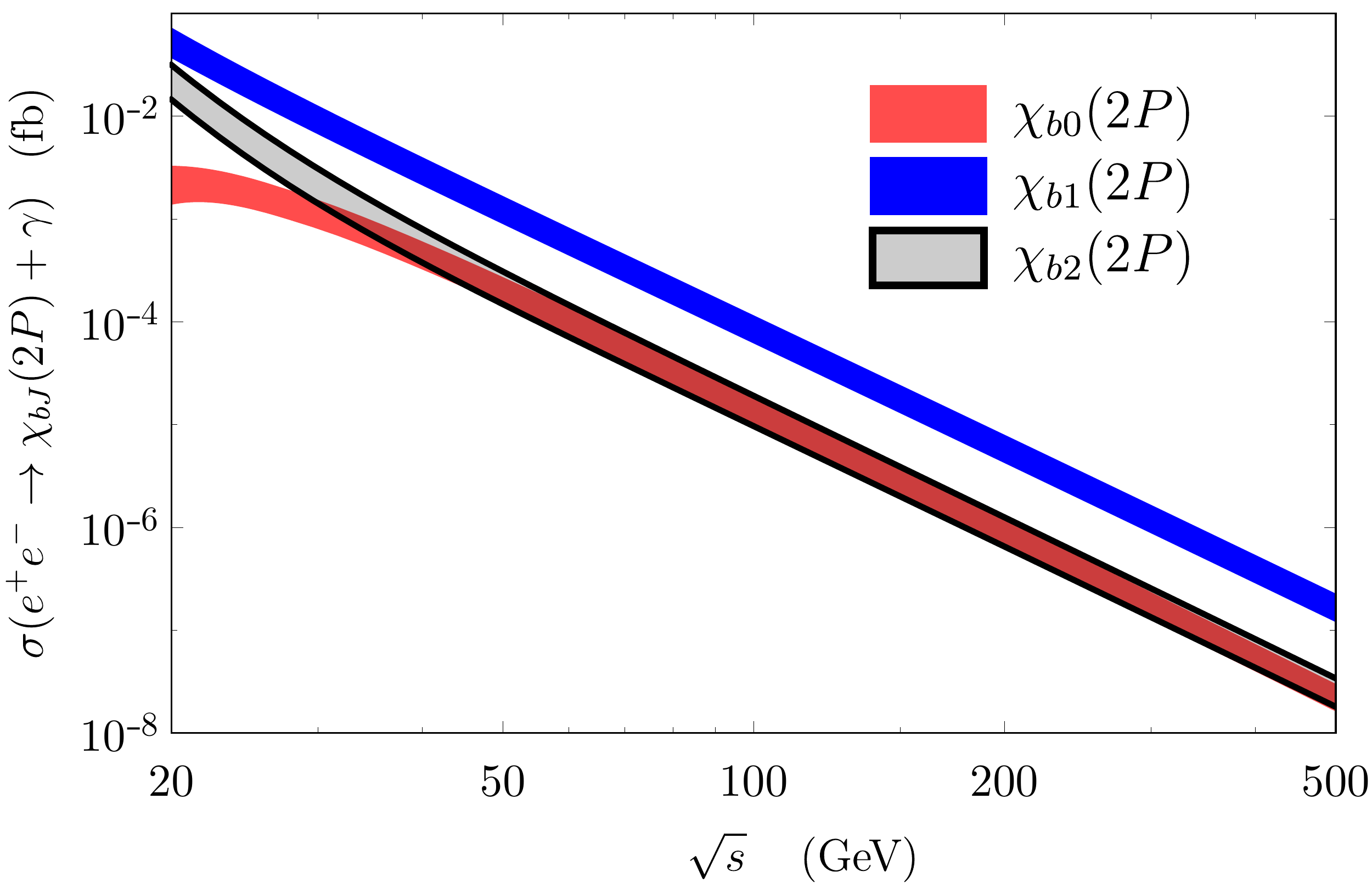}
\caption{\label{fig:chib2Pcrosssection} 
  Predicted cross sections $\sigma(e^+ e^- \to \chi_{bJ}(2P) + \gamma)$ for $J=0$ (red band), $J=1$ (blue band), and $J=2$ (grey band with black lines).}
\end{figure}

\begin{figure}[ht]
  \centering
\includegraphics[width=.6\textwidth]{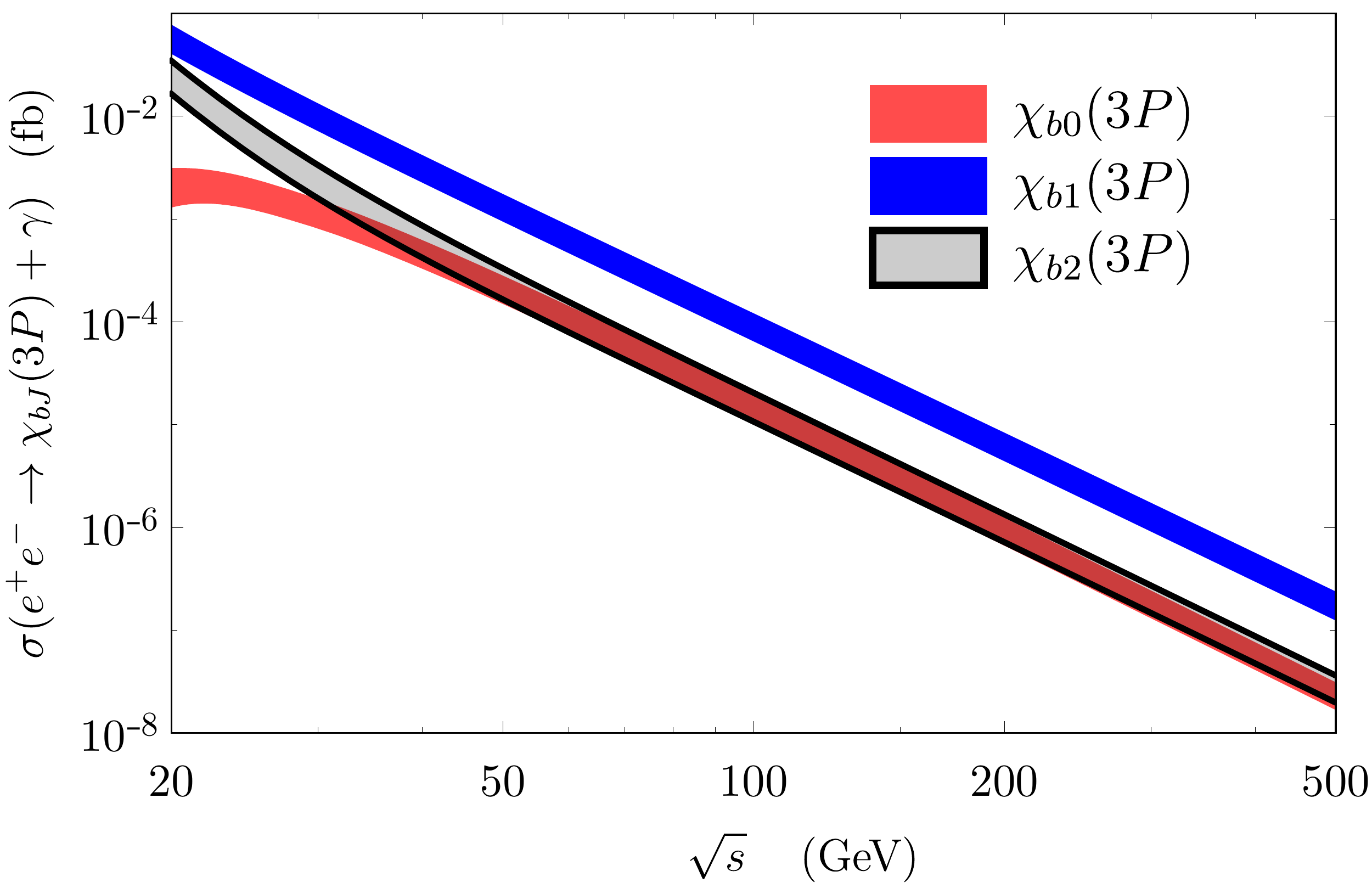}
\caption{\label{fig:chib3Pcrosssection} 
  Predicted cross sections $\sigma(e^+ e^- \to \chi_{bJ}(3P) + \gamma)$ for $J=0$ (red band), $J=1$ (blue band), and $J=2$ (grey band with black lines).}
\end{figure}

Using the same input as for the two photon decay widths, we can also make predictions for the cross sections $\sigma(e^+ e^- \to \chi_{bJ}(nP)+\gamma)$.
As has been pointed out in this context in ref.~\cite{Sang:2009jc} and mentioned at the end of appendix~\ref{app:NRQCDfact},
the perturbative expression of the electromagnetic cross section becomes singular when the center of mass energy approaches the heavy quark-antiquark pair production threshold.
In the bottomonium case, this threshold is around 10~GeV.
Therefore, in order to make predictions for $\sigma(e^+ e^- \to \chi_{bJ}(nP)+\gamma)$ using the factorization formulas provided in appendix~\ref{app:NRQCDfact},
the center of mass energy has to be significantly larger than 10~GeV.
We look at the energy range 20~GeV$<\sqrt{s}<$500~GeV that encompasses the energies of a possible future $e^+e^-$ collider. 
We evaluate $\alpha_{\rm s}$ at the scale $\sqrt{s}/2$. 
Furthermore, we fix $\alpha = 1/128$ and neglect the running, as the running of $\alpha$ only affects the cross section by less than a few percent,
which is negligible compared to other uncertainties.
The results for the electromagnetic cross sections of the $1P$, $2P$, and $3P$ bottomonium states are shown
in figures~\ref{fig:chib1Pcrosssection}, \ref{fig:chib2Pcrosssection} and~\ref{fig:chib3Pcrosssection}, respectively. 
The central values are obtained by averaging over the determinations of the
wavefunctions and binding energies from the different potential models
described in section~\ref{sec:potentials}.
The bands account for the uncertainties, which include
potential model dependence,  
uncertainties in ${\cal E}_1$ and $i{\cal E}_2$, 
uncertainties from uncalculated corrections of order $v^2$ and $\alpha_{\rm s}^2$, which we estimate to be 0.1 and $\alpha_{\rm s}^2$ times the central values, 
and uncertainties coming from varying $\alpha_{\rm s}$ between $\alpha_{\rm s}(\sqrt{s}/4)$ and $\alpha_{\rm s}(\sqrt{s})$.
We add these uncertainties in quadrature.
In particular, at $\sqrt{s} = 20$~GeV the cross sections are
\begin{eqnarray}
\sigma(e^+ e^- \to \chi_{b0}(1P) + \gamma) &=& (2.47 \pm 0.83 \pm 0.56) \times 10^{-3} \textrm{~fb}\,,
\label{sigmachib01Presult20}\\
\sigma(e^+ e^- \to \chi_{b1}(1P) + \gamma) &=& (47.8 \pm 12.5 \pm 11.4) \times 10^{-3} \textrm{~fb}\,,
\label{sigmachib11Presult20}\\
\sigma(e^+ e^- \to \chi_{b2}(1P) + \gamma) &=& (19.1 \pm 4.8 \pm 5.9) \times 10^{-3} \textrm{~fb}\,,
\label{sigmachib21Presult20}\\
\nonumber\\
\sigma(e^+ e^- \to \chi_{b0}(2P) + \gamma) &=& (2.33 \pm 0.78 \pm 0.55) \times 10^{-3} \textrm{~fb}\,,
\label{sigmachib02Presult20}\\
\sigma(e^+ e^- \to \chi_{b1}(2P) + \gamma) &=& (54.6 \pm 13.0 \pm 12.4) \times 10^{-3} \textrm{~fb}\,,
\label{sigmachib12Presult20}\\
\sigma(e^+ e^- \to \chi_{b2}(2P) + \gamma) &=& (22.9 \pm 5.3 \pm 6.6) \times 10^{-3} \textrm{~fb}\,,
\label{sigmachib22Presult20}\\
\nonumber\\
\sigma(e^+ e^- \to \chi_{b0}(3P) + \gamma) &=& (2.21 \pm 0.75 \pm 0.53) \times 10^{-3} \textrm{~fb}\,,
\label{sigmachib03Presult20}\\
\sigma(e^+ e^- \to \chi_{b1}(3P) + \gamma) &=& (58.9 \pm 13.2 \pm 13.0) \times 10^{-3} \textrm{~fb}\,,
\label{sigmachib13Presult20}\\
\sigma(e^+ e^- \to \chi_{b2}(3P) + \gamma) &=& (25.4 \pm 5.6 \pm 7.0) \times 10^{-3} \textrm{~fb}\,,
\label{sigmachib23Presult20}
\end{eqnarray}
where the first uncertainties come from the dependence on the potential models, 
and the second ones account for the other uncertainties that we have mentioned above eq.~\eqref{sigmachib01Presult20}.
Similarly at $\sqrt{s} = 90$~GeV we obtain 
\begin{eqnarray}
\sigma(e^+ e^- \to \chi_{b0}(1P) + \gamma) &=& (2.05 \pm 0.61 \pm 0.26) \times 10^{-5} \textrm{~fb}\,,
\label{sigmachib01Presult90}\\
\sigma(e^+ e^- \to \chi_{b1}(1P) + \gamma) &=& (12.1 \pm 3.5 \pm 2.0) \times 10^{-5} \textrm{~fb}\,,
\label{sigmachib11Presult90}\\
\sigma(e^+ e^- \to \chi_{b2}(1P) + \gamma) &=& (1.89 \pm 0.45 \pm 0.49)\times 10^{-5} \textrm{~fb}\,,
\label{sigmachib21Presult90}\\
\nonumber\\
\sigma(e^+ e^- \to \chi_{b0}(2P) + \gamma) &=& (2.20 \pm 0.60 \pm 0.28) \times 10^{-5} \textrm{~fb}\,,
\label{sigmachib02Presult90}\\
\sigma(e^+ e^- \to \chi_{b1}(2P) + \gamma) &=& (13.3 \pm 3.4 \pm 2.1) \times 10^{-5} \textrm{~fb}\,,
\label{sigmachib12Presult90}\\
\sigma(e^+ e^- \to \chi_{b2}(2P) + \gamma) &=& (2.17 \pm 0.47 \pm 0.54)\times 10^{-5} \textrm{~fb}\,,
\label{sigmachib22Presult90}\\
\nonumber\\
\sigma(e^+ e^- \to \chi_{b0}(3P) + \gamma) &=& (2.27 \pm 0.63 \pm 0.29) \times 10^{-5} \textrm{~fb}\,,
\label{sigmachib03Presult90}\\
\sigma(e^+ e^- \to \chi_{b1}(3P) + \gamma) &=& (13.9 \pm 3.6 \pm 2.1) \times 10^{-5} \textrm{~fb}\,,
\label{sigmachib13Presult90}\\
\sigma(e^+ e^- \to \chi_{b2}(3P) + \gamma) &=& (2.35 \pm 0.48 \pm 0.56) \times 10^{-5} \textrm{~fb}\,.
\label{sigmachib23Presult90}
\end{eqnarray}

\subsection[$P$-wave charmonium and bottomonium decay into light hadrons]
{\boldmath $P$-wave charmonium and bottomonium decay into light hadrons}
\label{sec:e3}
In this section, we analyze inclusive $P$-wave quarkonium decays into light hadrons (LH) at leading order in the velocity expansion.
The NRQCD factorization formula has been first derived in~\cite{Bodwin:1992ye} and we reproduce it in eq.~\eqref{factNRQCDPhad}.
It depends on two LDMEs. 
The color singlet LDME has been factorized in strongly coupled pNRQCD at leading order in $v$ in eq.~\eqref{eq:o1mathad}.
The color octet matrix element can be written in strongly coupled pNRQCD at leading order in $v$ as~\cite{Brambilla:2001xy}
\begin{equation}
\label{eq:O83S1pNRQCD}
\langle \chi_{QJ}(nP)|  {\cal O}_8(^1S_0) | \chi_{QJ}(nP) \rangle = \frac{2 T_F}{9N_c m^2}\,  \frac{3N_c}{2\pi}|R^{(0)\,\prime}_{n1}({0})|^2 {\cal E}_3\,,
\end{equation}
where the gluonic correlator ${\cal E}_3$ has been defined in eq.~\eqref{corr2}.
The expression of the decay width of a $P$-wave quarkonium into light hadrons under the conditions of validity of strongly coupled pNRQCD 
reads therefore at leading order in $v$:
\begin{equation}
\Gamma(\chi_{QJ}(nP) \rightarrow {\rm LH}) =  \frac{3N_c}{2\pi} \, |R_{n1}'(0)|^2  \left[ 32 \frac{ {\rm Im} f_1 ({}^3P_J)(\mu_\Lambda)}{M_{nP_Q}^4}
+ 32 \frac{ {\rm Im} f_8 ({}^3S_1)}{M_{nP_Q}^4} \frac{2 T_F}{9 N_c} {\cal E}_3(\mu_\Lambda) \right],
\label{GammachicJhadpNRQCD}
\end{equation}
where we have expressed the heavy quark pole mass in terms of the spin averaged $nP_Q$ mass
(analogous to the spin averaged mass defined in eq.~\eqref{Mspinaverage}) at leading order in the velocity.
The NRQCD short distance coefficients up to order $\alpha_{\rm s}^3$ accuracy are listed in eqs.~\eqref{f13P0}-\eqref{f83S1}. 

In eq.~\eqref{GammachicJhadpNRQCD} we have emphasized that both ${\rm Im} f_1 ({}^3P_J)(\mu_\Lambda)$ and ${\cal E}_3(\mu_\Lambda)$ depend on a cutoff $\mu_\Lambda$.
We use the $\overline{\rm MS}$ scheme for both quantities.
The correlator ${\cal E}_3(\mu_\Lambda)$ is dimensionless, and depends logarithmically on the scale~$\mu_\Lambda$. 
This dependence cancels in the decay width \eqref{GammachicJhadpNRQCD} 
against the $\mu_\Lambda$ dependence of the short distance coefficient ${\rm Im} f_1 ({}^3P_J)(\mu_\Lambda)$~\cite{Brambilla:2001xy}.
Also the one loop running with respect to the scale $\mu_\Lambda$ of ${\rm Im} f_1 ({}^3P_J)(\mu_\Lambda)$ and  ${\cal E}_3(\mu_\Lambda)$ is known.

We determine ${\cal E}_3(\mu_\Lambda)$ from a least squares fit to the ratios of decay rates
$\Gamma(\chi_{c0}(1P) \rightarrow {\rm LH})/\Gamma(\chi_{c1}(1P) \rightarrow{\rm LH})$, ~~$\Gamma(\chi_{c1}(1P) \rightarrow {\rm LH})/\Gamma(\chi_{c2}(1P)\rightarrow {\rm LH})$,~~
$\Gamma(\chi_{c0}(1P) \rightarrow$ $ {\rm LH})/$ $\Gamma(\chi_{c0}(1P)$ $ \rightarrow $ $\gamma \gamma)$, and $\Gamma(\chi_{c2}(1P) \rightarrow {\rm LH})/\Gamma(\chi_{c2}(1P) \rightarrow \gamma \gamma)$
at leading order in $v$. 
The theoretical expressions for the decay rates that we use are valid up to next-to-leading order in $\alpha_{\rm s}$, except for $\Gamma(\chi_{c1}(1P) \rightarrow {\rm LH})$,
which is known only at leading order in $\alpha_{\rm s}$. 
The decay rates $\Gamma(\chi_{cJ} (1P) \to {\rm LH})$ have been obtained by subtracting radiative decay rates and transition rates into other charmonia
from the total widths of $\chi_{cJ} (1P)$ given in ref.~\cite{Tanabashi:2018oca}. 
Among the subtracted rates, only the radiative transition into $J/\psi + \gamma$ makes a significant contribution. 
The experimental values of $\Gamma(\chi_{cJ}(1P) \rightarrow {\rm LH})$ that we use are 
\begin{eqnarray}
\Gamma(\chi_{c0} (1P) \to {\rm LH}) \big|_{\text{from PDG}} &=& 10.6 \pm 0.6 \textrm{~MeV}, \label{chic0LHexp}\\
\Gamma(\chi_{c1} (1P) \to {\rm LH}) \big|_{\text{from PDG}} &=& 0.552 \pm 0.041 \textrm{~MeV}, \label{chic1LHexp}\\
\Gamma(\chi_{c2} (1P) \to {\rm LH}) \big|_{\text{from PDG}} &=& 1.60 \pm 0.09 \textrm{~MeV}. \label{chic2LHexp}
\end{eqnarray}
We set $m_c = M_{1P_c}/2$ and $\mu_\Lambda = 1$~GeV. 
We use $\alpha_{\rm s}(m_c) = 0.282$ and $n_f = 3$. 
We take the theoretical uncertainties of the ratios 
$\Gamma(\chi_{c0}(1P) \rightarrow {\rm LH})/\Gamma(\chi_{c1}(1P) \rightarrow {\rm LH})$ and $\Gamma(\chi_{c1}(1P) \rightarrow {\rm LH})/\Gamma(\chi_{c2}(1P) \rightarrow {\rm LH})$
to be $0.3$ times the central values for the uncalculated order $v^2$ corrections, 
and $\alpha_{\rm s}$ times the central values for corrections of higher orders in $\alpha_{\rm s}$. 
For the ratios 
$\Gamma(\chi_{c0}(1P) \rightarrow {\rm LH})/\Gamma(\chi_{c0}(1P) \rightarrow \gamma \gamma)$ and $\Gamma(\chi_{c2}(1P) \rightarrow {\rm LH})/\Gamma(\chi_{c2}(1P) \rightarrow \gamma \gamma)$, 
we take the theoretical uncertainties to be $0.3$ times the central values for ignoring the order $v^2$ corrections,
and $\alpha_{\rm s}^2$ times the central values for corrections of higher orders in $\alpha_{\rm s}$.
At leading order in $v$ the ratios do not depend on the quarkonium wavefunctions.
We obtain, in the $\overline{\rm MS}$ scheme, 
\begin{equation}
\label{eq:e3result}
{\cal E}_3 (1\textrm{~GeV}) = 2.05 {}^{+0.94}_{-0.65}. 
\end{equation}
This result is compatible, within errors, with a previous determination in ref.~\cite{Brambilla:2001xy}. 
Nevertheless, we note that the uncertainties are smaller, despite the determination in~\cite{Brambilla:2001xy} did not include theoretical uncertainties.\footnote{
Note that the quantity ${\cal E} (\mu)$ in ref.~\cite{Brambilla:2001xy} corresponds to $N_c {\cal E}_3 (\mu)$ in this paper.}
From eq.~\eqref{eq:e3result}, we can compute ${\cal E}_3 (\mu_\Lambda)$ at different scales by using the one loop renormalization group improved expression~\cite{Brambilla:2001xy}:
\begin{equation}
\label{eq:e3evolution}
{\cal E}_3 (\mu_\Lambda) = {\cal E}_3 (\mu_\Lambda') + \frac{24 C_F}{\beta_0} \log \frac{\alpha_{\rm s} (\mu_\Lambda')}{\alpha_{\rm s}(\mu_\Lambda)}, 
\end{equation}
where $\beta_0 = 11 N_c/3 - 4 T_F n_f/3$.

\begin{table}[ht]
\centering
\begin{tabular}{|c|c|c|c|c|}
\hline
Potential model & A & B & C & D \\
\hline
$\langle {\cal O}_8(^1S_0) \rangle_{\chi_{cJ}} \times 10^3$~(GeV$^3$) & $4.59{}^{+2.10}_{-1.45}$ & $4.54{}^{+2.08}_{-1.44}$ & $2.63{}^{+1.20}_{-0.83}$ & $2.39{}^{+1.09}_{-0.76}$ \\ 
\hline
\end{tabular}
\caption{\label{tab:chico8result}
Results for the matrix element $\langle \chi_{cJ}(1P)| {\cal O}_8(^1S_0) |\chi_{cJ}(1P) \rangle$ at the scale $\mu_\Lambda = 1$~GeV, 
indicated with $\langle {\cal O}_8(^1S_0) \rangle_{\chi_{cJ}}$ for short.
The wavefunctions at the origin have been computed within the potential models
of section~\ref{sec:potentials}.}
\end{table}

From eq.~\eqref{eq:e3result} and eq.~\eqref{eq:O83S1pNRQCD}, we can compute the matrix element $\langle \chi_{cJ}(1P)|$ $  {\cal O}_8(^1S_0) $ $| \chi_{cJ}(1P) \rangle$. 
The results at the scale $\mu_\Lambda=1$~GeV for each potential model are listed in Table~\ref{tab:chico8result}.
If we average over them, we obtain 
\begin{equation}
\langle \chi_{cJ}(1P)|  {\cal O}_8(^1S_0) | \chi_{cJ}(1P) \rangle = (3.53{}^{+1.05}_{-1.15} {}^{+1.62}_{-1.12}) \times 10^{-3} \textrm{~GeV}^3,
\end{equation}
where the first uncertainty comes from the potential model dependence, and the second one is the average of the uncertainties in Table~\ref{tab:chico8result}.

From eq.~\eqref{eq:e3result} and eq.~\eqref{GammachicJhadpNRQCD}, we can compute the decay widths of the $\chi_{cJ}(1P)$ states into light hadrons. 
Rather than computing the decay rates directly using eq.~\eqref{GammachicJhadpNRQCD},
we determine $\Gamma(\chi_{cJ}(1P) \to {\rm LH})$ for $J=0$ and $2$ states by
combining the ratios $\Gamma(\chi_{cJ}(1P) \rightarrow$ $ {\rm LH})/$ $\Gamma(\chi_{cJ}(1P)$ $\rightarrow $ $\gamma \gamma)$ 
with the two photon decay widths computed in eq.~\eqref{Gammachic0result} and eq.~\eqref{Gammachic2result}. 
Similarly, we compute $\Gamma(\chi_{c1}(1P) \to {\rm LH})$
by combining these determinations of $\Gamma(\chi_{c0}(1P) \to {\rm LH})$ and $\Gamma(\chi_{c2}(1P) \to {\rm LH})$
with the ratios $\Gamma(\chi_{c0}(1P) \to {\rm LH})/\Gamma(\chi_{c1}(1P) \to {\rm LH})$ and $\Gamma(\chi_{c1}(1P) \to {\rm LH})/\Gamma(\chi_{c2}(1P) \to {\rm LH})$. 
This approach has the advantage that the ratios do not depend on the choice of potential model, a fact that reduces significantly the uncertainties.
Our results for the ratios $\Gamma(\chi_{cJ}(1P) \rightarrow$ $ {\rm LH})/$ $\Gamma(\chi_{cJ}(1P)$ $\rightarrow $ $\gamma \gamma)$ for $J=0$ and $2$ are 
\begin{eqnarray}
&& \frac{\Gamma(\chi_{c0}(1P) \rightarrow {\rm LH})}{\Gamma(\chi_{c0}(1P)\rightarrow \gamma \gamma)} = (2.96 ^{+0.92}_{-0.92}) \times 10^3, 
\label{eq:chic0gratioresult}\\
\nonumber\\  
&& \frac{\Gamma(\chi_{c2}(1P) \rightarrow {\rm LH})}{\Gamma(\chi_{c2}(1P)\rightarrow \gamma \gamma)} = (2.48 ^{+0.86}_{-0.77}) \times 10^3, 
\label{eq:chic2gratioresult}
\end{eqnarray}
where the uncertainties come from uncalculated corrections of relative order $v^2$ and $\alpha_{\rm s}^2$,
which are taken to be $0.3$ and $\alpha_{\rm s}^2$ times the central values, respectively.
These uncertainties are added in quadrature. 
Since the uncertainty in ${\cal E}_3$ is dominated by uncertainties from uncalculated higher order corrections to the theoretical expressions of the ratios,
we do not include the uncertainty in ${\cal E}_3$ to avoid double counting. 
Using eqs.~\eqref{Gammachic0result}-\eqref{Gammachic2result} and eqs.~\eqref{eq:chic0gratioresult}-\eqref{eq:chic2gratioresult},
we obtain for the inclusive decay widths into light hadrons:
\begin{eqnarray}
&& \Gamma(\chi_{c0}(1P) \rightarrow {\rm LH}) = 8.3^{+3.0}_{-3.1} \textrm{~MeV}, 
\label{eq:chic0LHth}\\
&& \Gamma(\chi_{c2}(1P) \rightarrow {\rm LH}) = 1.4^{+0.6}_{-0.6} \textrm{~MeV}. 
\label{eq:chic2LHth}
\end{eqnarray}
Comparing these results with the experimental determinations shown in eqs.~\eqref{chic0LHexp} and \eqref{chic2LHexp}, we see that they are consistent within errors. 
We also determine $\Gamma(\chi_{c1}(1P) \rightarrow {\rm LH})$ from the results in \eqref{eq:chic0LHth}-\eqref{eq:chic2LHth}
and the ratios $\Gamma(\chi_{c0}(1P) \rightarrow {\rm LH})/\Gamma(\chi_{c1}(1P) \rightarrow {\rm LH})$ and $\Gamma(\chi_{c1}(1P) \rightarrow {\rm LH})/\Gamma(\chi_{c2}(1P) \rightarrow {\rm LH})$.
The numerical results for these ratios are 
\begin{eqnarray}
&& \frac{\Gamma(\chi_{c0}(1P) \rightarrow {\rm LH})}{\Gamma(\chi_{c1}(1P) \rightarrow {\rm LH})} = 23.7 ^{+9.8}_{-9.8}\,, 
\label{eq:chic01ratioresult}\\
\nonumber\\
&& \frac{\Gamma(\chi_{c1}(1P) \rightarrow {\rm LH})}{\Gamma(\chi_{c2}(1P) \rightarrow {\rm LH})} = 0.33^{+0.16}_{-0.16}\,, 
\label{eq:chic12ratioresult}
\end{eqnarray}
where the uncertainties come from uncalculated corrections of relative order $v^2$ and $\alpha_{\rm s}$,
which are taken to be $0.3$ and $\alpha_{\rm s}$ times the central values, respectively.
These uncertainties are added in quadrature. 
If we use eq.~\eqref{eq:chic0LHth} and eq.~\eqref{eq:chic01ratioresult}, we obtain $\Gamma(\chi_{c1}(1P) \rightarrow {\rm LH}) = 0.35^{+0.28}_{-0.16}$~MeV, 
and if we use eq.~\eqref{eq:chic2LHth} and eq.~\eqref{eq:chic12ratioresult}, we obtain $\Gamma(\chi_{c1}(1P) \rightarrow {\rm LH}) = 0.48^{+0.28}_{-0.28}$~MeV. 
The average of the two determinations reads
\begin{equation}
\Gamma(\chi_{c1}(1P) \rightarrow {\rm LH}) = 0.42 ^{+0.06}_{-0.06} {}^{+0.28}_{-0.22} \textrm{~MeV}, 
\label{eq:chic1LHth}
\end{equation}
where the first uncertainty comes from the deviation of the central value of each determination from the average,
and the second is the average of the uncertainties in each determination. 
This result is also consistent with the experimental determination in eq.~\eqref{chic1LHexp} within errors.

We can also predict the decay widths of the $\chi_{bJ}(nP)$ states into light hadrons.
To compute $\Gamma(\chi_{bJ}(nP) \rightarrow {\rm LH})$ we take $m_b = M_{nP_b}/2$, with $\alpha_{\rm s} (m_b) = 0.200$, and $n_f=4$.
The short-distance coefficients in eqs.~\eqref{f13P0}-\eqref{f13P2} contain two scales, $m_b$ and $\mu_\Lambda$.
Since a value of $\mu_\Lambda$ that is too small compared to $m_b$ may spoil the convergence of the perturbation series,
we resum the leading logarithms of $\mu_\Lambda/M_{nP_b}$ that appear in the short-distance coefficients.
At the current level of accuracy, this is equivalent to computing ${\cal E}_3 (\mu_\Lambda)$ at the scale $\mu_\Lambda = M_{nP_b}$ using the formula in eq.~\eqref{eq:e3evolution}
and setting $\mu_\Lambda = M_{nP_b}$ in the short-distance coefficients \eqref{f13P0}-\eqref{f13P2}. 
As we have done for the $\chi_{cJ}(1P)$ states, we compute the decay widths $\Gamma(\chi_{bJ}(nP) \rightarrow {\rm LH})$
from the ratios $\Gamma(\chi_{b0}(nP) \rightarrow$ $ {\rm LH})/$ $\Gamma(\chi_{b0}(nP)$ $\rightarrow $ $\gamma \gamma)$,
$\Gamma(\chi_{b2}(nP) \rightarrow$ $ {\rm LH})/$ $\Gamma(\chi_{b2}(nP)$ $\rightarrow $ $\gamma \gamma)$, 
$\Gamma(\chi_{b0}(nP) \to {\rm LH})/\Gamma(\chi_{b1}(nP) \to {\rm LH})$ and $\Gamma(\chi_{b1}(nP) \to {\rm LH})/\Gamma(\chi_{b2}(nP) \to {\rm LH})$,
and the two photon widths determined in eqs.~\eqref{Gammachib01Presult}-\eqref{Gammachib23Presult}. 
Our results for the ratios $\Gamma(\chi_{b0}(nP) \to {\rm LH})/\Gamma(\chi_{b0}(nP) \to \gamma \gamma)$ and $\Gamma(\chi_{b2}(nP) \to {\rm LH})/\Gamma(\chi_{b2}(nP) \to \gamma \gamma)$ are 
\begin{eqnarray}
&& \frac{\Gamma(\chi_{b0}(1P) \rightarrow {\rm LH})}{\Gamma(\chi_{b0}(1P) \rightarrow \gamma \gamma)} = (23.0 ^{+2.5}_{-2.5}) \times 10^3, 
\label{eq:chib0g1Pratioresult}\\
\nonumber \\
&& \frac{\Gamma(\chi_{b2}(1P) \rightarrow {\rm LH})}{\Gamma(\chi_{b2}(1P) \rightarrow \gamma \gamma)} = (29.7^{+4.5}_{-3.6}) \times 10^3, 
\label{eq:chib2g1Pratioresult}\\
\nonumber \\
\nonumber \\
&& \frac{\Gamma(\chi_{b0}(2P) \rightarrow {\rm LH})}{\Gamma(\chi_{b0}(2P) \rightarrow \gamma \gamma)} = (23.0 ^{+2.5}_{-2.5}) \times 10^3, 
\label{eq:chib0g2Pratioresult}\\
\nonumber \\
&& \frac{\Gamma(\chi_{b2}(2P) \rightarrow {\rm LH})}{\Gamma(\chi_{b2}(2P) \rightarrow \gamma \gamma)} = (29.9^{+4.5}_{-3.6}) \times 10^3, 
\label{eq:chib2g2Pratioresult}\\
\nonumber \\
\nonumber \\
&& \frac{\Gamma(\chi_{b0}(3P) \rightarrow {\rm LH})}{\Gamma(\chi_{b0}(3P) \rightarrow \gamma \gamma)} = (23.0 ^{+2.5}_{-2.5}) \times 10^3, 
\label{eq:chib0g3Pratioresult} \\
\nonumber \\
&& \frac{\Gamma(\chi_{b2}(3P) \rightarrow {\rm LH})}{\Gamma(\chi_{b2}(3P) \rightarrow \gamma \gamma)} = (29.9^{+4.5}_{-3.7}) \times 10^3, 
\label{eq:chib2g3Pratioresult}
\end{eqnarray}
where the uncertainties come from the uncertainty in ${\cal E}_3$, and from the uncalculated corrections of order $v^2$ and of order $\alpha_{\rm s}^2$,
which are taken to be $0.1$ and $\alpha_{\rm s}^2$ times the central values, respectively.
These uncertainties are added in quadrature.
Using eqs.~\eqref{Gammachib01Presult}-\eqref{Gammachib23Presult} and eqs.~\eqref{eq:chib0g1Pratioresult}-\eqref{eq:chib2g3Pratioresult}, we obtain 
\begin{eqnarray}
&& \Gamma(\chi_{b0}(1P) \rightarrow {\rm LH}) = 1.07^{+0.33}_{-0.37} \textrm{~MeV}, 
\label{eq:chib01PLHth}\\
&& \Gamma(\chi_{b2}(1P) \rightarrow {\rm LH}) = 0.27^{+0.08}_{-0.10} \textrm{~MeV}, 
\label{eq:chib21PLHth} \\
\nonumber \\  
&& \Gamma(\chi_{b0}(2P) \rightarrow {\rm LH}) = 1.08^{+0.33}_{-0.35} \textrm{~MeV}, 
\label{eq:chib02PLHth} \\
&& \Gamma(\chi_{b2}(2P) \rightarrow {\rm LH}) = 0.28^{+0.09}_{-0.10} \textrm{~MeV}, 
\label{eq:chib22PLHth} \\
\nonumber \\
&& \Gamma(\chi_{b0}(3P) \rightarrow {\rm LH}) = 1.06^{+0.33}_{-0.33} \textrm{~MeV}, 
\label{eq:chib03PLHth} \\
&& \Gamma(\chi_{b2}(3P) \rightarrow {\rm LH}) = 0.28^{+0.09}_{-0.10} \textrm{~MeV}. 
\label{eq:chib23PLHth}
\end{eqnarray}
Now we determine the decay rate $\Gamma(\chi_{b1} (nP) \to {\rm LH})$
using the determinations of $\Gamma(\chi_{b0} (nP)$ $\to {\rm LH})$ and $\Gamma(\chi_{b2} (nP) \to {\rm LH})$ in eqs.~\eqref{eq:chib01PLHth}-\eqref{eq:chib23PLHth}
and the ratios $\Gamma(\chi_{b0}(nP) \to {\rm LH})/$ $\Gamma(\chi_{b1}(nP)$ $\to {\rm LH})$ and $\Gamma(\chi_{b1}(nP) \to {\rm LH})/\Gamma(\chi_{b2}(nP) \to {\rm LH})$.
Our results for the ratios are 
\begin{eqnarray}
&&\frac{\Gamma(\chi_{b0}(nP) \rightarrow {\rm LH})}{\Gamma(\chi_{b1}(nP) \rightarrow {\rm LH})} = 7.9 \pm 2.1\,, 
\label{eq:chib01ratioresult}\\
\nonumber\\
&&\frac{\Gamma(\chi_{b1}(nP) \rightarrow {\rm LH})}{\Gamma(\chi_{b2}(nP) \rightarrow {\rm LH})} = 0.54\pm 0.13\,, 
\label{eq:chib12ratioresult}
\end{eqnarray}
for $n=1,2,$ and 3, the differences between the results for different $n$ being negligible. 
The uncertainties come from the uncertainty in ${\cal E}_3$, and the uncalculated corrections of relative order $v^2$ and of relative order $\alpha_{\rm s}$,
which are taken to be $0.1$ and $\alpha_{\rm s}$ times the central values, respectively.
These uncertainties are added in quadrature.
Using the numerical results for the decay rates $\Gamma(\chi_{b0} (nP) \to {\rm LH})$ in eqs.~\eqref{eq:chib01PLHth}, \eqref{eq:chib02PLHth} and \eqref{eq:chib03PLHth},
and the ratio $\Gamma(\chi_{b0}(nP) \rightarrow {\rm LH})/\Gamma(\chi_{b1}(nP) \rightarrow {\rm LH})$ in eq.~\eqref{eq:chib01ratioresult},
we obtain the following determination for $\Gamma(\chi_{b1} (nP) \to {\rm LH})$: 
\begin{equation}
\Gamma(\chi_{b1}(nP) \rightarrow {\rm LH}) = 0.14 \pm 0.06 \textrm{~MeV}, 
\label{eq:chib1nPLHth}
\end{equation}
where, again, we find negligible differences between the results for $n=1,2,$ and 3.
If we compute this decay rate by using the values of $\Gamma(\chi_{b2} (nP) \to {\rm LH})$ in eqs.~\eqref{eq:chib21PLHth}, \eqref{eq:chib22PLHth} and \eqref{eq:chib23PLHth},
and the ratio $\Gamma(\chi_{b1}(nP) \rightarrow {\rm LH})/\Gamma(\chi_{b2}(nP) \rightarrow {\rm LH})$ in eq.~\eqref{eq:chib12ratioresult}, 
we find the same result as in eq.~\eqref{eq:chib1nPLHth}.

The predictions for the widths $\Gamma(\chi_{bJ}(1P) \rightarrow {\rm LH})$ given in eqs.~\eqref{eq:chib01PLHth}, \eqref{eq:chib1nPLHth} and~\eqref{eq:chib21PLHth}
are compatible with the total widths of the $\chi_{bJ}(1P)$ states recently computed in ref.~\cite{Segovia:2018qzb} from the electric dipole transition widths.
For the total width of the $\chi_{b0}(1P)$ state, the Belle collaboration has determined an upper limit, $\Gamma_{\chi_{b0}(1P)} < 2.4$~MeV, in ref.~\cite{Abdesselam:2016xbr},
which is also compatible with the result in eq.~\eqref{eq:chib01PLHth}.
Finally, our predictions for $\Gamma(\chi_{bJ}(nP) \rightarrow {\rm LH})$ support the hypothesis made in ref.~\cite{Han:2014kxa}
that the total widths of the $\chi_{bJ}(nP)$ states are approximately independent of the radial excitation $n$.
This hypothesis was then used to compute the feeddown contributions in the inclusive production cross sections of $\Upsilon(nS)$ from $\chi_{bJ}(3P)$ decays at the LHC.

\subsection[$\Upsilon(2S)$ and $\Upsilon(3S)$ decay into lepton pairs]
{\boldmath $\Upsilon(2S)$ and $\Upsilon(3S)$ decay into lepton pairs}
\label{sec:upsilon}
The NRQCD factorization formula for the decay width of a vector $S$-wave quarkonium state into a lepton pair at relative order $v^2$ is given by eq.~\eqref{factNRQCDSlep}.  
It depends on two LDMEs: $\langle V_Q(nS)|{\cal O}_1^{\rm em} ({}^3S_1)|V_Q(nS)\rangle$, whose factorization in strongly coupled pNRQCD at relative order $(\Lambda_{\rm QCD}/m)^2$ is in~\eqref{Oem3S1}, 
and $\langle V_Q(nS)|{\cal P}_1^{\rm em}(^3S_1)|V_Q(nS)\rangle$, whose factorization in strongly coupled pNRQCD at leading order is in~\eqref{P3S1}.

At relative order $(\Lambda_{\rm QCD}/m)^2$ the matrix element $\langle V_Q(nS)|{\cal O}_1^{\rm em} ({}^3S_1)|V_Q(nS)\rangle$ depends, 
besides on ${\cal E}_1$ and ${\cal E}_3$, also on a correlator involving four chromoelectric fields and a correlator involving two chromomagnetic fields. 
In this section, also to avoid dealing with correlators about which practically nothing is known,
we will explore the leptonic decays of the bottomonium states $\Upsilon(2S)$ and $\Upsilon(3S)$ assuming that these states satisfy the kinematical condition $m_bv \gg \Lambda_{\rm QCD} \gg m_bv^2$.
Under this condition the matrix element $\langle \Upsilon(nS)|{\cal O}_1^{\rm em} ({}^3S_1)|\Upsilon(nS)\rangle$ for $n=2$, $3$ can be written in strongly coupled pNRQCD at relative order $v^2$ as
\begin{equation}
\langle \Upsilon(nS)|{\cal O}_1^{\rm em} ({}^3S_1)|\Upsilon(nS)\rangle = \frac{N_c}{2\pi}  |R_{n101}(0)|^2 \left[ 1 - \frac{\varepsilon_{n0}^{(0)}}{m_b} \frac{2 {\cal E}_3}{9} \right],
\label{O1em3S1semistrong}
\end{equation}
as all other contributions in eq.~\eqref{Oem3S1} are of relative order  $(\Lambda_{\rm QCD}/m)^2$ and, therefore, suppressed. 

Under the assumption that the states $\Upsilon(2S)$ and $\Upsilon(3S)$ satisfy the condition $m_bv$ $\gg \Lambda_{\rm QCD}$ $\gg m_bv^2$, 
their decay width into a lepton pair can be written up to relative order $v^2$ in strongly coupled pNRQCD as
\begin{equation}
\label{eq:upsdecay}
\Gamma(\Upsilon(nS) \rightarrow e^+ e^-) = \frac{8 \pi e_b^2 \alpha^2}{3 M^2_{\Upsilon(nS)}} \,\frac{N_c}{2 \pi} |R_{n101}(0)|^2 
\left[ 1 - 2 C_F \frac{\alpha_{\rm s}}{\pi}- \frac{\varepsilon_{n0}^{(0)}}{3 M_{\Upsilon(nS)}} - \frac{\varepsilon_{n0}^{(0)}}{M_{\Upsilon(nS)}} \frac{2 {\cal E}_3}{9}  \right]^2,
\end{equation}
where $e_b=-1/3$. 
We have neglected corrections of order $\Lambda_{\rm QCD}^2/m^2$ compared to order $v^2$ corrections
and used the expressions of the short distance coefficients given in eqs.~\eqref{Imf3S1ee} and~\eqref{Img3S1ee}.
For the short distance coefficient in~\eqref{Img3S1ee} we only use the leading order expression.\footnote{
The next-to-leading order expression of ${\rm Im}\, g_{ee}({}^3S_1)$ contributes at relative order $\alpha_{\rm s}v^2$, which is beyond our accuracy. 
It is worth noting, however, that the next-to-leading order expression of ${\rm Im}\, g_{ee}({}^3S_1)$ depends on the cutoff $\mu_\Lambda$
and that this dependence cancels against the $\mu_\Lambda$ dependence of ${\cal E}_3$ in the expression of the dilepton decay width of $S$-wave quarkonia:
$$
{\rm Im}\, f_{ee}({}^3S_1) \left(-\frac{\varepsilon_{n0}^{(0)}}{m} \frac{2 {\cal E}_3}{9} \right)
+ {\rm Im}\, g_{ee}({}^3S_1) \frac{\varepsilon_{n0}^{(0)}}{m}
\sim
- \frac{\varepsilon_{n0}^{(0)}}{m}  \frac{2}{9} \left( 12 C_F \frac{\alpha_{\rm s}}{\pi} \log \frac{\mu_\Lambda}{m} \right)
-\frac{4}{3} \left( -C_F \frac{\alpha_{\rm s}}{\pi} 2 \log \frac{\mu_\Lambda}{m} \right)\frac{\varepsilon_{n0}^{(0)}}{m}
= 0.
$$
\label{footgee}}  
As done before for all electromagnetic processes, the formula follows from the
factorization at the amplitude level, see discussion in section~\ref{sec:e1e2}.
Finally, in eq.~\eqref{eq:upsdecay} we have expressed the bottom mass in terms of the $\Upsilon(nS)$ mass, $M_{\Upsilon(nS)}$, 
according to $M_{\Upsilon(nS)} = 2 m_b + \varepsilon_{n0}^{(0)}$, which is valid up to order $v^2$, 
and expanded in the leading order binding energy, $\varepsilon_{n0}^{(0)}$, up to relative order $v^2$.

Hence, under the above assumptions, the decay widths $\Gamma(\Upsilon(nS) \rightarrow e^+ e^-)$ for $n=2$, $3$  
depend on the $\Upsilon(nS)$ wavefunctions at the origin, the binding energies and the chromoelectric correlator ${\cal E}_3$.
The wavefunctions at the origin and the binding energies have been computed in the potential models of section~\ref{sec:potentials} and are listed in Table~\ref{tab:sbpotential}.
The correlator ${\cal E}_3$ has been computed in the previous section from the decays of $P$-wave charmonia and its value at 1~GeV is given in eq.~\eqref{eq:e3result}. 

\begin{table}[ht]
\centering
\begin{tabular}{|c|c|c|c|c|c|}
\hline
Potential model & A & B & C & D & E \\
\hline
$\Gamma_{\Upsilon(2S)}^{e^+ e^-}$~(keV) & 
$0.91{}^{+0.1}_{-0.1}$ & $0.46{}^{+0.05}_{-0.05}$ & $0.55{}^{+0.06}_{-0.06}$ & $0.54{}^{+0.06}_{-0.06}$ & $0.69{}^{+0.07}_{-0.07}$ \\ 
$\Gamma_{\Upsilon(3S)}^{e^+ e^-}$~(keV) & 
$0.57{}^{+0.06}_{-0.06}$ & $0.30{}^{+0.03}_{-0.03}$ & $0.35{}^{+0.04}_{-0.04}$ & $0.35{}^{+0.04}_{-0.04}$ & $0.44{}^{+0.05}_{-0.05}$ \\ 
\hline
\end{tabular}
\caption{\label{tab:upspotential} 
Results for the leptonic decay widths of the states $\Upsilon(2S)$ and $\Upsilon(3S)$, indicated with $\Gamma_{\Upsilon(2S)}^{e^+e^-}$ and $\Gamma_{\Upsilon(3S)}^{e^+e^-}$ for short. 
Wavefunctions at the origin and binding energies have been computed within the potential models of section~\ref{sec:potentials}.}
\end{table}

We take $\alpha = 1/131$ and compute $\alpha_{\rm s}$ at the scale of the meson mass, 
which gives $\alpha_{\rm s}(M_{\Upsilon(2S)}) = 0.177$ for the $2S$ state and $\alpha_{\rm s}(M_{\Upsilon(3S)}) = 0.176$ for the $3S$ state. 
Similarly to what we have done for $\Gamma(\chi_{bJ}(nP) \to {\rm LH})$, we compute ${\cal E}_3 (\mu_\Lambda)$ at the scale $\mu_\Lambda = M_{\Upsilon(nS)}$
using the expression at leading logarithmic accuracy given in eq.~\eqref{eq:e3evolution}, which, at the current level of accuracy,
is equivalent to resumming the leading logarithms of $\mu_\Lambda/M_{\Upsilon(nS)}$ in the short distance coefficients (in this case, the short distance coefficient~\eqref{Img3S1ee}).
The obtained leptonic widths of the bottomonium states $\Upsilon(2S)$ and $\Upsilon(3S)$ for the different potential model
determinations of the wavefunctions at the origin and binding energies are shown in Table~\ref{tab:upspotential}. 
The uncertainties are computed combining the uncertainties coming from uncalculated order $v^2$ corrections, estimated to be 0.1 times the central values, 
with the uncertainties coming from the neglected corrections of higher orders in $\alpha_{\rm s}$, estimated to be $\alpha_{\rm s}^2$ of the central values,  and with the uncertainty of ${\cal E}_3$.
The uncertainties are added in quadrature.

The present experimental values of the $\Upsilon(2S)$ and $\Upsilon(3S)$ leptonic decay widths are~\cite{Tanabashi:2018oca} 
\begin{eqnarray}
\Gamma(\Upsilon(2S) \rightarrow e^+e^-)\big|_{\rm PDG} &=& 0.612 \pm 0.011 \textrm{~keV}\,,
\label{Gamma2Seeexp}\\
\Gamma(\Upsilon(3S) \rightarrow e^+e^-)\big|_{\rm PDG} &=& 0.443 \pm 0.008 \textrm{~keV}\,.
\label{Gamma3Seeexp}
\end{eqnarray}
Few remarks concerning the determinations in Table~\ref{tab:upspotential}.
First, we recall that the central value of $\Gamma(\Upsilon(3S) \rightarrow e^+e^-)$ in model E coincides with the measurement, 
because the parameters of model E have been chosen to precisely reproduce it.
Second, even though the parameters of model D have been determined to reproduce the measured leptonic width of the $\Upsilon(3S)$, 
the model does not reproduce the measured rate when eq.~\eqref{eq:upsdecay} is used, 
because the contribution from ${\cal E}_3$ was not included in ref.~\cite{Chung:2010vz}. 
Taking the averages over the five determinations in Table~\ref{tab:upspotential} gives 
\begin{eqnarray}
\Gamma(\Upsilon(2S) \rightarrow e^+e^-) &=& 0.63{}^{+0.28}_{-0.17} {}^{+0.07}_{-0.07} \textrm{~keV}, 
\label{Gamma2Seethfinal}\\
\Gamma(\Upsilon(3S) \rightarrow e^+e^-) &=& 0.40{}^{+0.17}_{-0.11} {}^{+0.04}_{-0.05} \textrm{~keV}, 
\label{Gamma3Seethfinal}
\end{eqnarray}
where the first uncertainties are from the potential model dependence, and the second ones are the averages of the uncertainties in Table~\ref{tab:upspotential}.
The theoretical determinations \eqref{Gamma2Seethfinal} and \eqref{Gamma3Seethfinal} agree well, within uncertainties, with the data~\eqref{Gamma2Seeexp} and~\eqref{Gamma3Seeexp}.

\begin{table}[ht]
\centering
\begin{tabular}{|c|c|c|c|c|c|}
\hline
Potential model & A & B & C & D & E \\
\hline
$\langle {\cal O}_1^{\rm em} ({}^3S_1)\rangle_{\Upsilon(2S)}$~(GeV$^3$) & $2.45{}^{+0.25}_{-0.25}$ & $1.24{}^{+0.13}_{-0.13}$ & $1.46{}^{+0.15}_{-0.15}$ & $1.48{}^{+0.15}_{-0.15}$ & $1.88{}^{+0.19}_{-0.19}$ \\ 
$\langle {\cal P}_1^{\rm em} ({}^3S_1)\rangle_{\Upsilon(2S)}$~(GeV$^5$) & $5.72{}^{+0.57}_{-0.57}$ & $3.21{}^{+0.32}_{-0.32}$ & $2.00{}^{+0.20}_{-0.20}$ & $3.96{}^{+0.40}_{-0.40}$ & $4.54{}^{+0.45}_{-0.45}$ \\ 
\hline
$\langle {\cal O}_1^{\rm em} ({}^3S_1)\rangle_{\Upsilon(3S)}$~(GeV$^3$) & $1.70{}^{+0.18}_{-0.18}$ & $0.89{}^{+0.09}_{-0.1}$ & $1.03{}^{+0.10}_{-0.11}$ & $1.05{}^{+0.11}_{-0.11}$ & $1.32{}^{+0.14}_{-0.14}$ \\ 
$\langle {\cal P}_1^{\rm em} ({}^3S_1)\rangle_{\Upsilon(3S)}$~(GeV$^5$) & $8.09{}^{+0.81}_{-0.81}$ & $4.42{}^{+0.44}_{-0.44}$ & $3.63{}^{+0.36}_{-0.36}$ & $5.44{}^{+0.54}_{-0.54}$ & $6.30{}^{+0.63}_{-0.63}$ \\ 
\hline
\end{tabular}
\caption{\label{tab:upsmatrixelements}
Results for the matrix elements $\langle \Upsilon(nS)|{\cal O}_1^{\rm em} ({}^3S_1)|\Upsilon(nS)\rangle$ and $\langle \Upsilon(nS)|$ ${\cal P}_1^{\rm em}(^3S_1)$ $|\Upsilon(nS)\rangle$ at the scale $\mu_\Lambda = M_{\Upsilon(nS)}$,
indicated with $\langle {\cal O}_1^{\rm em} ({}^3S_1)\rangle_{\Upsilon(nS)}$ and $\langle {\cal P}_1^{\rm em} ({}^3S_1)\rangle_{\Upsilon(nS)}$ for short. 
Wavefunctions at the origin and binding energies have been computed within the
potential models of section~\ref{sec:potentials}.}
\end{table}

From eq.~\eqref{O1em3S1semistrong} we can compute the LDME $\langle \Upsilon(nS)|{\cal O}_1^{\rm em} ({}^3S_1)|\Upsilon(nS)\rangle$ 
(which is equal to $\langle \Upsilon(nS)|{\cal O}_1({}^3S_1)|\Upsilon(nS)\rangle$ at relative order $v^2$ and under the assumed kinematical conditions) 
and from eq.~\eqref{P3S1} $\langle \Upsilon(nS)|$ ${\cal P}_1^{\rm em}(^3S_1)$ $|\Upsilon(nS)\rangle$ (which is equal to $\langle \Upsilon(nS)|$ ${\cal P}_1(^3S_1)$ $|\Upsilon(nS)\rangle$ at leading order in $v$) 
for $n=2$, $3$, by using the determination of the correlator ${\cal E}_3$  at the scale $\mu_\Lambda = M_{\Upsilon(nS)}$,
and the potential model results for the wavefunctions at the origin and the binding energies.
The results are shown in Table~\ref{tab:upsmatrixelements}. 
The theoretical uncertainties from uncalculated corrections of higher orders in
$v$ are taken to be $0.1$ times the central values. 
In $\langle \Upsilon(nS)|{\cal O}_1^{\rm em} ({}^3S_1)|\Upsilon(nS)\rangle$,  the uncertainty from ${\cal E}_3$ is also included.
The uncertainties are added in quadrature. 
The averages of the determinations in Table~\ref{tab:upsmatrixelements} read
\begin{eqnarray}
\langle \Upsilon(2S)|{\cal O}_1^{\rm em} ({}^3S_1)|\Upsilon(2S)\rangle  = 1.70{}^{+0.75}_{-0.46} {}^{+0.17}_{-0.17} \textrm{~GeV}^3,
\\
\langle \Upsilon(2S)| {\cal P}_1^{\rm em}(^3S_1) |\Upsilon(2S)\rangle = 3.88{}^{+1.83}_{-1.89} {}^{+0.39}_{-0.39} \textrm{~GeV}^5,
\\
\nonumber\\
\langle \Upsilon(3S)|{\cal O}_1^{\rm em} ({}^3S_1)|\Upsilon(3S)\rangle = 1.20{}^{+0.50}_{-0.31} {}^{+0.12}_{-0.13} \textrm{~GeV}^3,
\\
\langle \Upsilon(3S)| {\cal P}_1^{\rm em}(^3S_1) |\Upsilon(3S)\rangle = 5.58{}^{+2.51}_{-1.95} {}^{+0.56}_{-0.56} \textrm{~GeV}^5,
\end{eqnarray}
where the first uncertainties are from the potential model dependence, and the second ones are the averages of the uncertainties in Table~\ref{tab:upspotential}.
The matrix elements are evaluated at the scale $\mu_\Lambda = M_{\Upsilon(nS)}$.

Under the same kinematical conditions considered above, we can also compute the inclusive decay widths of the $\Upsilon(2S)$ and $\Upsilon(3S)$ states into light hadrons.
The NRQCD factorization formula valid at relative order $v^2$ is eq.~\eqref{factNRQCDShad}.
It depends on the LDMEs: $\langle \Upsilon(nS)|{\cal O}_1({}^3S_1)|\Upsilon(nS)\rangle$ and $\langle \Upsilon(nS)|{\cal P}_1(^3S_1)|\Upsilon(nS)\rangle$ for $n=2$, $3$.
At relative order $v^2$ and under the condition $m_bv$ $\gg \Lambda_{\rm QCD}$ $\gg m_bv^2$, 
from the comparison of eq.~\eqref{O3S1} with eq.~\eqref{Oem3S1} it follows that $\langle \Upsilon(nS)|{\cal O}_1({}^3S_1)|\Upsilon(nS)\rangle$  is equal 
to $\langle \Upsilon(nS)|{\cal O}_1^{\rm em}({}^3S_1)|\Upsilon(nS)\rangle$. 
Moreover, $\langle \Upsilon(nS)|{\cal P}_1(^3S_1)|\Upsilon(nS)\rangle$ is equal to  $\langle \Upsilon(nS)|{\cal P}_1^{\rm em}(^3S_1)|\Upsilon(nS)\rangle$ 
at leading order in the velocity and $\Lambda_{\rm QCD}/m$ expansion.
Hence, by using the same strongly coupled pNRQCD factorization formulas for LDMEs employed above, 
we can write at relative order $v^2$ and neglecting corrections of order $\Lambda_{\rm QCD}^2/m^2$ 
\begin{eqnarray}
\Gamma(\Upsilon(nS)\rightarrow {\rm LH}) &=& \frac{N_c}{2 \pi} |R_{n101}(0)|^2 \left[ 
8\frac{{\rm Im} f_1 ({}^3S_1)}{M^2_{\Upsilon(nS)}}
 \left( 1 + \frac{2 \varepsilon_{n0}^{(0)}}{M_{\Upsilon(nS)}}- \frac{2 \varepsilon_{n0}^{(0)}}{M_{\Upsilon(nS)}} \frac{2 {\cal E}_3}{9} \right) \right.
\nonumber\\
&& \hspace{2.5cm}
\left. + 8\frac{{\rm Im} g_1 ({}^3S_1)}{M^2_{\Upsilon(nS)}} \frac{2 \varepsilon_{n0}^{(0)}}{M_{\Upsilon(nS)}} \right]. 
\label{eq:upsdecayLH}
\end{eqnarray}
The expressions for the short distance coefficients are in eq.~\eqref{Imf13S1} and eq.~\eqref{Img13S1}, 
and we have again expressed the bottom mass in terms of the $\Upsilon(nS)$ mass and the corresponding binding energy up to relative order $v^2$.\footnote{
Using the same reasoning of footnote~\ref{footgee}, from requiring that the right-hand side of eq.~\eqref{eq:upsdecayLH} is independent of the factorization scale $\mu_\Lambda$,
it follows that ${\rm Im} g_1 ({}^3S_1)$ must develop a $\mu_\Lambda$ dependence at order $\alpha_{\rm s}^4$ that exactly cancels the one in ${\cal E}_3$.
The $\mu_\Lambda$ dependent part of ${\rm Im} g_1 ({}^3S_1)$ at order $\alpha_{\rm s}^4$ then reads
$$
{\rm Im} g_1 ({}^3S_1)\big|_{\mu_\Lambda} = \frac{16}{27}(\pi^2-9)C_F^2(N_c^2-4)\left(\frac{N_c}{2}-C_F\right)^2\frac{\alpha_{\rm s}^4}{\pi}\log \frac{\mu_\Lambda}{m}.
$$
}  

If we consider the ratios of the leptonic decay widths, eq.~\eqref{eq:upsdecay}, 
with the corresponding decay widths into light hadrons, eq.~\eqref{eq:upsdecayLH}, the wavefunction dependence drops out
and the ratio depends only on the binding energies, whose potential model dependent values are in Table~\ref{tab:sbpotential}, 
and on the chromoelectric field correlator ${\cal E}_3$, which has been determined in eq.~\eqref{eq:e3result} 
and whose running at leading logarithmic accuracy is described by eq.~\eqref{eq:e3evolution}.
Furthermore, if we expand the ratio in powers of $v^2$, the dependence on ${\cal E}_3$ also drops out, and we obtain an expression that depends only on the binding energy.
To relative order $v^2$ accuracy, it reads
\begin{equation}
\label{eq:upsratio_th}
\frac{\Gamma(\Upsilon(nS) \rightarrow e^+ e^-)}{\Gamma(\Upsilon(nS)\rightarrow {\rm LH})} = 
\frac{{\rm Im} f_{ee} (^3S_1)}{{\rm Im} f_1  (^3S_1)}
+ \left( \frac{{\rm Im} g_{ee} (^3S_1)}{{\rm Im} f_1 (^3S_1)} - \frac{{\rm Im} g_1 (^3S_1) \;  {\rm Im} f_{ee} (^3S_1)}{({\rm Im} f_1 (^3S_1))^2}\right) \frac{2 \varepsilon_{n0}^{(0)}}{M_{\Upsilon(nS)}},
\end{equation}
where ${\rm Im} f_{ee} (^3S_1)$ and ${\rm Im} g_{ee} (^3S_1)$ can be read off eqs.~\eqref{Imf3S1ee} and~\eqref{Img3S1ee}, respectively. 
If we compute this ratio for the states $\Upsilon(2S)$ and $\Upsilon(3S)$,
the correction of relative order $v^2$ coming from the term proportional to the binding energy in eq.~\eqref{eq:upsratio_th} is almost half of the leading order contribution.
This may question the reliability of this expression to get accurate determinations of the ratios $\Gamma(\Upsilon(nS) \rightarrow e^+ e^-)/\Gamma(\Upsilon(nS)\rightarrow {\rm LH})$.
If we consider, instead, the ratio~\cite{Brambilla:2002nu}
\begin{eqnarray}
\label{eq:upsratio2_th}
R_{\Upsilon(2S)/\Upsilon(3S)} &\equiv& \frac{\Gamma(\Upsilon(2S) \rightarrow e^+ e^-)/\Gamma(\Upsilon(2S)\rightarrow {\rm LH})} {\Gamma(\Upsilon(3S) \rightarrow e^+ e^-)/\Gamma(\Upsilon(3S)\rightarrow {\rm LH})} 
\nonumber \\
&=& 
1 + \left( \frac{{\rm Im} g_{ee} (^3S_1)}{{\rm Im} f_{ee} (^3S_1)}-\frac{{\rm Im} g_{1} (^3S_1)}{{\rm Im} f_1 (^3S_1)} \right) 
\left( \frac{2 \varepsilon_{20}^{(0)}}{M_{\Upsilon(2S)}} -\frac{2 \varepsilon_{30}^{(0)}}{M_{\Upsilon(3S)}} \right),
\end{eqnarray}
we get an expression that is valid up to relative order $v^2$ and whose order $v^2$ correction is better under control. 
The results for this ratio are listed in Table~\ref{tab:BrbSwave} for each potential model determination of the binding energies.
The uncertainties are computed, as in the case of the leptonic widths,  
combining the uncertainties coming from uncalculated order $v^2$ corrections, estimated to be 0.1 times the central values, 
with the uncertainties coming from the neglected corrections of higher orders
in $\alpha_{\rm s}$, estimated to be $\alpha_{\rm s}^2$ of the central values. 
The uncertainties are combined in quadrature.
The number of flavors is taken to be $n_f=4$. 

\begin{table}[ht]
\centering
\begin{tabular}{|c|c|c|c|c|c|}
\hline
Potential model \!\!\! & A & B & C & D & E \\
\hline
$\!\!\!\!R_{\Upsilon(2S)/\Upsilon(3S)}\!\!\!\!$ & $\!\! 0.670 \pm 0.070 \!\!$ & $\!\! 0.672 \pm 0.070 \!\!$ & $\!\! 0.684 \pm 0.072\!\!$ & $\!\! 0.686 \pm 0.072\!\!$ & $\!\! 0.674 \pm 0.071$ \\
\hline
\end{tabular}
\caption{\label{tab:BrbSwave}
  Results for the ratio $R_{\Upsilon(2S)/\Upsilon(3S)}$, where the binding
energies are computed within the potential models of section~\ref{sec:potentials}.}
\end{table}

The average over the potential models in Table~\ref{tab:BrbSwave} gives 
\begin{eqnarray}
\label{eq:upsratio_averages}
R_{\Upsilon(2S)/\Upsilon(3S)} = 0.677{}^{+0.007}_{-0.007} {}^{+0.071}_{-0.071}\,, 
\end{eqnarray}
where the first uncertainties are from the potential model dependence, and the second ones are the averages of the uncertainties in Table~\ref{tab:BrbSwave}.
Since $R_{\Upsilon(2S)/\Upsilon(3S)}$ is one at leading order in $v$, the order $v^2$ correction amounts to about one third of the leading order contribution;
hence, the order $v^2$ correction in $R_{\Upsilon(2S)/\Upsilon(3S)}$ is in better control compared to the one in the ratios $\Gamma(\Upsilon(nS) \rightarrow e^+ e^-)/\Gamma(\Upsilon(nS)\rightarrow {\rm LH})$. 
In order to compare the result in eq.~\eqref{eq:upsratio_averages} with measurements,
we compute $\Gamma(\Upsilon(2S) \rightarrow {\rm LH})$ and $\Gamma(\Upsilon(3S) \rightarrow {\rm LH})$ by subtracting the radiative decay widths and the transition widths into other bottomonia 
from the total $\Upsilon(2S)$ and $\Upsilon(3S)$ widths given in ref.~\cite{Tanabashi:2018oca}; 
also the decay widths into $e^+ e^-$ are taken from ref.~\cite{Tanabashi:2018oca}. 
We obtain the following experimental value 
\begin{equation}
R_{\Upsilon(2S)/\Upsilon(3S)} \big|_{\text{from PDG}} = 0.761 ^{+0.110}_{-0.100}\,. 
\end{equation}
Compared to eq.~\eqref{eq:upsratio_averages}, the theoretical result is compatible with the experimental value within uncertainties. 
The result \eqref{eq:upsratio_averages} improves a more qualitative determination that can be found in ref.~\cite{Brambilla:2002nu}.

Other quarkonium $S$-wave vector states are the $\psi(2S)$, the  $\Upsilon(1S)$ and the $\Upsilon(4S)$.
In the present analysis, we have not considered the states $\psi(2S)$ and $\Upsilon(4S)$, because they are very close or above the open flavor threshold, respectively.
Effects due to degrees of freedom that are relevant above or close to the open flavor threshold have not been included in the pNRQCD Hamiltonian~\eqref{HpNRQCD}.
Hence, the effective field theory as formulated in this work is not suited to treat quarkonia like the $\psi(2S)$ and $\Upsilon(4S)$.
In this analysis, we have not considered the $\Upsilon(1S)$ too, because the hierarchy $mv^2 \ll \Lambda_{\rm QCD}$ is unlikely to be realized for this state,
which is commonly treated assuming  $mv^2 \gtrsim \Lambda_{\rm QCD}$ (see, for instance, refs.~\cite{Brambilla:2004wf,Brambilla:2010cs,Pineda:2011dg}). 
Hence, the only $S$-wave quarkonia that satisfy possibly the condition $mv$ $\gg \Lambda_{\rm QCD}$ $\gg mv^2$ within the effective field theory~\eqref{HpNRQCD} are the $\Upsilon(2S)$ and $\Upsilon(3S)$.
If they really realize this kinematical condition may be eventually established only at the hand of phenomenological analyses of the kind presented here.

\section{Summary and conclusion} 
\label{sec:summary}
In the paper, we have computed decay widths and exclusive electromagnetic production cross sections of charmonia and bottomonia based on strongly coupled pNRQCD.
In strong\-ly coupled pNRQCD, nonperturbative LDMEs are expressed in terms of quarkonium wavefunctions, binding energies and gluonic correlators.
Wavefunctions and binding energies are the solutions of the equation of motion of pNRQCD.
The gluonic correlators are nonperturbative parameters, which are independent of the quarkonium state and of the flavor of the heavy quark.
Owing to the universal nature of the gluonic correlators,  
the number of nonperturbative unknowns needed in pNRQCD to describe decay widths and exclusive electromagnetic production cross sections of charmonia and bottomonia
is smaller than the number of LDMEs needed in NRQCD, as they depend on the quarkonium state and on the flavor of the heavy quark.
This enables specific predictions in pNRQCD that are not possible in NRQCD.
Since strongly coupled pNRQCD is suited to describe non Coulombic quarkonium,
we have restricted our applications to charmonium $1P$ states and bottomonium $2S$, $3S$, $1P$, $2P$ and $3P$ states, which are possibly non Coulombic bound states.

The calculation of the NRQCD LDMEs in strongly coupled pNRQCD was first done in refs.~\cite{Brambilla:2001xy,Brambilla:2002nu}.
We have computed new corrections to $P$-wave LDMEs and revised the computation of $S$-wave LDMEs by adding some new contributions proportional to $\Lambda_{\rm QCD}^2/m^2$.
The newly computed corrections are expressed in terms of gluonic correlators.
Our results for $P$-wave and $S$-wave LDMEs in the strongly coupled pNRQCD factorization are listed in sections~\ref{sec:pwave} and~\ref{sec:swave}, respectively.
The LDMEs that we have computed satisfy the Gremm--Kapustin relations, as discussed in section~\ref{sec:GK}.

We have applied strongly coupled pNRQCD for the first time to exclusive electromagnetic production cross sections.
In particular, we have computed the cross sections $e^+e^- \to \chi_{cJ}(1P) + \gamma$ in section~\ref{sec:e1e2} 
and $e^+e^- \to \chi_{bJ}(nP) + \gamma$ for $n=1$, $2$ and $3$ in section~\ref{sec:Pdecayproduction}.
Although straightforward in the case of exclusive electromagnetic production, the application of strongly coupled pNRQCD has never been attempted before for any production process.
This has enabled us to make first and so far unique predictions for exclusive electromagnetic production cross sections of bottomonia.
Furthermore, in section~\ref{sec:e1e2} we have computed the decay widths $\chi_{c0}(1P) \to \gamma \gamma$ and $\chi_{c2}(1P) \to \gamma \gamma$,
in section~\ref{sec:Pdecayproduction} the decay widths $\chi_{b0}(nP) \to \gamma \gamma$ and $\chi_{b2}(nP) \to \gamma \gamma$ for $n=1$, $2$ and $3$
and in section~\ref{sec:upsilon}, under some assumptions, the dilepton decay widths $\Upsilon(2S) \to e^+e^-$ and $\Upsilon(3S) \to e^+e^-$.
We have also considered hadronic annihilations. 
In section~\ref{sec:e3}, we have computed the inclusive annihilation widths into light hadrons (LH) of charmonium spin triplet $1P$ states and bottomonium spin triplet $1P$, $2P$ and $3P$ states.
Finally, in section~\ref{sec:upsilon} we have also computed the particular ratio of decay widths
$[\Gamma(\Upsilon(2S) \rightarrow e^+ e^-)\Gamma(\Upsilon(3S)\rightarrow {\rm LH})]/[\Gamma(\Upsilon(2S)\rightarrow {\rm LH})\Gamma(\Upsilon(3S) \rightarrow e^+ e^-)]$
that involves the inclusive annihilation widths into light hadrons of the $2S$ and $3S$ bottomonium states.

From the theoretical side, our expressions are generally accurate up to relative order $v^2$ in the velocity expansion,
with the exception of the $P$-wave inclusive annihilation widths into light hadrons, where we have truncated our expansions in $v$ at leading order. 
The results that we obtain are in agreement, within errors, with experimental data, when available.
The determinations of the $e^+e^- \to \chi_{c0}(1P) + \gamma$, $e^+e^- \to \chi_{c2}(1P) + \gamma$,  $e^+e^- \to \chi_{bJ}(nP) + \gamma$  cross sections,
and of the $\chi_{b0}(nP) \to \gamma \gamma$, $\chi_{b2}(nP) \to \gamma \gamma$ and $\chi_b(nP)\rightarrow {\rm LH}$ decay widths for $n=1$, $2$ and $3$ are predictions.
These predictions were made possible by the universal nature of the potential and gluonic correlators that determine the LDMEs in strongly coupled pNRQCD. 
The gluonic correlators can be computed, in principle, in lattice QCD.
However, since lattice QCD determinations of the gluonic correlators are not available yet, we have determined them from the available data on decay and production of charmonia
and used to compute bottomonium observables.
This procedure should be contrasted with NRQCD, where one cannot, in general, infer the bottomonium LDMEs from the charmonium ones.

Our results rely on potential models for determining the quarkonium wavefunctions at the origin and the binding energies.
We rely on potential models because a first principle determination of the wavefunctions and binding energies from the equation of motion of pNRQCD
is hindered by the poor or incomplete knowledge of the corrections to the
wavefunction at higher orders in $v$ stemming from $1/m$ suppressed terms in the potential.
Difficulties come from the limited accuracy in our knowledge of the potential beyond the static term, 
and also from the renormalization of the divergences in the wavefunction at the origin due to the potential at short distances.
Concerning the $1/m$ corrections to the potential, we remark that, although these corrections can be expressed in terms of Wilson loops and gauge field strength insertions on them,
not all of them have been computed in lattice QCD and with enough precision.
Indeed, besides the static potential, no $1/m$ suppressed correction to the quarkonium potential has been computed in full (unquenched)~QCD.

The lack of a reliable determination of the quarkonium wavefunctions at the origin and the binding energies reflects in the wide spread of potential model results presented
and discussed in section~\ref{sec:potentials}.
As the quarkonium wavefunction enters most quarkonium observables at leading order,
the poor knowledge of the quarkonium wavefunction is the main source of uncertainty for most observables computed in NRQCD,
and, in particular, for the observables computed in this work that are not ratios of suitably chosen decay widths.
Parametrically, the uncertainty in the wavefunction affects these observables at least at relative order $v^2$ assuming a perfect knowledge of the leading order potential.
It can be argued that this is the case if the leading order potential coincides with the static potential;
the relative order $v^2$ uncertainty stems, then, from the $1/m$ suppressed terms in the potential that have not been included neither in an accurate nor in a complete form.
The uncertainty may be larger, however, if the term $V^{(1)}/m$ contributes at leading order to the potential (see the discussion in section~\ref{sec:stcopNRQCD}).

The poor knowledge of the quarkonium wavefunctions at the origin and the binding energies is the main limitation in the phenomenological analyses done in the present (and similar) works.
It could be overcome by improving our knowledge of the quarkonium potential, ideally via lattice QCD,
but also by using all available short distance and long distance information on the potential in a comprehensive analysis.
Wavefunctions at the origin and binding energies could be determined, in principle, also from a global fit of quarkonium observables versus data.
The fit would then determine these parameters together with the field strength correlators encoding the universal non perturbative parts of the LDMEs.
As we discussed in section~\ref{sec:e1e2}, for the set of observables considered in the present work, it is not possible, even in principle, 
to disentangle the wavefunction at the origin from all field strength correlators.
It remains an open and interesting question to answer if an enlarged set of observables may be able to solve the problem and fix on the data all non perturbative parameters.

Finally, we note that the strategy used in this work to compute the 
nonperturbative LDMEs could be possibly applied to study inclusive hadroproduction of heavy quarkonia too. 
This may improve our understanding of the inclusive production mechanism of heavy quarkonium that remains elusive to this day. 
In particular, the pNRQCD calculations of color-octet LDMEs could lead to a reduction of the number of nonperturbative unknowns. 

\acknowledgments
A.~V. thanks Antonio Pineda and Joan Soto for discussions.
The research of N.~B. is supported by the DFG Grant No. BR 4058/2-2.
N.~B., A.~V and H.~S.~C. acknowledge support from the Deutsche Forschungsgemeinschaft (DFG, German Research Foundation) cluster of excellence ``ORIGINS''
under Germany's Excellence Strategy - EXC-2094 - 390783311.
H.~S.~C. also acknowledges support from the Alexander von Humboldt Foundation.

\appendix 

\section{Four-fermion operators}
\label{app4fermion}
In the following, we list the four-fermion operators in the NRQCD Lagrangian relevant for the present analysis.
Extensive lists of four-fermion operators can be found in~\cite{Brambilla:2006ph,Brambilla:2008zg,Berwein:2018fos}.
The specified quantum numbers identify the state on which the operator projects dominantly.
The electromagnetic (em) operators are
\begin{eqnarray} 
{\cal O}_1^{\rm em}(^1S_0) &=& \psi^\dagger \chi |{\rm vac}\rangle \langle {\rm vac}| \chi^\dagger \psi,
\label{EM1S0}\\
{\cal O}_1^{\rm em}(^3S_1) &=& \psi^\dagger \bm{\sigma} \chi |{\rm vac}\rangle \cdot \langle {\rm vac}|\chi^\dagger \bm{\sigma} \psi,
\label{EM3S1}\\
{\cal O}_1^{\rm em}({}^3P_{0}) &=&  \frac{1}{3} \; \psi^\dagger  \left(- \frac{i}{2} \overleftrightarrow{\bm{D}} \cdot \bm{\sigma}\right) \chi |{\rm vac}\rangle
                          \langle {\rm vac}|\chi^\dagger  \left(- \frac{i}{2} \overleftrightarrow{\bm{D}} \cdot \bm{\sigma}\right) \psi,
\label{EM3P0}\\
{\cal O}_1^{\rm em}({}^3P_{1}) &=&  \frac{1}{2} \;\psi^\dagger   \left(- \frac{i}{2} \overleftrightarrow{\bm{D}} \times \bm{\sigma}\right) \chi |{\rm vac}\rangle  \cdot
\langle {\rm vac}| \chi^\dagger \left(- \frac{i}{2} \overleftrightarrow{\bm{D}} \times \bm{\sigma}\right) \psi,
\label{EM3P1} \\
{\cal O}_1^{\rm em}({}^3P_{2}) &=& \psi^\dagger \left(- \frac{i}{2} \overleftrightarrow{D}^{(i} \sigma^{j)}\right) \chi|{\rm vac}\rangle
\langle {\rm vac}|\chi^\dagger  \left(- \frac{i}{2} \overleftrightarrow{D}^{(i} \sigma^{j)}\right)  \psi,
\label{EM3P2}\\ 
{\cal P}_1^{\rm em}(^1S_0) &=& \frac{1}{2} \psi^\dagger \chi |{\rm vac}\rangle 
  \langle {\rm vac}|\chi^\dagger \left(- \frac{i}{2} \overleftrightarrow{\bm{D}}\right)^2 \psi + {\rm H.c.},
\label{EMP1S0}\\
{\cal P}_1^{\rm em}(^3S_1) &=& \frac{1}{2} \psi^\dagger \bm{\sigma} \chi |{\rm vac}\rangle \cdot
  \langle {\rm vac}|\chi^\dagger \bm{\sigma} \left(- \frac{i}{2} \overleftrightarrow{\bm{D}}\right)^2 \psi + {\rm H.c.},
\label{EMP3S1}\\
\mathcal P_1^{\rm em}(^3P_0) &=& \frac{1}{6} \psi^{\dag} \left( -\frac{i}{2}\overleftrightarrow{\bm{D}} \cdot \bm{\sigma} \right) 
\left(-\frac{i}{2}\overleftrightarrow{\bm{D}}\right)^2 \chi |{\rm vac} \rangle 
\langle {\rm vac}|\chi^{\dag}\left(-\frac{i}{2}\overleftrightarrow{\bm{D}} \cdot \bm{\sigma}\right)\psi 
+ \textrm{H.c.},
\label{EMP3P0}\\
\mathcal P_1^{\rm em}(^3P_1) &=& \frac{1}{4} \psi^{\dag} \left( -\frac{i}{2}\overleftrightarrow{\bm{D}} \times \bm{\sigma} \right) 
\left(-\frac{i}{2}\overleftrightarrow{\bm{D}}\right)^2 \chi |{\rm vac} \rangle \cdot 
\langle {\rm vac}| \chi^{\dag}\left(-\frac{i}{2}\overleftrightarrow{\bm{D}} \times \bm{\sigma}\right)\psi 
+ \textrm{H.c.},
\nonumber\\
\label{EMP3P1}\\
\mathcal P_1^{\rm em}(^3P_2) &=& \frac{1}{2}\, \psi^{\dag}\left(-\frac{i}{2} \overleftrightarrow{D}^{(i} \sigma^{j)}\right)
\left(- \frac{i}{2} \overleftrightarrow{\bm{D}} \right)^2 \chi |{\rm vac} \rangle 
\langle {\rm vac}| \chi^{\dag}\left(-\frac{i}{2}\overleftrightarrow{D}^{(i}\sigma^{j)}\right)\psi 
                                   + \textrm{H.c.},
\nonumber\\
\label{EMP3P2}\\
\mathcal{T}_{8}^{\rm em}(^3 P_0) &=&  \frac{1}{3} \psi^\dagger   \left(- i g \bm{E} \cdot \bm{\sigma}\right) \chi |{\rm vac}\rangle 
\langle {\rm vac}| \chi^\dagger \left(- \frac{i}{2} \overleftrightarrow{\bm{D}} \cdot \bm{\sigma}\right) \psi  + {\rm H.c.},
\label{T8em3P0}\\
\mathcal{T}_{8}^{\rm em}(^3 P_1) &=&  \frac{1}{2} \psi^\dagger   \left(- i g \bm{E} \times \bm{\sigma}\right) \chi |{\rm vac}\rangle \cdot 
\langle {\rm vac}| \chi^\dagger \left(- \frac{i}{2} \overleftrightarrow{\bm{D}} \times \bm{\sigma}\right) \psi  + {\rm H.c.},
\label{T8em3P1}\\
\mathcal{T}_{8}^{\rm em}(^3 P_2) &=&  \psi^\dagger   \left(- i g E^{(i} \sigma^{j)} \right) \chi |{\rm vac}\rangle 
\langle {\rm vac}| \chi^\dagger \left(- \frac{i}{2} \overleftrightarrow{D}^{(i} \sigma^{j)}\right) \psi  + {\rm H.c.},
\label{T8em3P2}
\end{eqnarray} 
where the fields are defined as in section~\ref{sec:stcoNRQCD}, 
$|{\rm vac}\rangle \langle {\rm vac}|$ projects on the QCD vacuum state, $T^{(ij)} \equiv (T^{ij}+T^{ji})/2 - T^{kk}\delta^{ij}/3$, and H.c. stands for Hermitian conjugate.
The subscript ``1'' labels four-fermion operators that project dominantly on a color singlet heavy quark-antiquark pair,
whereas the subscript ``8'' labels four-fermion operators that project dominantly on a color octet heavy quark-antiquark pair.
The relevant four-fermion hadronic operators are 
\begin{eqnarray}
{\cal O}_1({}^1S_0) &=& \psi^\dagger \chi \, \chi^\dagger \psi ,
\label{1S0} \\
{\cal O}_1({}^3S_1) &=& \psi^\dagger \bm{\sigma} \chi \cdot \chi^\dagger \bm{\sigma} \psi ,
\label{3S1} \\
{\cal O}_8({}^1S_0) &=& \psi^\dagger T^a \chi \, \chi^\dagger T^a \psi ,
\label{Oct1S0} \\
{\cal O}_8({}^3S_1) &=& \psi^\dagger \bm{\sigma} T^a \chi \cdot\chi^\dagger \bm{\sigma} T^a \psi,
\label{Oct3S1}\\
{\cal O}_1({}^3P_{0}) &=&  \frac{1}{3} \;
                    \psi^\dagger  \left(- \frac{i}{2} \overleftrightarrow{\bm{D}} \cdot \bm{\sigma}\right) \chi \, 
\chi^\dagger  \left(- \frac{i}{2} \overleftrightarrow{\bm{D}} \cdot \bm{\sigma}\right) \psi,
\label{3P0} \\
{\cal O}_1({}^3P_{1}) &=&  \frac{1}{2}
                   \;\psi^\dagger   \left(- \frac{i}{2} \overleftrightarrow{\bm{D}} \times \bm{\sigma}\right) \chi \cdot 
\chi^\dagger \left(- \frac{i}{2} \overleftrightarrow{\bm{D}} \times \bm{\sigma}\right) \psi,
\label{3P1} \\
{\cal O}_1({}^3P_{2}) &=& \psi^\dagger \left(- \frac{i}{2} \overleftrightarrow{D}^{(i} \sigma^{j)}\right) \chi \, 
\chi^\dagger  \left(- \frac{i}{2} \overleftrightarrow{D}^{(i} \sigma^{j)}\right)  \psi,
\label{3P2} \\ 
{\cal P}_1(^1S_0) &=& \frac{1}{2} \psi^\dagger \chi \chi^\dagger \left(- \frac{i}{2} \overleftrightarrow{\bm{D}}\right)^2 \psi + {\rm H.c.},
\label{PP1S0}\\
{\cal P}_1(^3S_1) &=& \frac{1}{2} \psi^\dagger \bm{\sigma} \chi \cdot \chi^\dagger \bm{\sigma} \left(- \frac{i}{2} \overleftrightarrow{\bm{D}}\right)^2 \psi + {\rm H.c.},
\label{PP3S1}\\
{\cal O}_8({}^1P_1) &=& \psi^\dagger \left(-\frac{i}{2} \overleftrightarrow{\bm{D}} \right) T^a\chi \cdot \chi^\dagger ( -\frac{i}{2}  \overleftrightarrow{\bm{D}}) T^a\psi ,
\label{Oct1P1}\\
{\cal O}_8({}^3P_{0}) &=&  \frac{1}{3} \;\psi^\dagger  \left(- \frac{i}{2} \overleftrightarrow{\bm{D}} \cdot \bm{\sigma}\right) T^a\chi \, 
\chi^\dagger  \left(- \frac{i}{2} \overleftrightarrow{\bm{D}} \cdot \bm{\sigma}\right) T^a\psi,
\label{Oct3P0}\\
{\cal O}_8({}^3P_{1}) &=&  \frac{1}{2}\;\psi^\dagger   \left(- \frac{i}{2} \overleftrightarrow{\bm{D}} \times \bm{\sigma}\right) T^a\chi \cdot 
\chi^\dagger \left(- \frac{i}{2} \overleftrightarrow{\bm{D}} \times \bm{\sigma}\right) T^a\psi,
\label{Oct3P1}\\
{\cal O}_8({}^3P_{2}) &=& \psi^\dagger \left(- \frac{i}{2} \overleftrightarrow{D}^{(i} \sigma^{j)}\right) T^a\chi \, 
\chi^\dagger  \left(- \frac{i}{2} \overleftrightarrow{D}^{(i} \sigma^{j)}\right)  T^a\psi.
\label{Oct3P2} 
\end{eqnarray}

\section{NRQCD factorization formulas}
\label{app:NRQCDfact}
In this appendix, we list the NRQCD factorization formulas for quarkonium decay widths and exclusive electromagnetic production cross sections
for which we provide the strongly coupled pNRQCD factorization formulas in the main body of the paper.
We list first the decay formulas and conclude with the electromagnetic production ones.

Leptonic decay widths of $S$-wave quarkonium vector states are described in NRQCD up to relative order $v^2$ by two LDMEs~\cite{Bodwin:1994jh}:
\begin{eqnarray}
  \Gamma(V_Q (nS) \rightarrow e^+e^-) &=& \frac{2\,{\rm Im\,}f_{ee}(^3 S_1)}{m^2}\ \langle V_Q(nS)|{\cal O}_1^{\rm em}(^3S_1)|V_Q(nS)\rangle
\nonumber\\
                                    && + \frac{2\,{\rm Im\,}g_{ee}(^3 S_1)}{m^4} \langle V_Q(nS)|{\cal P}_1^{\rm em}(^3S_1)|V_Q(nS)\rangle.
\label{factNRQCDSlep}  
\end{eqnarray}
The short distance coefficients, which are the imaginary parts of the coefficients of the corresponding four fermion operators in the NRQCD Lagrangian,
read at next-to-leading order~\cite{Barbieri:1975ki,Celmaster:1978yz,Luke:1997ys, Bodwin:2008vp}
(the next-to-next-to-leading order correction to ${\rm Im\,}f_{ee}(^3 S_1)$ has been computed in refs.~\cite{Czarnecki:1997vz,Beneke:1997jm}, 
and the next-to-next-to-next-to leading order correction to ${\rm Im\,}f_{ee}(^3 S_1)$ has been computed in ref.~\cite{Marquard:2014pea}) 
\begin{eqnarray}
{\rm Im\,}f_{ee}(^3 S_1) &=& \frac{\pi e_Q^2 \alpha^2}{3} \left( 1 - 4 C_F \frac{\alpha_{\rm s}}{\pi} \right),
\label{Imf3S1ee}  \\
{\rm Im\,}g_{ee}(^3S_1) &=& - \frac{4}{9} \pi e_Q^2 \alpha^2 
\left[ 1 - C_F \frac{\alpha_s}{\pi} 
\left( \frac{23}{6} + 2 \log \frac{\mu_\Lambda}{m} \right)
\right],
\label{Img3S1ee}
\end{eqnarray}
where $e_Q$ is the fractional electric charge of a heavy quark of flavor $Q$, $\alpha$ is the fine structure constant,
$\alpha_{\rm s}$ is the strong coupling in the $\overline{\rm MS}$ scheme 
and $C_F=(N_c^2-1)/(2N_c)$ ($=4/3$ in QCD) is the Casimir of the fundamental representation of SU$(N_c)$.
The infrared cutoff $\mu_\Lambda$ arises from renormalization of the short
distance coefficients in NRQCD. 
In here and in the following, 
the short distance coefficients are renormalized in
the $\overline{\rm MS}$ scheme.

Two photon decay widths of spin one and $J=0,2$ $P$-wave quarkonia are described in NRQCD up to relative order $v^2$ by the factorization formulas~\cite{Ma:2002eva,Brambilla:2006ph}:
\begin{eqnarray}
\Gamma (\chi_{QJ}(nP) \rightarrow \gamma \gamma) &=& \frac{2 \, \textrm{Im} \, f_{\textrm{em}} (^3 P_J)}{m^4} \langle \chi_{QJ}(nP)| \mathcal{O}_{1}^{\rm em} (^3 P_J)| \chi_{QJ}(nP) \rangle
  \nonumber\\
 && + \frac{2 \, \textrm{Im} \, g_{\textrm{em}} (^3 P_J)}{m^6} \langle \chi_{QJ}(nP) | \mathcal{P}_{1}^{\rm em} (^3 P_J)| \chi_{QJ}(nP) \rangle
  \nonumber \\
&& + \frac{2 \, \textrm{Im} \, t_{8 \, \textrm{em}} (^3 P_J)}{m^5}  \langle \chi_{QJ}(nP)| \mathcal{T}_{8 \, \textrm{em}} (^3 P_J)| \chi_{QJ}(nP) \rangle.
   \label{factNRQCDPem}
\end{eqnarray}
The short distance coefficients read 
\begin{eqnarray}
\textrm{Im} \, f_{\textrm{em}} (^3 P_0) &=& 3 \alpha^2 e_Q^4 \pi \left( 1 + \frac{3\pi^2-28}{12} C_F \frac{\alpha_{\rm s}}{\pi} \right), \\
\textrm{Im} \, f_{\textrm{em}} (^3 P_2) &=& \frac{4}{5} \alpha^2 e_Q^4 \pi \left( 1 - 4 C_F \frac{\alpha_{\rm s}}{\pi} \right), \\
\textrm{Im} \, g_{\textrm{em}} (^3 P_0) &=& - 7 \alpha^2 e_Q^4 \pi,\\
\textrm{Im} \, g_{\textrm{em}} (^3 P_2) &=& -\frac{8}{5} \alpha^2 e_Q^4 \pi, \\
\textrm{Im} \, t_{8\, \textrm{em}} (^3 P_0) &=& -\frac{3}{2} \alpha^2 e_Q^4 \pi, \\
\textrm{Im} \, t_{8\, \textrm{em}} (^3 P_2) & =& 0.
\end{eqnarray}
The coefficients $\textrm{Im} \, f_{\textrm{em}} (^3 P_J)$ for $J=0$, $2$ have been computed at order $\alpha_{\rm s}$ in refs.~\cite{Barbieri:1975am,Barbieri:1980yp},  
and the coefficients $\textrm{Im} \, g_{\textrm{em}} (^3 P_J)$ and $\textrm{Im} \, t_{8\, \textrm{em}} (^3 P_J)$ for $J=0$, $2$ have been computed at leading order in ref.~\cite{Brambilla:2006ph}.

Inclusive decay widths into light hadrons (LH) of $S$-wave quarkonium vector states are described in NRQCD up to relative order $v^2$ by the factorization formula~\cite{Brambilla:2002nu}:
\begin{eqnarray}
\Gamma(V_Q (nS) \rightarrow {\rm LH}) &=&
 \frac{2\,{\rm Im\,}f_1(^3 S_1)}{m^2} \langle V_Q(nS)|{\cal O}_1(^3S_1)|V_Q(nS)\rangle
\nonumber\\
&&\hspace{-2.5cm}
+ \frac{2\,{\rm Im\,}f_8(^3 S_1)}{m^2} \langle V_Q(nS)|{\cal O}_8(^3S_1)|V_Q(nS)\rangle
+ \frac{2\,{\rm Im\,}f_8(^1 S_0)}{m^2} \langle V_Q(nS)|{\cal O}_8(^1S_0)|V_Q(nS)\rangle 
\nonumber\\
&&\hspace{-2.5cm}
+ \frac{2\,{\rm Im\,}g_1(^3 S_1)}{m^4}\langle V_Q(nS)|{\cal P}_1(^3S_1)|V_Q(nS)\rangle
+ \frac{2\,{\rm Im\,}f_8(^3 P_0)}{m^4}\langle V_Q(nS)|{\cal O}_8(^3P_0)|V_Q(nS)\rangle
\nonumber\\
&&\hspace{-2.5cm}
+ \frac{2\,{\rm Im\,}f_8(^3 P_1)}{m^4}\langle V_Q(nS)|{\cal O}_8(^3P_1)|V_Q(nS)\rangle
+ \frac{2\,{\rm Im\,}f_8(^3 P_2)}{m^4}\langle V_Q(nS)|{\cal O}_8(^3P_2)|V_Q(nS)\rangle
\nonumber\\
&&\hspace{-2.5cm}
\approx \frac{2\,{\rm Im\,}f_1(^3 S_1)}{m^2} \langle V_Q(nS)|{\cal O}_1(^3S_1)|V_Q(nS)\rangle
+ \frac{2\,{\rm Im\,}g_1(^3 S_1)}{m^4}\langle V_Q(nS)|{\cal P}_1(^3S_1)|V_Q(nS)\rangle.
   \nonumber\\
\label{factNRQCDShad}  
\end{eqnarray}
The approximation in the last line holds when $mv \gg \Lambda_{\rm QCD} \gg mv^2$
and when neglecting, at relative order $v^2$, all contributions that scale like $(\Lambda_{\rm QCD}/m)^2$
(one should consider in the power counting of strongly coupled pNRQCD the expressions of the LDMEs in eqs.~\eqref{O3S1}, \eqref{P3S1}, \eqref{O83S1}, \eqref{O83PJ}
and the expression of $\langle V_Q(nS)|{\cal O}_8(^3S_1)|V_Q(nS)\rangle$ given in ref.~\cite{Brambilla:2002nu}).\footnote{
The last line also provides $\Gamma(V_Q (nS) \rightarrow {\rm LH})$ up to relative order $v^2$ in the power counting of ref.~\cite{Bodwin:1994jh}.}
The short distance coefficients appearing in the last line of \eqref{factNRQCDShad} read
\begin{eqnarray}
  \textrm{Im} \, f_1 (^3 S_1) &=& \frac{2}{9} (\pi^2-9) C_F (N_c^2-4) \left( \frac{N_c}{2} -C_F \right)^2 \alpha_{\rm s}(m)^3 
\nonumber\\
&& \hspace{3cm}
\times \left\{ 1 + \frac{\alpha_{\rm s}}{\pi}\left[ -9.46(2)\, C_F + 4.13(17)\, N_c - 1.161(2)\, n_f \right]\right\}
\nonumber \\
&& + \pi e_Q^2 \, \left( \sum_{k=1}^{n_f} e_{q_k}^2\right)\, \alpha^2 \left\{ 1 - \frac{13}{4} C_F \frac{\alpha_{\rm s}}{\pi}\right\}, 
\label{Imf13S1}\\
\textrm{Im} \, g_1(^3 S_1) &=&
   - \frac{19\pi^2 -132}{54} C_F (N_c^2 -4) \left(\frac{N_c}{2} -C_F \right)^2   \alpha_{\rm s}^3, 
\label{Img13S1}
\end{eqnarray}
where $e_{q_k}$ is the fractional electric charge of a massless quark of flavor $q_k$.
The expression of the  short distance coefficient $\textrm{Im} \, f_1 (^3 S_1)$, which is accurate up to order $\alpha_{\rm s}^4$ and  $\alpha^2\alpha_{\rm s}$, 
comes from refs.~\cite{Bodwin:1994jh,Mackenzie:1981sf},
and the expression of $\textrm{Im} \, g_1(^3 S_1)$, which is accurate at leading order, comes from ref.~\cite{Gremm:1997dq}.

Inclusive decay widths of spin one $P$-wave quarkonia into light hadrons are described in NRQCD at leading order in the velocity expansion 
by the factorization formula~\cite{Bodwin:1992ye}:
\begin{eqnarray}
\Gamma(\chi_{QJ}(nP)  \rightarrow {\rm LH}) &=& 
\frac{2\,{\rm Im \,}  f_1(^3P_J)}{m^4}  \langle \chi_{QJ}(nP) | {\cal O}_1(^3P_J ) | \chi_{QJ}(nP) \rangle
\nonumber\\
&&+ \frac{2\,{\rm Im \,}  f_8(^3S_1)}{m^2} \langle \chi_{QJ}(nP) | {\cal O}_8(^1S_0 ) | \chi_{QJ}(nP) \rangle.
\label{factNRQCDPhad}
\end{eqnarray}
The short distance coefficients ${\rm Im} f_1 ({}^3P_J)$ and ${\rm Im} f_8 ({}^3S_1)$ are known up to order $\alpha_{\rm s}^3$ accuracy~\cite{Petrelli:1997ge}: 
\begin{eqnarray}
2 \, {\rm Im} f_1 ({}^3P_0) &=& 
\frac{3 \pi C_F}{N_c} \alpha_{\rm s}(m)^2 \left\{ 1 + \frac{\alpha_{\rm s}}{\pi} \left[ C_F \left( - \frac{7}{3} +\frac{\pi^2}{4} \right)
+ N_c \left( \frac{427}{81}- \frac{\pi^2}{144} \right) \right.\right.
\nonumber \\
&& \hspace{3.3cm} - \beta_0 \log 2\bigg] \bigg\} + n_f \frac{4C_F}{9 N_c} \alpha_{\rm s}^3 \left( - \frac{29}{6}-\log \frac{\mu_\Lambda}{2 m} \right), 
\label{f13P0}\\
2 \, {\rm Im} f_1 ({}^3P_1) &=& 
\frac{C_F}{2} \alpha_{\rm s}^3 \left( \frac{587}{27} - \frac{317}{144} \pi^2 \right) 
+ \frac{4}{9} n_f \frac{C_F}{N_c} \alpha_{\rm s}^3 \left( - \frac{4}{3}-\log \frac{\mu_\Lambda}{2 m} \right), 
\label{f13P1}\\
2 \, {\rm Im} f_1 ({}^3P_2) &=&
\frac{4 \pi  C_F}{5 N_c} \alpha_{\rm s}(m)^2 \left\{ 1 + \frac{\alpha_{\rm s}}{\pi} \left[ -4 C_F
+ N_c \left( \frac{2185}{216} - \frac{337 \pi^2}{384} + \frac{5}{3} \log 2 \right) \right.\right.
\nonumber \\
&& \hspace{3.3cm}  - \beta_0 \log 2 \bigg] \bigg\} + n_f \frac{4C_F}{9N_c} \alpha_{\rm s}^3 \left( - \frac{29}{15}-\log \frac{\mu_\Lambda}{2 m} \right), 
\label{f13P2}\\
2 \, {\rm Im} f_8 ({}^3S_1) &=& 
\frac{\pi n_f}{3} \alpha_{\rm s}(m)^2 \left\{ 1+ \frac{\alpha_{\rm s}}{\pi} \left[ - \frac{13}{4} C_F 
+ N_c \left( \frac{133}{18} + \frac{2}{3} \log 2 - \frac{\pi^2}{4} \right) - \frac{10}{9} n_f T_F \right.\right. 
\nonumber \\
&& \hspace{3.3cm} - \beta_0 \log 2\bigg] \bigg\} + \frac{5}{3} \alpha_{\rm s}^3 \left( -\frac{73}{4} + \frac{67 \pi^2}{36} \right),
\label{f83S1}
\end{eqnarray}
where $\beta_0 = 11 N_c/3 - 4 T_F n_f/3$ and $\mu_\Lambda$ is an infrared cutoff.

Finally, we give the NRQCD factorization formula for the cross sections $\sigma(e^+ e^- \rightarrow \chi_{QJ}(nP) + \gamma)$ describing exclusive electroproduction of spin one $P$-wave quarkonia.
The formula is valid up to relative order $v^2$; it reads~\cite{Brambilla:2017kgw}
\begin{eqnarray}
&& \hspace{-1cm} \sigma(e^+ e^- \rightarrow \chi_{QJ}(nP) + \gamma)  = 
\sigma_{QJ}^{(0)}(m,s,r) \biggl \{ c_{J}^{(O_1)}(r) \langle \chi_{QJ}(nP)| \mathcal{O}_1 (^3 P_J)| \chi_{QJ}(nP)\rangle 
\nonumber \\
&& \hspace{-0.8cm} + \frac{c_{J}^{(P)}(r)}{m^2} \langle \chi_{QJ}(nP)| \mathcal{P}_1 (^3 P_J)|\chi_{QJ}(nP)\rangle 
+ \frac{c_{J}^{(T)}(r)}{m} \langle \chi_{QJ}(nP)| \mathcal{T}_8 (^3 P_J)| \chi_{QJ}(nP)\rangle  \biggr \},
\label{factNRQCDPempro}
\end{eqnarray}
where $r=4 m^2/s$ and $s$ is the square of the center of mass energy.
The factors $\sigma_{QJ}^{(0)}(m,s,r)$ are the production cross sections of color singlet heavy quark-antiquark pairs of flavor $Q$ 
in a $^3P_J$ state at leading order in $\alpha_{\rm s}$ and $v$.
They are given by~\cite{Chung:2008km}
\begin{eqnarray}
\sigma_{Q0}^{(0)}(m,s,r) &=& \frac{(4 \pi)^3 \alpha^3 e_Q^4 (1-3 r)^2}{18 \pi m^3 s^2 (1-r)}, \label{sigmaQ0}  \\
\sigma_{Q1}^{(0)}(m,s,r) &=& \frac{(4 \pi)^3 \alpha^3 e_Q^4 (1+r)}{3 \pi m^3 s^2 (1-r)}, \label{sigmaQ1}  \\
\sigma_{Q2}^{(0)}(m,s,r) &=& \frac{(4 \pi)^3 \alpha^3 e_Q^4 (1+3 r+6 r^2)}{9 \pi m^3 s^2 (1-r)}. \label{sigmaQ2} 
\end{eqnarray}
The short distance coefficients $c_{J}^{(O_1)}(r)$ have been computed at next-to-leading order in $\alpha_{\rm s}$ in refs.~\cite{Sang:2009jc,Li:2009ki}:
$c_{J}^{(O_1)}(r) = 1 +  c_{J}^{(O_1){\rm NLO}}(r) \, (\alpha_{\rm s}/\pi)$. The coefficients $c_{J}^{(O_1){\rm NLO}}(r)$ are of the form     
\begin{eqnarray}
c_{0}^{(O_1){\rm NLO}} (r) &=& C_0^0(r), \label{c0O1}\\
c_{1}^{(O_1){\rm NLO}} (r) &=& \frac{C_1^0(r)+r C_1^1(r)}{1+r}, \label{c1O1}\\
c_{2}^{(O_1){\rm NLO}} (r) &=& \frac{C_2^0(r)+3 r C_2^1(r)+6 r^2 C_2^2(r)}{1+3 r+6 r^2}, \label{c2O1}
\end{eqnarray}
where the explicit expressions of the coefficients $C_i^j(r)$ can be found in ref.~\cite{Sang:2009jc}.
The short distance coefficients $c_J^{(T)} (r)$ and $c_J^{(P)} (r)$ have been computed at leading order in $\alpha_{\rm s}$
in ref.~\cite{Brambilla:2017kgw} and in refs.~\cite{Li:2013nna,Chao:2013cca}, respectively. 
They read
\begin{eqnarray}
c_{0}^{(T)} (r) &=&  \frac{(3 - 2 r + 3 r^2)}{4 (1 - 4r + 3r^2)},\label{c0T}\\
c_{1}^{(T)} (r) &=& - \frac{(3+r^2)}{4 (1 - r^2)} ,\label{c1T}\\
c_{2}^{(T)} (r) &=&  - \frac{(3 + 2 r - 3 r^2 + 18 r^3)}{4 (1-r) (1 + 3 r + 6 r^2)},\label{c2T}\\
c_{0}^{(P)} (r) &=& - \frac{13 - 18 r + 25 r^2}{10 (1 - 4 r + 3 r^2)},\label{c0P}\\
c_{1}^{(P)} (r) &=& -\frac{11 - 20 r - 11 r^2}{10 (1-r^2)},\label{c1P}\\
c_{2}^{(P)} (r) &=& -\frac{7 + 6 r - 83 r^2 - 30 r^3}{10 (1-r) (1 + 3 r + 6 r^2)}. \label{c2P}
\end{eqnarray}
We observe that the cross sections are singular in the limit $r \to 1$, i.e., when the center of mass energy approaches the heavy quark-antiquark pair production threshold.

\section{Derivation of some LDMEs in strongly coupled pNRQCD} 
\label{sec:LDMEapp}
We give here few more details on the computation of the electromagnetic
$P$-wave LDMEs appearing in section~\ref{sec:pwave}.

\subsection[$\langle \chi_{QJ}(nP) | {\cal O}_1^{\rm em} ({}^3P_J) | \chi_{QJ}(nP) \rangle$]
{\boldmath $\langle \chi_{QJ}(nP) | {\cal O}_1^{\rm em} ({}^3P_J) | \chi_{QJ}(nP) \rangle$}
\label{sec:NRQCDME_O1}
We start by computing the matrix element 
$\displaystyle {}^{(0)}\langle\underline{\rm n};\bm{x}_{1},\bm{x}_{2}| \int d^3x \, {\cal O}_1^{\rm em} ({}^3P_J) (\bm{x}) |\underline{\rm k};\bm{x}_1',\bm{x}_2'\rangle^{(0)}$:
\begin{eqnarray}
&& 
{}^{(0)}\langle\underline{\rm n};\bm{x}_{1},\bm{x}_{2}| \int d^3x \, {\cal O}_1^{\rm em} ({}^3P_J) (\bm{x})|\underline{\rm k};\bm{x}_1',\bm{x}_2'\rangle^{(0)} 
\nonumber \\ && \hspace{1cm}
= - \frac{1}{4} T_{1J}^{ij}\, 
\int d^3x \; {}^{(0)}\langle n ;\bm{x}_{1},\bm{x}_{2}| \chi^\dag (\bm{x}_2) \psi(\bm{x}_1) \, (\psi^\dag \overleftrightarrow{D}^i \chi)(\bm{x})  | {\rm vac} \rangle 
\nonumber \\ && \hspace{6cm}
\times \langle {\rm vac}  | (\chi^\dag \overleftrightarrow{D}^j \psi)(\bm{x}) \psi^\dag (\bm{x}_1') \chi (\bm{x}_2') | k ;\bm{x}_1',\bm{x}_2'\rangle^{(0)}
\nonumber \\ && \hspace{1cm}
= - \frac{1}{4} T_{1J}^{ij} \,
\int d^3x \; {}^{(0)}\langle n ;\bm{x}_{1},\bm{x}_{2}| \delta^{(3)} (\bm{x}_1 - \bm{x}) \overleftrightarrow{D}^i (\bm{x}) 
\delta^{(3)} (\bm{x}_2 - \bm{x}) | {\rm vac}  \rangle 
\nonumber \\ && \hspace{6cm}
\times  \langle {\rm vac}  |  \delta^{(3)} (\bm{x}_2' - \bm{x}) \overleftrightarrow{D}^j (\bm{x}) \delta^{(3)} (\bm{x}_1' - \bm{x}) | k ;\bm{x}_1',\bm{x}_2'\rangle^{(0)} 
\nonumber \\ && \hspace{1cm}
= - \frac{1}{4} N_c T_{1J}^{ij} \,
{}^{(0)} \langle n | (D_1 - D_{c2} )^i | 0 \rangle^{(0)} \delta^{(3)} (\bm{r}) {}^{(0)} \langle 0 | (D_1 - D_{c2} )^j | k \rangle^{(0)} 
\nonumber \\ && \hspace{6cm}
\times \delta^{(3)} (\bm{x}_1 - \bm{x}_1') \delta^{(3)} (\bm{x}_2 - \bm{x}_2'),
\end{eqnarray}
where we have used eq.~\eqref{eq:gluonstate} for the first equality, Wick's theorem for the second one
and $| 0\rangle^{(0)} \delta^{(3)} (\bm{r}) = \mathbb{1}_c | {\rm vac} \rangle/\sqrt{N_c} \delta^{(3)} (\bm{r})$ for the third one.
The spin projectors $T_{1J}^{ij}$ are defined as
\begin{eqnarray}
T_{10}^{ij} &=& \frac{1}{3}\sigma^i \otimes \sigma^j,
\label{T10ij}\\
T_{11}^{ij} &=& \frac{1}{2}\epsilon_{kim}\epsilon_{kjn}\sigma^m \otimes \sigma^n,
\label{T11ij}\\
T_{12}^{ij} &=& \left(\frac{\delta_{im}\sigma^n + \delta_{in}\sigma^m}{2} -
\frac{\delta_{mn}}{3} \sigma^i\right)  \otimes \left(\frac{\delta_{jm}\sigma^n + \delta_{jn}\sigma^m}{2} - \frac{\delta_{mn}}{3} \sigma^j\right).
\label{T12ij}
\end{eqnarray}
For $n=k=0$ and from eqs.~\eqref{eq:Didentity}, it follows 
\begin{eqnarray}
&& 
{}^{(0)}\langle\underline{\rm 0};\bm{x}_{1},\bm{x}_{2}| \int d^3x \, {\cal O}_1^{\rm em} ({}^3P_J) (\bm{x}) |\underline{\rm 0};\bm{x}_1',\bm{x}_2'\rangle^{(0)}
\nonumber \\ && \hspace{1cm} = 
- \frac{1}{4} N_c T_{1J}^{ij}\, (\nabla_1 - \nabla_2)^i \delta^{(3)} (\bm{r}) (\nabla_1 - \nabla_2)^j \delta^{(3)} (\bm{x}_1 - \bm{x}_1') \delta^{(3)} (\bm{x}_2 - \bm{x}_2') 
\nonumber \\ && \hspace{1cm} =  
- N_c T_{1J}^{ij} \, \nabla_{\bm{r}}^i \delta^{(3)} (\bm{r}) \nabla_{\bm{r}}^j \delta^{(3)} (\bm{x}_1 - \bm{x}_1') \delta^{(3)} (\bm{x}_2 - \bm{x}_2').
\label{eq:o1matLO}
\end{eqnarray}
For $n \neq 0$, $k = 0$ and from eqs.~\eqref{eq:Didentities_neq}, it follows 
\begin{eqnarray}
&& 
{}^{(0)}\langle\underline{\rm n};\bm{x}_{1},\bm{x}_{2}| \int d^3x\, {\cal O}_1^{\rm em} ({}^3P_J) (\bm{x})|\underline{\rm 0};\bm{x}_1',\bm{x}_2'\rangle^{(0)}
\nonumber \\ && \hspace{1cm} = 
- N_c T_{1J}^{ij} \, \frac{{}^{(0)} \langle n | g E^i_1 | 0 \rangle^{(0)} }{E_n^{(0)} - E_0^{(0)}} \delta^{(3)} (\bm{r}) \nabla_{\bm{r}}^j 
\delta^{(3)} (\bm{x}_1 - \bm{x}_1') \delta^{(3)} (\bm{x}_2 - \bm{x}_2'), 
\label{eq:o1matLO2}
\end{eqnarray}
where we have also used the fact that 
${}^{(0)} \langle n | g \bm{E}_1 | 0 \rangle^{(0)}$ and ${}^{(0)} \langle n | g \bm{E}_2^T | 0 \rangle^{(0)}$ are equal at $\bm{r} = 0$. 
Finally, for completeness we give the result for $n \neq 0$ and $k \neq 0$, which reads
\begin{eqnarray}
&& \hspace{-5mm}
{}^{(0)}\langle\underline{\rm n};\bm{x}_{1},\bm{x}_{2}| \int d^3x\, {\cal O}_1^{\rm em} ({}^3P_J) (\bm{x}) |\underline{\rm k};\bm{x}_1',\bm{x}_2'\rangle^{(0)}
\nonumber \\ && \hspace{5mm} = 
- N_c T_{1J}^{ij} \, \frac{{}^{(0)} \langle n | g E^i_1 | 0 \rangle^{(0)} } {E_n^{(0)} - E_0^{(0)}} 
\delta^{(3)} (\bm{r}) \frac{{}^{(0)} \langle 0 | g E^j_1 | k \rangle^{(0)} } {E_0^{(0)} - E_k^{(0)}} 
\delta^{(3)} (\bm{x}_1 - \bm{x}_1') \delta^{(3)} (\bm{x}_2 - \bm{x}_2'). 
\end{eqnarray}

We also need the correction to the matrix element \eqref{eq:o1matLO} stemming from next-to-leading order in the quantum-mechanical expansion in $1/m$.
In particular, we need the part that projects on $P$ waves. It is given by
\begin{eqnarray}
\label{eq:o1matNLO}
&& 
{}_{P\text{-wave}}\hspace{-4mm}{}^{(1)}\langle\underline{\rm 0};\bm{x}_{1},\bm{x}_{2}|\int d^3x\, {\cal O}_1^{\rm em} ({}^3P_J) (\bm{x})|\underline{\rm 0};\bm{x}_1',\bm{x}_2'\rangle^{(0)} + {\rm H.c.}
\nonumber \\ && \hspace{1cm} =  
N_c T_{1J}^{ij} \, \sum_{k \neq 0} \left( \bm{\nabla}_1 \cdot \frac{{}^{(0)}\langle 0| g\bm{E}_1|k\rangle^{(0)}}{(E^{(0)}_{0}-E^{(0)}_{k})^2}
- \bm{\nabla}_2 \cdot \frac{{}^{(0)}\langle 0| g\bm{E}_2^T|k\rangle^{(0)}}{(E^{(0)}_{0}-E^{(0)}_{k})^2}\right)
\nonumber \\ && \hspace{4cm}
\times \frac{{}^{(0)} \langle k | g E^i_1 | 0 \rangle^{(0)}} {E_k^{(0)} - E_0^{(0)}} \delta^{(3)} (\bm{r}) 
\nabla_{\bm{r}}^j \delta^{(3)} (\bm{x}_1 - \bm{x}_1')\delta^{(3)} (\bm{x}_2 - \bm{x}_2') + {\rm H.c.} 
\nonumber \\ && \hspace{1cm} = 
- \frac{2}{3} N_c T_{1J}^{ij} \, \nabla_{\bm{r}}^i i {\cal E}_2\delta^{(3)} (\bm{r}) \nabla_{\bm{r}}^j \delta^{(3)} (\bm{x}_1 - \bm{x}_1')\delta^{(3)} (\bm{x}_2 - \bm{x}_2'),
\end{eqnarray}
where we have used eq.~\eqref{eq:state_pwave}, eq.~\eqref{eq:o1matLO2} and, in the last equality, the definition~\eqref{corr1}.

Plugging eqs.~\eqref{eq:o1matLO} and \eqref{eq:o1matNLO} into eq.~\eqref{matchinglocal} one gets $V^{(4)}_{{\cal O}_1^{\rm em} ({}^3P_J)}$:
\begin{equation}
V^{(4)}_{{\cal O}_1^{\rm em} ({}^3P_J)} = N_c T_{1J}^{ij} \, \nabla_{\bm{r}}^i \, \left(1 + \frac{2}{3}  \frac{i {\cal E}_2}{m}   \right) \, \delta^{(3)} (\bm{r}) \nabla_{\bm{r}}^j\,.
\label{V4O1em}
\end{equation}
In turn, $V^{(4)}_{{\cal O}_1^{\rm em} ({}^3P_J)}$ is the fundamental ingredient to obtain from eq.~\eqref{eq:master} 
the LDME $\langle \chi_{QJ}(nP) | {\cal O}_1^{\rm em} ({}^3P_J) | \chi_{QJ}(nP) \rangle$.
The result is in eq.~\eqref{eq:o1mat}. 
Comments on and implications of eq.~\eqref{eq:o1mat} are in the main body of
the paper in section~\ref{sec:pwave}.

\subsection[$\langle \chi_{QJ}(nP) | {\cal T}_8^{\rm em} ({}^3P_J) | \chi_{QJ}(nP) \rangle$]
{\boldmath $\langle \chi_{QJ}(nP) | {\cal T}_8^{\rm em} ({}^3P_J) | \chi_{QJ}(nP) \rangle$}
\label{sec:NRQCDME_T8}
We compute here the matrix element 
$\displaystyle \langle\underline{\rm 0};\bm{x}_{1},\bm{x}_{2}|\int d^3x \, {\cal T}_8^{\rm em}({}^3P_J) (\bm{x}) |\underline{\rm 0};\bm{x}_1',\bm{x}_2'\rangle$.
Differently from the matrix elements computed in the previous and in the next section, at leading order in the $1/m$ expansion this matrix element does not contribute to $P$-wave quarkonium states.
Hence, the first non vanishing contribution of the matrix element of the color octet operator ${\cal T}_8^{\rm em}({}^3P_J)$ on a $P$-wave state appears only at next-to-leading order in $1/m$. 

Indeed, at leading order the matrix element reads
\begin{eqnarray}
\label{eq:t8matLO}
&&
{}^{(0)}\langle\underline{\rm 0};\bm{x}_{1},\bm{x}_{2}|\int d^3x \, {\cal T}_8^{\rm em} ({}^3P_J) (\bm{x}) |\underline{\rm 0};\bm{x}_1',\bm{x}_2'\rangle^{(0)}
\nonumber \\ && \hspace{1cm}
= 2 N_c T_{1J}^{ij} \, \left( \nabla_{\bm{r}}^i E_0^{(0)} (\bm{x}_1, \bm{x}_2) \right) \delta^{(3)} (\bm{r}) \nabla_{\bm{r}}^j 
\delta^{(3)} (\bm{x}_1 - \bm{x}_1') \delta^{(3)} (\bm{x}_2 - \bm{x}_2'),
\end{eqnarray}
where we have used eqs.~\eqref{eq:Eidentity}. 
The function $\delta^{(3)} (\bm{r})$ brings the derivative of the static energy, $\bm{\nabla}_{\bm{r}} E_0^{(0)} (\bm{x}_1, \bm{x}_2)$, into the perturbative regime. 
At $\bm{r}=\bm{0}$ this vanishes in dimensional regularization, because it is scaleless. 
Moreover, eq.~\eqref{eq:t8matLO} defines through~\eqref{matchinglocal} a contact term $V^{(5)}_{{\cal T}_8^{\rm em} ({}^3P_J)}$ with only one derivative left to act on the wavefunctions.
This is not enough in eq.~\eqref{eq:master} to produce a non vanishing contribution for $P$-wave states. At least two derivatives are necessary.

At next-to-leading order in the quantum-mechanical $1/m$ expansion, the missing derivative comes from the $1/m$ correction to the state shown in eq.~\eqref{eq:state_pwave}:
\begin{eqnarray}
\label{eq:t8matNLO}
&& 
{}_{P\text{-wave}}\hspace{-4mm}{}^{(1)}\langle\underline{\rm 0};\bm{x}_{1},\bm{x}_{2}| \int d^3x \, {\cal T}_8^{\rm em} ({}^3P_J) (\bm{x}) |\underline{\rm 0};\bm{x}_1',\bm{x}_2'\rangle^{(0)}+ {\rm H.c.}
\nonumber \\ && \hspace{1cm}
= - \frac{4}{3} N_c {\cal T}^{ij}_{1J} \, \nabla_{\bm{r}}^i {\cal E}_1 \delta^{(3)} (\bm{r}) \nabla_{\bm{r}}^j \delta^{(3)} (\bm{x}_1 - \bm{x}_1') \delta^{(3)} (\bm{x}_2 - \bm{x}_2').
\end{eqnarray}
Plugging eq.~\eqref{eq:t8matNLO} into eq.~\eqref{matchinglocal} one gets a contact term $V^{(5)}_{{\cal T}_8^{\rm em} ({}^3P_J)}$ with two derivatives left to act on the wavefunctions:
\begin{equation}
V^{(5)}_{{\cal T}_8^{\rm em} ({}^3P_J)} = N_c T_{1J}^{ij} \, \nabla_{\bm{r}}^i \, \frac{4}{3} \frac{{\cal E}_1}{m} \, \delta^{(3)} (\bm{r}) \nabla_{\bm{r}}^j\,.
\label{V5T8em}
\end{equation}
From $V^{(5)}_{{\cal T}_8^{\rm em} ({}^3P_J)}$ and eq.~\eqref{eq:master} the expression of the octet LDME, $\langle \chi_{QJ}(nP) | {\cal T}_8^{\rm em} ({}^3P_J)$ $| \chi_{QJ}(nP) \rangle$, 
given in eq.~\eqref{eq:t8mat} follows.

\subsection[$\langle \chi_{QJ}(nP) | {\cal P}_1^{\rm em} ({}^3P_J) | \chi_{QJ}(nP) \rangle$]
{\boldmath $\langle \chi_{QJ}(nP) | {\cal P}_1^{\rm em} ({}^3P_J) | \chi_{QJ}(nP) \rangle$}
\label{sec:NRQCDME_P1}
Finally, we compute the matrix element 
$\displaystyle \langle\underline{\rm 0};\bm{x}_{1},\bm{x}_{2}| \int d^3x \, {\cal P}_1^{\rm em} ({}^3P_J) (\bm{x}) |\underline{\rm 0};\bm{x}_1',\bm{x}_2'\rangle$.
For our purposes, it is sufficient to consider it at leading order in the quantum-mechanical $1/m$ expansion:
\begin{eqnarray}
\label{eq:p1matLO}
&& 
{}^{(0)}\langle\underline{\rm 0};\bm{x}_{1},\bm{x}_{2}| \int d^3x \, {\cal P}_1^{\rm em} ({}^3P_J) (\bm{x}) |\underline{\rm 0};\bm{x}_1',\bm{x}_2'\rangle^{(0)}
\nonumber \\ && \hspace{1cm}
= N_c T_{1J}^{ij} \, \nabla_{\bm{r}}^i \delta^{(3)} (\bm{r}) \left( \bm{\nabla}_{\bm{r}}^2 + \frac{5}{3} {\cal E}_1 \right) \nabla_{\bm{r}}^j 
\delta^{(3)} (\bm{x}_1 - \bm{x}_1') \delta^{(3)} (\bm{x}_2 - \bm{x}_2').
\end{eqnarray}
The corresponding contact term from eq.~\eqref{matchinglocal} reads
\begin{equation}
V^{(6)}_{{\cal P}_1^{\rm em} ({}^3P_J)} = N_c T_{1J}^{ij} \, \nabla_{\bm{r}}^i \delta^{(3)} (\bm{r}) \, \left( - \bm{\nabla}_{\bm{r}}^2 - \frac{5}{3} {\cal E}_1 \right) \, \nabla_{\bm{r}}^j\,.
\label{V6P1em}
\end{equation}
From $V^{(6)}_{{\cal P}_1^{\rm em} ({}^3P_J)}$ and eq.~\eqref{eq:master} the expression of the LDME, 
$\langle \chi_{QJ}(nP) | {\cal P}_1^{\rm em} ({}^3P_J)$ $| \chi_{QJ}(nP) \rangle$, 
written in eq.~\eqref{eq:p1mat} follows if we replace the Laplacian acting on the wavefunction at the origin with the expression given in eq.~\eqref{eq:laplacian2}.
Note that the two derivatives in \eqref{V6P1em}, $\nabla_{\bm{r}}^i$ and $\nabla_{\bm{r}}^j$,  
need to act on the wavefunctions to ensure a non vanishing contribution for $P$-wave states to the right-hand side of eq.~\eqref{eq:master}.

\end{document}